\documentclass[11pt]{article}
\usepackage{amsthm}
\usepackage{amssymb}
\usepackage{amsmath}
\usepackage{eucal}
\usepackage{mathrsfs}
\usepackage[all]{xy}
\usepackage{graphicx}
\usepackage{bbm}

\newcommand{\M}{\mathcal{M}} 
\newcommand{\X}{\mathcal{X}} 
\newcommand{\g}{\mathrm{g}} 
\newcommand{\J}{\mathrm{J}}%
\newcommand{\D}{\mathrm{D}}
\newcommand{\I}{\mathrm{I}}%
\newcommand{\Po}{\mathcal{P}}
\newcommand{\C}{\mathcal{C}} 
\newcommand{\Mk}{\mathbb{M}^4} 
\newcommand{\dc}{\mathcal{O}} 
\newcommand{\K}{\mathcal{K}}  
\newcommand{\Kr}{\mathcal{K}_{\diamond}}  
\newcommand{\A}{\mathcal{A}} 
\newcommand{\Al}{\mathscr{A}} 
\newcommand{\si}{\sigma}

\newcommand{\ga}{\gamma}
\newcommand{\be}{\beta}

\newcommand{\eps}{\varepsilon}
\newcommand{\io}{\iota}
\newcommand{\al}{\alpha}
\newcommand{\la}{\lambda}
\newcommand{\Si}{\Sigma}

\newcommand {\con}[1]{\overline{#1}}
\newcommand{\Zu}{\mathcal{Z}^1(\Al_\K)}         
\newcommand{\Zut}{\mathcal{Z}^1_t(\Al_\K)}      
\newcommand{\Zutx}{\mathcal{Z}^1_t(\Al_{\K_x})} 
\newcommand{\Zutr}{\mathcal{Z}^1_t(\Al_{\Kr})} 
\newcommand{\Zup}{\mathcal{Z}^1(\Al_\Po)}       
\newcommand{\Zutp}{\mathcal{Z}^1_t(\Al_\Po)}    
\newcommand{\Bh}{\mathfrak{B}(\mathcal{H}_o)} 

\newcommand{\amp}{\mathsf{A}}   
\newcommand{\ctr}{\mathsf{C}}   

\newcommand{\f}{\mathrm{f}}
\newcommand{\F}{\mathrm{F}}
\newcommand{\RE}{\mathrm{R}}

\newcommand {\ordpos}{\trianglelefteq}

\newcommand {\defi}{\equiv}

\newcommand {\norm}[1]{\Vert{#1}\Vert}

\newcommand{\R}{\mathbb{R}} 

\newcommand{\Hil}{\mathcal{H}} 

\author{Giuseppe Ruzzi \\
\small{Dipartimento di Matematica, Universit\`a di Roma ``Tor Vergata'' }\\
\small{Via della Ricerca Scientifica, 00133, Roma,  Italy}  \\
\small{\texttt{ruzzi@mat.uniroma2.it}} 
\date{}}

\title{Homotopy of posets,  net-cohomology and 
       superselection sectors in 
       globally hyperbolic spacetimes}

\begin{document}

   \maketitle

\begin{abstract}
We study sharply localized sectors, known as sectors of DHR-type,   
of a net of local observables, in  arbitrary
globally hyperbolic spacetimes with dimension $\geq 3$. 
We show that these sectors define, has it happens in Minkowski space,   
a $\mathrm{C}^*-$category in which the charge structure 
manifests itself by the existence of a tensor product, 
a permutation symmetry and a conjugation. The mathematical framework 
is that of the net-cohomology of posets according to J.E. Roberts. 
The net of local observables is indexed by a poset formed by a basis 
for the topology of the spacetime ordered under inclusion.  
The category of sectors, is equivalent to the category of 
1-cocycles of the poset with values in the net. 
We succeed to  analyze  the structure of this category
because we  show how topological properties of the spacetime are encoded 
in the poset used as index set:
the first homotopy group of a poset is introduced and it is shown that   
the fundamental group of the poset and the one of the underlying 
spacetime are isomorphic;
any  1-cocycle defines a unitary representation 
of these fundamental groups. Another important result is the invariance 
of the net-cohomology under a suitable change of index set of the net. 
\end{abstract}

  \theoremstyle{plain}
  \newtheorem{df}{Definition}[section]
  \newtheorem{teo}[df]{Theorem}
  \newtheorem{prop}[df]{Proposition}
  \newtheorem{cor}[df]{Corollary}
  \newtheorem{lemma}[df]{Lemma}

  \theoremstyle{definition}
  \newtheorem{oss}[df]{Remark}

\tableofcontents
\markboth{Contents}{Contents}
\section{Introduction}
\label{Int}
The present paper is concerned with the study of charged 
superselection sectors in globally hyperbolic spacetimes in the
framework  of the algebraic approach to quantum field theory \cite{Ha,HK}. 
The basic object of this approach is the abstract net 
\textit{of local observables} $\mathscr{R}_\K$, namely the correspondence 
\[
\K\ni\dc\rightarrow\mathcal{R}(\dc),
\] 
which associates to any element $\dc$ of a family $\K$ of relatively compact
open regions  of the spacetime $\M$, considered as a fixed background 
manifold,  
the $\mathrm{C}^*-$algebra $\mathcal{R}(\dc)$ 
generated by all the observables which are measurable within 
$\dc$.  Sectors are unitary equivalence classes of 
irreducible representations of this net,  
the labels distinguishing different classes are the quantum numbers.
The study of physically meaningful sectors of the net of local 
observables, and  how to select them,  
is the realm of the theory of superselection sectors.\\
\indent One of the main results of superselection sectors theory  
has been the demonstration that in  Minkowski space $\Mk$, 
among the representations of the net of local observables 
it is possible to select a family of sectors 
whose quantum numbers manifest the same properties 
as  the charges carried by elementary particles: 
a \textit{composition law}, the \textit{alternative of Bose and Fermi 
statistics} and the \textit{charge conjugation 
symmetry}. The first example of sectors manifesting these properties
has been provided in \cite{DHR1,DHR2}, known as DHR-analysis, 
where the authors investigated sharply localized sectors.
Namely, a representation $\pi$ of $\mathscr{R}_\K$ is a sector of 
DHR-type whenever its restriction to the spacelike complement $\dc^\perp$ 
of any element $\dc$ of $\K$ is unitary equivalent to the vacuum 
representation $\pi_o$ of the net,  in symbols
\begin{equation}
\label{Int:1}
\pi|_{\mathcal{R}(\dc^\perp)} \cong \pi_o|_{\mathcal{R}(\dc^\perp)}, 
\qquad \dc\in\K.
\end{equation}
Although no known charge present in nature is sharply localized,
the importance of the DHR-analysis resides in the following reasons.
First, it suggests the idea that  
physically charged sectors might be localized in a more generalized sense 
with respect to (\ref{Int:1}). Secondly,
the introduction of  powerful  techniques 
based only on the causal structure of the Minkowski space that can 
be used  to investigate other types of localized sectors.
A relevant example are the BF-sectors \cite{BF} which describe 
charges in purely massive theories. BF-sectors  are localized 
in spacelike cones, which  are a family of noncompact regions of $\Mk$. 
In curved spacetimes the nontrivial topology 
can induce superselection sectors, see \cite{AS} and references 
quoted therein. 
However, up until  now, the localization properties of these sectors are 
not known, hence it is still  not possible to investigate 
their charge structure. \\
\indent In the present paper we deal 
with the study of sectors of DHR-type in arbitrary globally hyperbolic 
spacetimes. Because of the sharp localization, sectors of DHR-type 
should be insensitive to the nontrivial topology of the spacetime,
and their quantum numbers should exhibit the same features as in 
Minkowski space. 
However, the first investigations \cite{GLRV,Rob3} 
have provided  only partial results in this direction,  and, 
in particular, they have pointed out that for particular 
classes of spacetimes the topology might affect the properties of 
sectors of DHR-type.  The aim 
of the present paper is to show how this type of sectors   and 
the topology of spacetime are related and that they manifest the 
properties described above also in an 
arbitrary globally hyperbolic spacetime. 
We want to stress that the results of this paper are confined 
to spacetimes whose dimension  is $\geq 3$.\\[3pt]
\indent Before entering  the theory of DHR-sectors 
in globally hyperbolic spacetimes,  a key fact has still to be 
mentioned. The DHR-analysis can be equivalently               
read in terms of \textit{net-cohomology of posets}, a cohomological 
approach initiated and developed by J.E. Roberts                
\cite{Rob1}, (see also \cite{Rob2,Rob3} and references therein). 
Such an approach makes clear 
that the spacetime information which is 
relevant for the analysis of the DHR-sectors 
is  the topological and the causal structure
of Minkowski space (the Poincar\'e symmetry enters the theory 
only in the definition of the vacuum representation).
In particular, the essential point is how these two 
properties are encoded in 
the structure of the index $\K$  as a partially ordered set 
(\textit{poset}) with respect to \textit{inclusion} order relation 
$\subseteq$.  
Representations satisfying 
(\ref{Int:1}) are, up to equivalence, in 1-1 correspondence  with 
1-cocycles $z$ of the poset $\K$ with values 
in the vacuum representation of the net $\Al_\K: \dc\rightarrow \A(\dc)$. 
Here $\A(\dc)$ is the von Neumann 
algebra obtained by taking the bicommutant $\pi_o(\mathcal{R}(\dc))''$
of $\pi_o(\mathcal{R}(\dc))$.  
These 1-cocycles, which  are nothing but  the charge transporters 
of DHR-analysis,  define a tensor $\mathrm{C}^*-$category $\Zut$ 
with a permutation symmetry and conjugation.\\
\indent The first investigation     of sectors 
of DHR-type in a globally hyperbolic spacetime $\M$ 
has been done in \cite{GLRV}.  
First, the authors consider 
the net of local observables $\mathscr{R}_{\Kr}$ indexed by 
the set $\Kr$ of regular diamonds of $\M$: a family of relatively compact
open sets codifying  the topological and the causal 
properties of $\M$. Secondly,  they  take a reference representation 
$\pi_o$ of $\mathscr{R}_{\Kr}$ on a Hilbert space $\Hil_o$ such that
the net $\Al_{\Kr}:\Kr\ni\dc\rightarrow \A(\dc)\defi \pi_o(\mathcal{R}(\dc))''$ 
satisfies Haag duality and the Borchers property (see Section \ref{C}). 
The reference representation $\pi_o$  plays for the theory 
the same role that the vacuum representation plays in the case 
of Minkowski space. Examples of physically meaningful nets of local algebras 
indexed by  regular diamonds have been given in \cite{Ve}. 
Finally,  the DHR-sectors are singled  out from the representations 
of the net $\mathscr{R}_{\Kr}$ by  
generalizing, in a suitable way, the criterion (\ref{Int:1}).
As in  Minkowski space, 
the physical content of DHR-sectors  is contained in 
the $\mathrm{C}^*-$category $\Zutr$ of 1-cocycles 
of $\Kr$ with values in $\Al_{\Kr}$, and when $\Kr$ is directed under 
inclusion, there exist  a tensor product, a symmetry and 
conjugated 1-cocycles. The analogy with the theory in the 
Minkowski space  breaks down when the $\Kr$ is not directed.  
In this situation only the introduction of a tensor product 
on $\Zutr$ and the existence of a symmetry have been achieved 
in \cite{GLRV}, although the definition of the tensor product
is not completely discussed (see below).\\
\indent  There are two well known 
topological conditions on the spacetime, implying that not only regular
diamonds but any reasonable set of indices for a net of local algebras is 
not directed: this happens when the spacetime is either nonsimply connected 
or has compact Cauchy surfaces (Corollary  \ref{Ada:7}  and 
Lemma \ref{Ba:2}). There arises, therefore, the necessity 
to understand the connection between net-cohomology and topology 
of the underlying spacetime. Progress in this direction 
has been achieved in \cite{Rob3}. The  homotopy of paths,
the net-cohomology under a change   of the index
set  are issues developed in that work that will turn out to be fundamental 
for our aim. Moreover, it has been shown that the statistics 
of sectors can be classified provided that the net 
satisfies punctured Haag duality (see Section \ref{C}).
However, no result concerning the conjugation has been achieved.\\
\indent  To see what is the main drawback caused by 
the non directness of the poset $\Kr$, we have to describe more in detail 
$\Zutr$. The elements $z$ of $\Zutr$ are 1-cocycles trivial in $\Bh$ or,
equivalently, path-independent on $\Kr$. The latter 
means that the evaluation of $z$ on a path of $\Kr$ 
depends only on the endpoints of the path. When $\Kr$ is 
directed any 1-cocycle is trivial in 
$\Bh$, but this might not be hold when $\Kr$ is not directed. 
The consequences can be easily showed: 
let $\widehat{\otimes}$ be the tensor product  introduced in \cite{GLRV}: 
for any $z,z_1\in\Zutr$,  it turns out that $z\widehat{\otimes} z_1$
is a 1-cocycle of $\Kr$ with values in $\Al_{\Kr}$, but it is not clear
whether  it is trivial in $\Bh$ (we will see 
in Remark \ref{Cc:5a} that this 1-cocycle is  trivial in $\Bh$).
Now, we know that the nonsimply connectedness
and the compactness of the Cauchy surfaces are topological 
obstructions to the directness of the index sets. The first aim 
of this paper is to understand whether these conditions are also 
obstructions to the triviality in $\Bh$ of 1-cocycles. This
problem is analyzed in great generality in Section \ref{A}. 
We introduce the notions of the first homotopy group and  
fundamental group for an abstract poset $\Po$
(Definition \ref{Ab:12}) and prove that any 1-cocycle $z$ of $\Po$, 
with values in a net of local algebras $\Al_\Po$ indexed by $\Po$,
defines a unitary representation of the fundamental group of $\Po$
(Theorem \ref{Ac:4}). In the case that $\Po$ is a basis for  a
topological space ordered under inclusion,
and whose elements are arcwise and simply connected sets,
then the fundamental group of $\Po$ is isomorphic
to the fundamental group of the underlying topological space 
(Theorem \ref{Ada:6}). This states that  
\textit{the only possible topological obstruction
to the triviality in $\Bh$ of 1-cocycles is 
the nonsimply connectedness} (Corollary \ref{Adb:2}).\\
\indent  Before studying superselection sectors 
in a  globally hyperbolic spacetime $\M$,
we have to point out  another problem arising 
in \cite{GLRV, Rob3}. Regular diamonds do not need to have 
arcwise connected causal complements. This, on the one hand 
creates some technical problems; on the other hand 
it is not clear whether it is justified to assume Haag 
duality on $\Al_{\Kr}$: the only known result 
showing that a net of local observables, 
in the presence of a nontrivial superselection structure, 
inherits Haag duality from fields makes use of  the arcwise connectedness 
of causal complements of the elements of the index set
\cite[Theorem 3.15]{GLRV}.
We start to deal with this problem 
 by showing that net-cohomology 
 is invariant under  a change of the index set
 (Theorem \ref{Adb:4}), provided the new index set  
 is a refinement  of $\Kr$ (see Definition \ref{Aca:1} and 
 Lemma \ref{Adb:3}a).
 In Section \ref{Bb} we introduce the set $\K$ of diamonds 
 of $\M$. $\K$ is a refinement of $\Kr$  and 
 any element of $\K$ has an arcwise connected causal complement.  
 Therefore adopting $\K$ as index set the cited problems
 are overcome.\\
\indent In Section \ref{C}, we consider an irreducible  net $\Al_\K$ 
satisfying the Borchers property and punctured Haag duality.
The key for studying superselection sectors of the net  $\Al_\K$,
namely the $\mathrm{C}^*-$category $\Zut$, 
is provided by the following fact. 
We introduce  the  causal puncture $\K_x$ of $\K$
induced by a point $x$ of $\M$ (\ref{Bba:1}) and 
consider the categories $\Zutx$ of 1-cocycles of
$\K_x$, trivial in $\Bh$, with values in the net $\Al_{\K_x}$. 
We show that  a  family $z_x\in\Zutx$ 
for any $x\in\M$ admits an extension 
to a 1-cocycle $z\in\Zut$ if, and only if, a suitable 
gluing condition is verified (Proposition \ref{Ca:2}).
A similar result holds for arrows (Proposition \ref{Ca:4}),
and can be easily generalized to functors.  
These results suggest that one could proceed as follows:
first, prove that  the categories $\Zutx$
have the right structure to describe the superselection 
theory (\textit{local theory}, 
Section \ref{Cb});
secondly, check that the constructions  we have made on $\Zutx$
satisfy the mentioned gluing condition, and consequently can be extended 
to $\Zut$ (\textit{global theory},  Section \ref{Cc}). 
This argument works. We will prove
that  $\Zut$ has a tensor product, a symmetry and that any  object
has left inverses. The full subcategory $\Zut_\f$ of $\Zut$ whose 
objects have finite statistics has conjugates (Theorem \ref{Cbc:11}).\\[5pt]
\indent In Appendix \ref{X} we give some basics definitions and results 
on tensor $\mathrm{C}^*-$categories. 
\section{Homotopy and net-cohomology of posets}
\label{A}
After some preliminaries,  the main topics 
are discussed in full generality in the first three sections:
the first homotopy group of a poset; the connection between homotopy 
and net-cohomology;  the behaviour  of net-cohomology under a change
of the index set. The remaining two sections 
are devoted to study the case that the poset is a basis for the topology 
of a topological space. We stress that the results 
obtained in the first three sections in terms of abstract posets 
can be applied, not only to sharply localized charges
which are the subject of the present investigation,  
but also to charges like those studied in \cite{BF, AS}.  
\subsection{Preliminaries: the simplicial set and net-cohomology}
\label{Aa}
In the present section we recall the definition of simplicial set 
of a poset and the notion of net-cohomology of a poset, 
thereby establishing our notations. 
References for this section are \cite{Rob2, GLRV, Rob3}.\\[5pt]
\textbf{The simplicial set - } A \textit{poset} $(\Po,\leq)$ is a partially ordered set. This means that $\leq$ is a binary relation on a nonempty set $\Po$,
satisfying 
\[
\begin{array}{lcll}
\mbox{for any } \dc\in\Po & \Rightarrow & \dc\leq \dc & reflexive\\
\dc_1\leq\dc_2 \mbox{ and } \dc_2\leq\dc_1 & \Rightarrow & 
 \dc_1=\dc_2 & antisymmetric\\
\dc_1\leq\dc_2 \mbox{ and } \dc_2\leq \dc_3 & \Rightarrow & 
\dc_1\leq \dc_3 & transitive
\end{array}
\]
A poset is said to be \textit{directed}
if for any pair $\dc_1,\dc_2\in\Po$ there exists $\dc_3\in\Po$ such that 
$\dc_1,\dc_2\leq \dc_3$. 
For our aim, important examples  of posets are provided by 
the standard simplices. 
A \textit{standard} $n-simplex$  is defined as
\[
\Delta_n \defi \big\{ (\la_0,\ldots,\la_n)\in \mathbb{R}^{n+1} \ 
  | \ \la_0+\cdots +\la_n=1, \ \ \la_i\in[0,1]\big\}. 
\]
It is clear that $\Delta_0$ is a point,  
$\Delta_1$ is a closed interval etc... 
The \textit{inclusion maps} $d^n_i$ between standard simplices 
are maps  $d^n_i:\Delta_{n-1}\longrightarrow \Delta_n$ defined as 
\[
d^n_i(\la_0,\ldots,\la_{n-1})= 
(\la_0,\la_1,\ldots,\la_{i-1}, 0, \la_{i},\ldots\la_{n-1}),
\]
for $n\geq 1$ and $0\leq i\leq n$.
Now, note that a  
standard n-simplex $\Delta_n$ can be regarded as a partially ordered 
set  with respect to the inclusion of  its subsimplices.   
A \textit{singular n-simplex} of a poset $\Po$ 
is an order preserving map $f:\Delta_n\longrightarrow \Po$.
We denote by $\Sigma_n(\Po)$ the collection of singular n-simplices
of $\Po$ and by $\Si_*(\Po)$ the collection of all singular simplices
of $\Po$. $\Si_*(\Po)$ is the \textit{simplicial set} of $\Po$.
The inclusion maps $d^n_i$ between 
standard simplices are extended  to maps
$\partial^n_i :\Sigma_n(\Po)\longrightarrow \Sigma_{n-1}(\Po)$, called 
\textit{boundaries},
between singular simplices by setting
$\partial^n_i f  \defi f \circ d^n_i$. One can easily check, by 
the definition of $d^n_i$,  that the following 
relations
\[
\partial^{n-1}_i\circ \partial^{n}_j = 
\partial^{n-1}_j \circ \partial^{n}_{i+1},\qquad  i\geq j, 
\]
hold. From now on, we will omit the superscripts from the symbol
$\partial^n_i$, and will denote: the composition 
$\partial_i\circ \partial_j$ by  the symbol $\partial_{ij}$;
0-simplices by the letter $a$; 1-simplices by $b$
and 2-simplices by $c$. Notice that a 0-simplex $a$ is nothing
but an element of $\Po$; 
a 1-simplex $b$ is formed by two 0-simplices 
$\partial_0b$, $\partial_1b$ and an element $|b|$ of $\Po$, called the 
\textit{support} of $b$, such that $\partial_0b$, $\partial_1b\leq |b|$.
Given $a_0,a_1\in\Si_0(\Po)$, a  \textit{path from $a_0$ to $a_1$}
is a finite ordered 
sequence $p=\{b_n,\ldots,b_1\}$ of 1-simplices  satisfying the relations
\[
 \partial_1b_1=a_0, \ \ \ \partial_0 b_{i} = \partial_1 b_{i+1} \  \mbox{ with } 
\ i\in\{1,\ldots,n-1\}, \ \ \  
                                          \partial_0b_n=a_1.
\]
The \textit{startingpoint} of $p$, written $\partial_1p$, 
is the 0-simplex $a_0$, while the \textit{endpoint} of $p$, 
written $\partial_0p$, is the 0-simplex  $a_1$.
We will denote by $\mathrm{P}(a_0,a_1)$ the set of paths from 
$a_0$ to $a_1$, and by $\mathrm{P}(a_0)$ the set of closed  paths
with endpoint $a_0$. 
$\Po$ is said to be \textit{pathwise connected} whenever
for any pair $a_0,a_1$ of 0-simplices  there exists a path 
$p\in \mathrm{P}(a_0,a_1)$. The \textit{support} of the 
path is the collection   $|p|\defi \{|b_i| \ | \ i=1,\ldots,n\}$, 
and we will write  $|p|\subseteq P$
if $P$ is a subset of $\Po$  with  $|b_i|\in P$ for any $i$. 
Furthermore, with an abuse of notation, 
we will write $|p|\subseteq\dc$ if  $\dc\in\Po$ 
with $|b_i|\leq \dc$ for any $i$.
\begin{figure}[!ht]
\begin{center}
\includegraphics[scale=0.6]{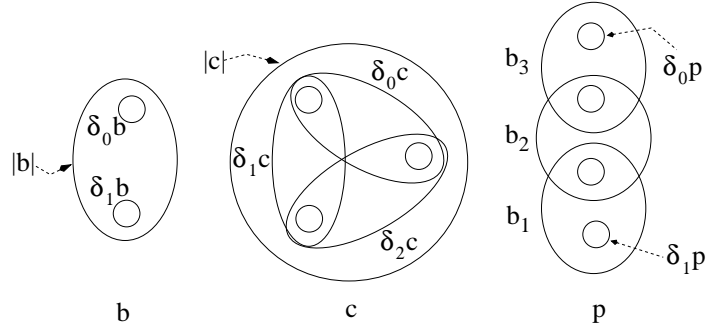}
     \caption{$b$ is a 1-simplex, $c$ is a 2-simplex, 
     $p=\{b_3,b_2,b_1\}$ is a path. 
     The symbol $\delta$ stands for $\partial$.} 
\end{center}
\end{figure}\\\\
\noindent \textbf{Causal disjointness and net of local algebras - } 
Given a poset $\Po$, a \textit{causal disjointness relation} on $\Po$ 
is a symmetric binary relation $\perp$  on $\Po$ 
satisfying the following properties: 
\begin{equation}
\label{Aa:0}
\begin{array}{ll}
\mbox{(i)} &  \dc_1\in\Po \ \Rightarrow \  \exists\dc_2\in\Po 
 \mbox{ such that }  \ \dc_1\perp \dc_2;\\
\mbox{(ii)} & \dc_1\leq \dc_2  \mbox{ and }
  \dc_2\perp \dc_3 \ \Rightarrow \ \dc_1\perp \dc_3;\\
\end{array}
\end{equation}
Given a subset $P\subseteq\Po$, the \textit{causal complement} 
of $P$ is the subset $P^\perp$ of $\Po$ defined as 
\[
 P^\perp \defi \{\dc\in\Po \ | \ \dc\perp\dc_1, \ \forall \dc_1\in P\}.
\]   
Note that   if $P_1\subseteq P$, then $P^\perp\subseteq P^\perp_1$.
Now, assume that $\Po$ is a pathwise connected poset equipped
with a causal disjointness relation $\perp$. A  
\textit{net of local algebras} indexed by  $\Po$ is a correspondence
\[
\Al_\Po: \Po\ni \dc\longrightarrow \A(\dc)\subseteq\Bh,
\]
associating to any  $\dc$  a von Neumann algebras $\A(\dc)$
defined on a fixed  Hilbert space $\mathcal{H}_o$,  and satisfying
\[
\begin{array}{lcll}
\dc_1\leq \dc_2 & \Rightarrow & \A(\dc_1)\subseteq \A(\dc_2), & 
\qquad isotony,\\
\dc_1\perp \dc_2   & \Rightarrow & \A(\dc_1)\subseteq \A(\dc_2)',& 
\qquad causality,
\end{array}
\]
where the prime over the algebra stands for the commutant of the 
algebra. The algebra $\A(\dc^\perp)$ associated with 
the causal complement $\dc^\perp$ of $\dc\in\Po$, 
is the $\mathrm{C}^*-$algebra generated 
by the algebras $\A(\dc_1)$ for any $\dc_1\in\Po$ with $\dc_1\perp\dc$.
The net  $\Al_\Po$ is said to be \textit{irreducible} whenever, 
given $T\in\Bh$ such that 
$T\in\A(\dc)'$ for any $\dc\in\Po$, 
then  $T=c\cdot \mathbbm{1}$.\\[5pt]
\textbf{The category of 1-cocycles - } We refer the reader 
to the Appendix for the definition of $\mathrm{C}^*-$category. 
Let $\Po$ be a poset with a 
causal disjointness relation $\perp$,
and let $\Al_\Po$ be an irreducible net of local algebras.  
A \textit{1-cocycle} $z$ of $\Po$ with values in $\Al_\Po$  is a field 
$z:\Si_1(\Po)\ni b\longrightarrow z(b)\in \Bh$ of  unitary operators 
satisfying the 1-cocycle identity:
\[
z(\partial_0 c) \cdot z(\partial_2 c) =  z(\partial_1 c),  
\qquad  c\in\Si_2(\Po),
\]
and the locality condition: 
$z(b)\in\A(|b|)$ for any  1-simplex $b$.
An \textit{intertwiner} $t\in(z,z_1)$  between a pair of 1-cocycles 
$z,z_1$ is a field  $t:\Si_0(\Po)\ni a\longrightarrow t_a\in\Bh$ 
satisfying the relation 
\[
  t_{\partial_0b}\cdot z(b) =  z_1(b)\cdot t_{\partial_1b},
\qquad b\in\Si_1(\Po),
\]
and the locality condition: $t_a\in\A(a)$ for any 0-simplex $a$. 
The \textit{category of 1-cocycles} $\Zup$   is the category 
whose objects are 1-cocycles and whose arrows are the corresponding 
set of intertwiners. The composition 
between $s\in(z,z_1)$ and $t\in(z_1,z_2)$ is the arrow 
$t\cdot s\in (z,z_2)$ defined as 
\[
  (t\cdot s)_a \defi t_a\cdot s_a, \qquad a\in\Si_0(\Po).
\]
Note that the arrow  $1_z$ of  $(z,z)$  defined as $(1_z)_a =\mathbbm{1}$, 
for any $a\in\Si_0(\Po)$, is the identity of $(z,z)$.  
Now, the set $(z,z_1)$ has a structure of complex vector space 
defined as 
\[
 (\alpha\cdot t + \be\cdot s)_a\defi \alpha\cdot t_a + \be\cdot s_a, \qquad 
 a\in\Si_0(\Po),
\]
for any $\al,\be\in\mathbb{C}$ and $t,s\in(z,z_1)$. 
With these operations and  the composition ``$\cdot$'', the set 
$(z,z)$ is an algebra with  identity $1_z$. 
The category $\Zup$ has an adjoint $*$, defined
on as the identity, $z^*=z$, on the objects,  while 
the adjoint $t^*\in (z_1,z)$ of on  arrows $t\in(z,z_1)$  is defined as  
\[
    (t^*)_a \defi (t_a)^*, \qquad a\in\Si_0(\Po),
\]
where $(t_a)^*$ stands for the adjoint in $\Bh$ of the operator
$t_a$. Now, let  $\norm{ \ }$ be the norm of $\Bh$. Given $t\in(z,z_1)$, 
we have that $\norm{t_a}=\norm{t_{a_1}}$ for any 
pair $a,a_1$ of 0-simplices because $\Po$ is pathwise connected. 
Therefore, by defining
\[
\norm{t} \defi \norm{t_a} \qquad a\in\Si_0(\Po)
\]
it turns out $(z,z_1)$ is a complex Banach space for any $z,z_1\in\Zup$, 
while $(z,z)$ is a $\mathrm{C}^*-$algebra for any $z\in\Zup$. 
This entails that  $\Zup$ is a  $\mathrm{C}^*-$category. 
Two 1-cocycles $z,z_1$ are  
\textit{equivalent} (or  \textit{cohomologous}) if there exists a
unitary arrow $t\in(z,z_1)$.  A 1-cocycle $z$ is 
\textit{trivial} if it is equivalent to the \textit{identity} cocycle 
$\io$ defined as $\io(b)=\mathbbm{1}$ for any 1-simplex $b$. 
Note that, since $\Al_{\Po}$ is irreducible, $\io$ is irreducible: 
$(\io,\io)= \mathbb{C}\mathbbm{1}$. \\[5pt]
\textbf{Equivalence in $\boldsymbol{\Bh}$ and path-independence - }
A weaker form of equivalence between 1-cocycles is the following: 
$z,z_1$ are said to be \textit{equivalent in} $\Bh$ if there exists a field 
$V:\Si_0(\Po)\ni a \longrightarrow V_a\in\Bh$ of unitary operators  such that 
\[
 V_{\partial_0b}\cdot z(b) =  z_1(b)\cdot V_{\partial_1b}, \qquad 
b\in\Si_1(\Po).
\]
Note that the field $V$ is not an arrow of $(z,z_1)$ because 
it is not required that $V$ satisfies the locality condition.
A 1-cocycle is \textit{trivial in} $\Bh$ 
if it is equivalent in $\Bh$ to the trivial 1-cocycle $\io$. 
We denote by $\Zutp$
the set of the  1-cocycles trivial in $\Bh$  and with the same 
symbol we denote the full $\mathrm{C}^*-$subcategory of $\Zup$ whose objects are the 
1-cocycles trivial in $\Bh$. Triviality in $\Bh$ is related to the notion
of path-independence. The evaluation of a 1-cocycle $z$ on a  path 
$p=\{b_n,\ldots,b_1\}$ is defined as 
\[
z(p)\defi z(b_n)\cdots z(b_2)\cdot z(b_1).
\]
$z$ is said to be \textit{path-independent} on a subset 
$P\subseteq\Po$  whenever 
\begin{equation}
\label{Aa:1}
 z(p)=z(q)  \mbox{ for any } p,q\in\mathrm{P}(a_0,a_1) \mbox{ such that }
 |p|,|q|\subseteq P.
\end{equation}
As $\Po$ is pathwise connected, a 1-cocycle is trivial in $\Bh$ if, 
and only if, it is path-independent on all $\Po$ \cite{GLRV}. 
For later purposes, we recall the following result:
assume that  $z$ is a 1-cocycle trivial in $\Bh$,
if the causal complement 
$\dc^\perp$ of $\dc$ is pathwise connected then 
\begin{equation}
\label{Aa:2}
 z(p)\cdot A\cdot z(p)^* = A \qquad A\in\A(\dc)
\end{equation}
for any path $p$ with $\partial_1p, \partial_0p\perp \dc$ 
\cite[Lemma 3A.5]{GLRV}. 
\subsection{The first homotopy group of a poset}
\label{Ab}
The logical steps necessary to define  the first homotopy group 
of posets are the same  as in the case  of  topological spaces.
We first recall the definition of a homotopy of paths; 
secondly, we introduce the reverse of a path, the composition of paths
and prove that they behave well under the homotopy equivalence relation;
finally we define the first homotopy group of a poset.\\[5pt]
\indent The definition of a homotopy of paths (\cite{Rob3}, p.322) needs 
some preliminaries.
An \textit{ampliation} of a 1-simplex $b$
is a 2-simplex $c$ such that $\partial_1c=b$. We denote by 
$\amp(b)$ the set of the ampliations of $b$. 
An \textit{elementary ampliation} of a  path 
$p=\{b_n,\ldots, b_1\}$, is a path $q$ of the form 
\begin{equation}     
\label{Ab:1}
q = \{b_n, \ldots, b_{j+1},\partial_0c,\partial_2c, b_{j-1},\ldots b_2,b_1\}
 \qquad c\in  \amp(b_j). 
\end{equation}      
Consider now a pair $\{b_2, b_1\}$ of 1-simplices satisfying 
$\partial_1b_2=\partial_0b_1$. A \textit{contraction} of 
$\{b_2,b_1\}$ is a 2-simplex $c$ satisfying 
$\partial_0c=b_2$, $\partial_2c=b_1$. We denote by 
$\ctr(b_2,b_1)$ the set of the contractions 
of $\{b_2,b_1\}$.  An \textit{elementary contraction} of a  
path $p=\{b_n,\ldots,b_1\}$ is a path $q$ of the form 
\begin{equation}           
\label{Ab:2}
q = \{b_n,\ldots, b_{j+2}, \partial_1c,
b_{j-1},\ldots, b_1\} \qquad c\in \ctr(b_{j+1},b_{j}).
\end{equation}
An \textit{elementary deformation}
of a path $p$ is a path $q$ which is either an elementary 
ampliation or an elementary contraction of $p$.\\
\begin{figure}[!ht]
\begin{center}
\includegraphics[scale=0.7]{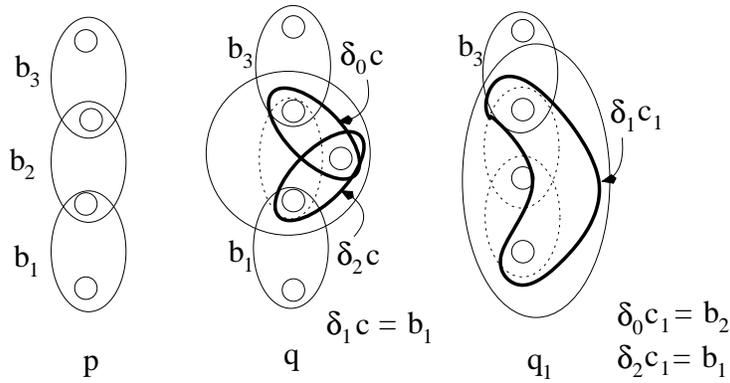}
     \caption{$q$ is an elementary ampliation of the path $p$. 
      $q_1$ is an elementary contraction of $p$. The symbol 
     $\delta$ stands for $\partial$.} 
\end{center}
\end{figure}\\
Note that 
a path $q$ is an elementary 
ampliation of a path $p$ if, and only if, $p$ is an 
elementary contraction of $q$. This can be easily seen by observing that 
if  $c\in\Si_2(\Po)$, then 
$c\in\amp(\partial_1c)$ and $c\in\ctr(\partial_0c, \partial_2c)$. 
This entails that deformation is a symmetric, reflexive binary relation 
on the set of paths with the same endpoints.  However, 
if $\Po$ is not directed,  deformation does not need to be  an 
equivalence relation on paths with the same endpoints, because transitivity 
might fail.\\  
\indent Given $a_0, a_1\in\Si_0(\Po)$, a \textit{homotopy of paths} 
in $\mathrm{P}(a_0,a_1)$ is a map
$h(i): \{1,2,\ldots,n\} \longrightarrow  \mathrm{P}(a_0,a_1)$    
such that $h(i)$ is an elementary  deformation of  $h(i-1)$  
for $1<i\leq n$.  We will say 
that two paths $p,q\in\mathrm{P}(a_0,a_1)$ are \textit{homotopic}, $p\sim q$, 
if there exists a homotopy of paths $h$  in $\mathrm{P}(a_0,a_1)$ 
such that $h(1)=q$ and $h(n)=p$. 
It is clear that  a homotopy of paths 
is an equivalence relation on  paths with the same endpoints.\\
\indent We now define the composition of paths and the reverse
of a path. Given 
$p=\{b_n,\ldots,b_1\}\in \mathrm{P}(a_0,a_1)$ and 
$q=\{b'_k,\ldots b'_1\}\in\mathrm{P}(a_1,a_2)$,  the \textit{composition}
of $p$ and $q$ is the path $p*q\in \mathrm{P}(a_0,a_2)$ defined as 
\begin{equation}
\label{Ab:4}
q*p\defi \{b'_k,\ldots, b'_1,b_n,\ldots,b_1 \}.
\end{equation}
Note that $p_1*(p_2*p_3)= (p_1*p_2)*p_3$,
if the composition is defined.
\begin{lemma}
\label{Ab:5}
Let $p_1,q_1\in\mathrm{P}(a_0,a_1)$, $p_2,q_2\in\mathrm{P}(a_1,a_2)$.  
If  $p_1\sim q_1$ and $p_2\sim q_2$, then $p_2*p_1\sim q_2* q_1$.
\end{lemma}
\begin{proof}
Let $h_1:\{1,\ldots,n\}\longrightarrow \mathrm{P}(a_0,a_1)$ 
and $h_2:\{1,\ldots,k\}\longrightarrow \mathrm{P}(a_1,a_2)$ be homotopies of 
paths such that 
$h_1(1)=p_1$, $h_1(n)=q_1$ and $h_2(1)=p_2$, $h_2(k)=q_2$. Define
\[
h(i)\defi \left\{
    \begin{array}{ll}
     h_2(1)*h_1(i) & i\in\{1,\ldots,n\} \\
     h_2(i-n)*h_1(n) & i\in\{n+1,\ldots,n+k\}      
    \end{array}\right.
\] 
Then $h:\{1,\ldots,n+k\}\longrightarrow \mathrm{P}(a_0,a_2)$ is a 
homotopy of paths such that $h(1)= p_2*p_1$ and $h(n+k)=q_2*q_1$,
completing the proof. 
\end{proof}
The \textit{reverse} of a 1-simplex $b$, is the 1-simplex $\con{b}$
defined as 
\begin{equation}
\label{Ab:6}
\partial_0\con{b} = \partial_1b, \ \ \  
\partial_1\con{b} = \partial_0b, \ \ \ 
|\con{b}| = |b|.
\end{equation}
So, the \textit{reverse} of a path 
$p=\{b_n,\ldots,b_1\}\in\mathrm{P}(a_0,a_1)$ is the path
$\con{p}\in\mathrm{P}(a_1,a_0)$ defined as  
$\con{p}\defi \{ \con{b_1},\ldots,\con{b_n}\}$.
It is clear that $\con{\con{p}}=p$. Furthermore
\begin{lemma}
\label{Ab:8}
 $p\sim q$  $ \ \  \Rightarrow \ \ $  $\con{p}\sim\con{q}$. 
\end{lemma}
\begin{proof}
The reverse  of a 2-simplex $c$ is
the 2-simplex $\con{c}$ defined as 
\[
\partial_1\con{c}=\con{\partial_1c}, \ \ 
\partial_0\con{c}=\con{\partial_2c}, \ \ 
\partial_2\con{c}=\con{\partial_0c}, \ \ 
|\con{c}|=|c|. 
\]
Note that, 
if $c\in\amp(b)$, then $\con{c}\in\amp(\con{b})$; 
if $c\in\ctr(b_2,b_1)$, 
then $\con{c}\in\ctr\big( \con{b_1},\con{b_2}\big)$. So, 
let $h:\{1,\ldots,n\}\longrightarrow \mathrm{P}(a_0,a_1)$ 
be a homotopy of paths. Then maps 
$\con{h}:\{1,\ldots,n\}\longrightarrow \mathrm{P}(a_1,a_0)$ defined 
as $\con{h}(i)\defi \con{h(i)}$ for any $i$ is a homotopy of paths,
completing the proof.
\end{proof}
A 1-simplex $b$ is said 
to be \textit{degenerate to} a 0-simplex $a_0$ whenever
\begin{equation}
\label{Ab:9}
 \partial_0 b =a_0 =\partial_1b, \ \ \ \  a_0=|b| 
\end{equation}
We will denote by $b(a_0)$ the 1-simplex degenerate to $a_0$.
\begin{lemma}
\label{Ab:10}
The following assertions hold:\\
(a) $p*b(\partial_1p)\sim p \sim b(\partial_0p)*p$;\\
(b) $p*\con{p}\sim b(\partial_0p)$ and $\con{p}*p\sim b(\partial_1p)$.
\end{lemma}
\begin{proof}
By Lemma \ref{Ab:5}  it is enough to prove the assertions
in the case that $p$ is a 1-simplex $b$. 
(a) Let $c_1$ be the 2-simplex defined as 
$\partial_2 c_1 = b(\partial_1b)$, $\partial_0c_1 = b$,  
$\partial_1c_1 = b$  and whose support $|c_1|$ equals $|b|$. Then
$c_1\in\ctr(b,b(\partial_1b))$ and  $b*b(\partial_1b) \sim b$. 
The other identity follows in a similar way.
(b)  Let $c_2$ be the 2-simplex defined as 
$\partial_0c_2=b$,  $\partial_2c_2=\con{b}$, 
$\partial_1c_2= b(\partial_0b)$ and whose support $|c_2|$ 
equals $|b|$. Then $c_2\in\ctr(b,\con{b})$ and 
$b*\con{b}\sim b(\partial_0b)$. 
The other identity follows in a similar way.
\end{proof}
We now are in a  position to define the first homotopy group
of a poset. Fix a base 0-simplex $a_0$ and consider the set of 
closed paths $\mathrm{P}(a_0)$. Note that the composition 
and the reverse 
are internal operations of $\mathrm{P}(a_0)$ and that 
$b(a_0)\in \mathrm{P}(a_0)$. We define 
\begin{equation}
\label{Ab:11}
  \pi_1(\Po,a_0) \defi  \mathrm{P}(a_0) / \sim
\end{equation}
where $\sim$ is the homotopy equivalence relation. Let 
$[p]$ denote the homotopy class of an element $p$ of 
$\mathrm{P}(a_0)$. Equip 
$\pi_1(\Po,a_0)$ with the product
\[
  \qquad [p]*[q] \defi [p*q] \qquad   [p],[q]\in  \pi_1(\Po,a_0).
\]  
$*$ is associative, and it easily follows from  previous lemmas that 
$\pi_1(\Po,a_0)$ with $*$ is a group: 
the identity $1$ of the group  is $[b(a_0)]$; 
the inverse $[p]^{-1}$ of $[p]$ is
$[\con{p}]$. Now, assume that $\Po$ is pathwise connected. Given 
a 0-simplex $a_1$, let $q$ be a path from  $a_0$ to $a_1$. Then
the map 
\[
\pi_1(\Po,a_0)\ni [p]\longrightarrow  [q*p*\con{q}]\in \pi_1(\Po,a_1)
\]
is a  group isomorphism. On the grounds of these facts, we give the following 
\begin{df}
\label{Ab:12}
We call $\pi_1(\Po,a_0)$ the \textbf{first homotopy group} of $\Po$ 
with base $a_0\in\Si_0(\Po)$. If $\Po$ is pathwise connected, 
we denote this group by $\pi_1(\Po)$ and call it 
the \textbf{fundamental group} of $\Po$. 
If $\pi_1(\Po)=1$ we will say that 
$\Po$ is \textbf{simply connected}. 
\end{df}
We have the following result 
\begin{prop}
\label{Ab:13}
If $\Po$ is directed, then $\Po$ is pathwise and simply connected. 
\end{prop}
\begin{proof}
Clearly $\Po$ is pathwise connected. 
Let $p= \{b_n,\ldots,b_1\} \in\mathrm{P}(a_0)$. As $\Po$ is directed, 
we can find $c_i\in\amp(b_i)$ for $i =2,\ldots, n-1$ with 
\[
 \partial_2c_2 = \con{b_1}, \ \ 
 \con{\partial_0c_{i-1}} = \partial_2c_{i} \ \mbox{ for } \ 
  i = 3,\ldots, n-1, \ \ \partial_0c_{n-1} =  \con{b_n}.
\]
One can easily deduce from these relations that $\partial_{02}c_i= a_0$
for any $i=2,\ldots, n-1$. By  Lemmas \ref{Ab:5}, \ref{Ab:8} and \ref{Ab:10}, 
we have 
\begin{align*}    
p & \sim b_n* \partial_0c_{n-1} * \partial_2c_{n-1}* 
            \partial_0c_{n-2} 
              *\cdots *  \partial_2c_{3}* 
\partial_0c_2 * \partial_2c_2* b_1 \\
 &  =  b_n * \con{b_n}* \partial_2c_{n-1}* \con{\partial_2c_{n-1}}*
  \cdots * \partial_2c_3*\con{\partial_2c_3}*\con{b_1}* b_1\\
  &  \sim b(a_0) * \cdots *b(a_0) \ \sim \ b(a_0),
\end{align*}
completing the proof.
\end{proof}
\subsection{Connection between homotopy and net-cohomology}
\label{Ac}
Let us consider a pathwise-connected poset $\Po$, equipped with 
a causal disjointness 
relation $\perp$, and let  $\Al_\Po$ be an irreducible net 
of local algebras. In this section we show the relation between 
$\pi_1(\Po)$ and the set $\Zup$.\\   
\indent To begin with, we prove the invariance of  1-cocycles  for 
homotopic paths. 
\begin{lemma}
\label{Ac:1}
Let $z\in\Zup$. For any  pair $p,q$ of paths with the same endpoints, 
if $p\sim q$, then  $z(p)=z(q)$. 
\end{lemma}
\begin{proof}
It is enough to check the invariance of $z$ for  elementary deformations.
For instance let 
$q=\{b_n, \ldots, b_{j+1},\partial_0c,\partial_2c,b_{j-1},\ldots, b_2, b_1\}$, 
with $c\in \amp(b_j)$, that is an elementary ampliation of $p$.
By definition of $\amp(b_j)$ and the 1-cocycle identity we have 
\begin{multline*}
z(q)  = z(b_n)\cdots  z(b_{j+1}) \cdot z(\partial_0c)
  \cdot z(\partial_2c)\cdot z(b_{j-1})\cdots z(b_1) \\
  = z(b_n)\cdots  z(b_{j+1}) \cdot z(\partial_1c)\cdot z(b_{j-1})
    \cdots z(b_1) = z(p).
\end{multline*}
The invariance for  elementary contractions follows in a similar way.
\end{proof}
\begin{lemma}
\label{Ac:2}
Let $z\in\Zup$. Then:\\
(a) $z(b(a))=\mathbbm{1}$ for any 0-simplex $a$;\\
(b) $z(\con{p})=z(p)^*$ for any path $p$.
\end{lemma}
\begin{proof}
(a) Let $c(a_0)$ be the 2-simplex degenerate to $a_0$, that is 
\[
 \partial_0c(a_0)=\partial_2c(a_0) =  \partial_1c(a_0) = b(a_0), \ \ 
          |c(a_0)|=a_0 
\]
where $b(a_0)$ is the 1-simplex degenerate to $a_0$.   
By the 1-cocycle identity we have: 
$z(\partial_0c(a))\cdot z(\partial_2c(a))$ $= z(\partial_1c(a))$ 
$\iff  \ \ z(b(a))\cdot z(b(a)) = z(b(a))$ $\iff$ 
$z(b(a))=\mathbbm{1}$. (b) follows from (a),  Lemma \ref{Ac:1} and Lemma
\ref{Ab:10}b.
\end{proof}
We now are in a position to show the connection between 
the fundamental group of $\Po$ and $\Zup$. 
Fix a base 0-simplex $a_0$.  Given  $z\in\Zup$, define  
\begin{equation}
\label{Ac:3}
\qquad \pi_z([p]) \defi z(p) \qquad \qquad  [p]\in\pi_1(\Po,a_0)
\end{equation}
This definition is well posed as $z$ is invariant for homotopic paths.
\begin{teo}
\label{Ac:4}
The correspondence $\Zup\ni z\longrightarrow \pi_z$  maps 
1-cocycles $z$, equivalent in $\Bh$, into equivalent
unitary representations $\pi_z$ of $\pi_1(\Po)$ in $\mathcal{H}_o$. 
Up to equivalence,  this map is injective. 
If $\pi_1(\Po)= 1$, then $\Zup=\Zutp$. 
\end{teo} 
\begin{proof}
First, recall that the identity $1$ of $\pi_1(\Po)$ is 
the equivalence class $[b(a_0)]$ associated with the 1-simplex
degenerate to $a_0$. By Lemma \ref{Ac:2} we have that $\pi_z(1)=1$
and that $\pi_z([p]^{-1})= \pi_z([p])^*$. Furthermore,
it is obvious from the definition of $\pi_z$, that 
$\pi_z([p]*[q])=\pi_z([p])\cdot \pi_z([q])$, therefore 
$\pi_z$ is a unitary representation of $\pi_1(\Po)$ in $\mathcal{H}_o$.
Note that if $z_1\in\Zup$ and $u\in(z,z_1)$ is unitary, 
then $u_{a_0}\cdot \pi_z([p]) = \pi_{z_1}([p])\cdot  u_{a_0}$.
Now, let $\pi$ be a unitary representation
of $\pi_1(\Po)$  on  $\mathcal{H}_o$. Fix a base 0-simplex $a_0$, and for 
any 0-simplex $a$,  denote by $p_a$ a path with 
$\partial_1p_a = a$ and $\partial_0p_a = a_0$. Let 
\[
z_\pi(b)\defi \pi([ p_{\partial_0b}*b * \con{p_{\partial_1b}}]) 
   \qquad b\in\Si_1(\Po) 
\] 
Given  a 2-simplex $c$,  we have 
\begin{align*}
z_\pi(\partial_0c)\cdot z_\pi(\partial_2c) &  
  = \pi([p_{\partial_{00}c} * \partial_0c*\con{p_{\partial_{10}c}} * 
  p_{\partial_{02}c}*\partial_2c *\con{p_{\partial_{12}c}}]) \\
 &  =\pi([p_{\partial_{00}c} * \partial_0c*\con{p_{\partial_{10}c}} * 
p_{\partial_{10}c}*\partial_2c *\con{p_{\partial_{12}c}}])\\
& =\pi([p_{\partial_{00}c} * \partial_0c*\partial_2c *
  \con{p_{\partial_{11}c}}])
 =\pi([p_{\partial_{01}c} * \partial_1c*
    \con{p_{\partial_{11}c}}])\\
& = z_\pi(\partial_1c).
\end{align*}
Hence $z_\pi$ satisfies the 1-cocycle identity but in general 
$z_\pi\not\in\Zu$ 
because $z_\pi(b)$ does not need to belong to $\A(|b|)$. However note that 
if we consider $\pi_{z_1}$ for some $z_1\in\Zu$,  then 
\[
z_{\pi_{z_1}}(b) = \pi_{z_1}([ p_{\partial_0b}*b * \con{p_{\partial_1b}}]) = 
z_1( p_{\partial_0b})\cdot z_1(b)\cdot z_1( p_{\partial_0b})^*.
\]
therefore $z_{\pi_{z_1}}$ is equivalent in $\Bh$  to $z_1$. This entails
that if $\pi_z$ is equivalent to $\pi_{z_1}$, then 
$z$ is equivalent in $\Bh$
to $z_1$. Finally, assume that $\pi_1(\Po)=1$,  
then $z(p)=\mathbbm{1}$ for any closed path $p$.
This entails that $z$ is path-independent on $\Po$, hence 
$z$ is trivial in $\Bh$. 
\end{proof}
\subsection{Change of index set}
\label{Aca}
The purpose is to show the invariance of 
net-cohomology under a suitable change of the index set. To begin with, 
by a \textit{subposet} of a poset $\Po$ we mean  a 
subset $\widehat{\Po}$ of $\Po$ equipped with 
the same order relation of $\Po$.
\begin{df}
\label{Aca:1}
Consider a subposet $\widehat{\Po}$ of $\Po$. We will say that 
$\hat{\Po}$ is a \textbf{refinement} of $\Po$, if
for any $\dc\in\Po$ there exists $\widehat{\dc}\in\widehat{\Po}$
such that $\widehat{\dc}\leq \dc$. A refinement $\widehat{\Po}$ of $\Po$
is said to be \textbf{locally relatively connected} 
if given $\dc\in\Po$, for any pair 
$\widehat{\dc}_1,\widehat{\dc}_2\in\widehat{\Po}$ with 
$\widehat{\dc}_1,\widehat{\dc}_2 \leq \dc$ there is a path $\hat{p}$ 
in $\widehat{\Po}$ from $\widehat{\dc}_1$ to $\widehat{\dc}_2$ such that  
$|\hat{p}|\leq\dc$.
\end{df}
\begin{lemma}
\label{Aca:2}
Let $\widehat{\Po}$ be a locally relatively connected refinement of $\Po$.\\
(a) $\Po$ is pathwise connected if, and only if, 
$\widehat{\Po}$ is pathwise connected.\\
(b) If $\perp$
is a causal disjointness relation for $\Po$, then the restriction of 
$\perp$ to $\widehat{\Po}$ is a causal disjointness relation.
\end{lemma}
\begin{proof}
(a) Assume that  $\Po$ is pathwise connected. It easily follows from 
the definition of a locally relatively connected refinement that 
$\widehat{\Po}$ is pathwise connected. Conversely, 
assume that $\widehat{\Po}$ is pathwise connected. Given $a_0,a_1\in\Po$, 
let  $\hat{a}_0,\hat{a}_1\in\widehat{\Po}$ be such that 
$\hat{a}_0\leq a_1$ and $\hat{a}_1\leq a_1$, and let $\hat{p}$ be a path
in $\widehat{\Po}$ from  $\hat{a}_0$ to $\hat{a}_1$.
Then, $b_1*p*b_0$ is a path from $a_0$ to $a_1$, where 
$b_0,b_1$ are 1-simplices of $\Po$  defined as follows:
$\partial_1b_0= a_0$, $\partial_0b_0= \hat{a}_0$;
$|b_0|=a_0$ and $\partial_0b_1= a_1$, $\partial_1b_1= \hat{a}_1$;
$|b_1|=a_1$. (b) It is clear that $\perp$, restricted to 
$\widehat{\Po}$, is a symmetric binary relation
satisfying the property (ii)  of the definition (\ref{Aa:0}).
Let $\widehat{\dc}\in\widehat{\Po}$. 
Since $\perp$ is a causal disjointness relation on $\Po$, 
we can find 
$\dc_1\in\Po$ with $\widehat{\dc}\perp \dc_1$. Since $\widehat{\Po}$
is a refinement of $\Po$, there exists $\widehat{\dc}_1\in\widehat{\Po}$ with 
$\widehat{\dc}_1\leq \dc_1$. Hence $\widehat{\dc}\perp \widehat{\dc}_1$,
completing the proof.
\end{proof}
Let $\Po$ be a pathwise connected poset and let 
$\perp$ be a causally disjointness relation for $\Po$. 
Let $\Al_{\Po}$ be an irreducible net of local 
algebras indexed by $\Po$ and defined on a Hilbert space 
$\mathcal{H}_o$. If $\widehat{\Po}$ is a locally
relatively connected refinement of $\Po$, then, by the previous lemma,
$\widehat{\Po}$ is pathwise connected and $\perp$ is a causal disjointness 
relation  on $\widehat{\Po}$.
Furthermore, the restriction 
of $\Al_{\Po}$ to $\widehat{\Po}$ is a net of local algebras  
$\Al_{\Po|\widehat{\Po}}$ indexed by $\widehat{\Po}$. 
Let $\mathcal{Z}^1_t(\Al_{\Po|\widehat{\Po}})$ be the category 
of 1-cocycles of $\widehat{\Po}$, trivial in $\Bh$, with values in 
the net $\Al_{\Po|\widehat{\Po}}$. Notice that 
$\Al_{\Po|\widehat{\Po}}$ might be not irreducible,
hence it is not clear, at a first sight, if the trivial 1-cocycle
$\hat{\io}$ of  
$\mathcal{Z}^1_t(\Al_{\Po|\widehat{\Po}})$ is irreducible or not.
This could create some  problems in the following, since 
the properties of tensor 
$\mathrm{C}^*-$categories whose identity is not irreducible
are quite complicated (see  \cite{LR, Vas, BL}). However, as a consequence 
of the fact that $\widehat{\Po}$ is a refinement of $\Po$, this is not the case
as shown by the following lemma.
\begin{lemma}
Let $\Al_{\Po}$ be an irreducible net of local 
algebras. For any locally relatively connected refinement $\widehat{\Po}$ 
of $\Po$, the trivial 1-cocycle $\hat{\io}$ of 
$\mathcal{Z}^1_t(\Al_{\Po|\widehat{\Po}})$ is irreducible.
\end{lemma}
\begin{proof}
Let $\hat{t}\in (\hat{\io},\hat{\io})$.
By the definition of $\hat{\io}$ we have that 
$\hat{t}_{\partial_1\hat{b}} = \hat{t}_{\partial_0\hat{b}}$ for any  
1-simplex $\hat{b}$ of $\widehat{\Po}$. 
Since $\widehat{\Po}$  is pathwise connected, we have that 
$\hat{t}_{\hat{a}} = \hat{t}_{\hat{a}_1}$ for any pair $\hat{a},\hat{a}_1$ 
of 0-simplices of $\widehat{\Po}$. By the localization properties 
of $\hat{t}$, it turns out that if  
we define  $T\defi \hat{t}_{\hat{a}}$ for some 0-simplex $\hat{a}$ of 
$\widehat{\Po}$, then   $T\in\A(\widehat{\dc})$ for any 
$\widehat{\dc}\in\widehat{\Po}$.
Now, observe that given  $\dc\in\Po$, by the definition of causal 
disjointness relation, there is $\dc_1\in\Po$ such that $\dc_1\perp\dc$. 
Since $\widehat{\Po}$ is a refinement of $\Po$, there is 
$\widehat{\dc}_1\in\widehat{\Po}$ such that $\widehat{\dc}_1\subseteq\dc_1$.
Hence $\widehat{\dc}_1\perp \dc$. Since $T\in\A(\widehat{\dc}_1)$
we have that $T\in\A(\dc)'$. But this holds 
for any $\dc\in\Po$, hence $T=c\cdot \mathbbm{1}$ because 
the net $\Al_\Po$ is irreducible.
\end{proof}
We now are ready to show the main result of this section. 
\begin{teo}
\label{Aca:3}
Let $\widehat{\Po}$ be locally relatively connected 
refinement of $\Po$. 
Then the categories  $\Zutp$ and 
$\mathcal{Z}^1_t(\Al_{\Po|\widehat{\Po}})$ are equivalent.
\end{teo}
\begin{proof}
For any $z\in\Zutp$ and for any 
$t\in(z,z_1)$ define
\[
\RE(z) \defi z\upharpoonright \Si_1(\widehat{\Po}), \qquad 
\RE(t) \defi t\upharpoonright \Si_0(\widehat{\Po}).
\]
It is clear that $\RE$ is a covariant and faithful functor 
from $\Zutp$ into $\mathcal{Z}^1_t(\Al_{\Po|\widehat{\Po}})$. 
We now define a functor from $\mathcal{Z}^1_t(\Al_{\Po|\widehat{\Po}})$
to $\Zutp$. To this purpose, 
we choose a function $\f:\Po\longrightarrow\widehat{\Po}$
satisfying the following properties: given $\dc\in\Po$, 
\[
 \mbox{ if } \dc\in\widehat{\Po}  \ \Rightarrow \    \f(\dc)= \dc, \ 
 \mbox{ otherwise }  \ \f(\dc)\leq \dc. 
\]
For any $b\in\Si_1(\Po)$ we denote by 
$\hat{p}(\f(\partial_0b),\f(\partial_1b))$ a path of $\widehat{\Po}$
from $\f(\partial_0b)$ to $\f(\partial_1b)$ whose support is contained 
in $|b|$, this is possible because $\widehat{\Po}$ 
is a locally relatively connected refinement of $\Po$.
For any  $\hat{z}\in\mathcal{Z}^1_t(\Al_{\Po|\widehat{\Po}})$ we define
\[ 
 \F(\hat{z})(b) \defi \hat{z}(\hat{p}(\f(\partial_0b),\f(\partial_1b))) 
 \qquad b\in\Si_1(\Po)
\]
By the properties of the path $\hat{p}(\f(\partial_0b),\f(\partial_1b))$ 
it follows that $\F(\hat{z})(b)\in\A(|b|)$. For any $c\in\Si_2(\Po)$,
by using the path-independence of $\hat{z}$ we have  
\begin{align*}
 \F(\hat{z})(\partial_0c)\cdot \F(\hat{z})(\partial_2c) & = 
   \hat{z}(\hat{p}(\f(\partial_{00}c),\f(\partial_{10}c)))\cdot 
     \hat{z}(\hat{p}(\f(\partial_{02}c),\f(\partial_{12}c))) \\ 
  & =   \hat{z}(\hat{p}(\f(\partial_{01}c),\f(\partial_{02}c)))\cdot 
     \hat{z}(\hat{p}(\f(\partial_{02}c),\f(\partial_{11}c))) \\
  &  = \hat{z}(\hat{p}(\f(\partial_{01}c),\f(\partial_{11}c))) 
     = \F(\hat{z})(\partial_1c)
\end{align*}
Hence $\F(\hat{z})$ satisfies the 1-cocycle identity, and it is  
trivial in $\Bh$ because so is $\hat{z}$. Therefore, 
$\F(\hat{z})\in\Zutp$. Now, for any $\hat{t}\in(\hat{z},\hat{z}_1)$, 
define 
\[
 \F(\hat{t})_a \defi \hat{t}_{\f(a)}, \qquad a\in\Si_0(\Po).  
\]
Clearly  $\F(\hat{t})_a\in\A(a)$. Moreover, for any $b\in\Si_1(\Po)$ 
we have
$\F(\hat{t})_{\partial_0b}\cdot \F(\hat{z})(b)$ 
$ = \hat{t}_{\f(\partial_0b)} \cdot 
      \hat{z}(\hat{p}(\f(\partial_0b),\f(\partial_1b)))$
$ = \hat{z}_1(\hat{p}(\f(\partial_0b),\f(\partial_1b)))\cdot 
     \hat{t}_{\f(\partial_1b)}$ 
$  = \F(\hat{z}_1)(b)\cdot \F(\hat{t})_{\partial_1b}$.
Therefore, $\F$ is a covariant functor from 
$\mathcal{Z}^1_t(\Al_{\Po|\widehat{\Po}})$ to $\Zutp$. \\
\indent Now, we show that the pair $\RE$, $\F$ states an equivalence 
between $\Zutp$ and $\mathcal{Z}^1_t(\Al_{\Po|\widehat{\Po}})$. 
Given $\hat{z}\in\mathcal{Z}^1_t(\Al_{\Po|\widehat{\Po}})$, for any 
$\hat{b}\in\Si_1(\widehat{\Po})$,  we have  that
\[ 
(\RE\circ \F)(\hat{z})(\hat{b}) = \F(\hat{z})(\hat{b}) = 
\hat{z}(\hat{p}(\f(\partial_0b),\f(\partial_1b))) = \hat{z}(\hat{b}),
\] 
because  
$\f(\partial_0\hat{b}) = \partial_0\hat{b}$ and $\f(\partial_1\hat{b})= 
\partial_1\hat{b}$. Clearly $(\RE\circ \F)(\hat{t})(\hat{a})= 
\hat{t}_{\hat{a}}$ for any 
$\hat{a}\in\Si_0(\widehat{\Po})$. Therefore,  
$\RE\circ \F = 1_{\mathcal{Z}^1_t(\Al_{\Po|\widehat{\Po}})}$, 
where $1_{\mathcal{Z}^1_t(\Al_{\Po|\widehat{\Po}})}$ is the 
identity functor of
$\mathcal{Z}^1_t(\Al_{\Po|\widehat{\Po}})$. 
The proof follows once we have shown that 
the functor $\F\circ\RE$ is  naturally isomorphic to $1_{\Zutp}$.   
To this end, for any $a\in\Si_0(\Po)$ 
let  $b(\f(a),a)\in\Si_1(\Po)$ defined as 
\[
  \partial_0 b(\f(a),a)= \f(a), \ \  \partial_1 b(\f(a),a)= a, \ \ 
   |b(\f(a),a)| = a.
\]
Given $z\in\Zutp$, let $u(z)_a \defi z(b(\f(a),a))$ for  any $a\in\Si_0(\Po)$. 
By definition  $u(z)_a\in\A(a)$. Furthermore, 
\begin{align*}
u(z)_{\partial_0b}\cdot z(b)
 & =  z(b(\f(\partial_0b),\partial_0b))\cdot 
      z(b)\\
 & =  z(p(\f(\partial_0b),\f(\partial_1b)))\cdot  
      z(b(\f(\partial_1b), \partial_0b))\cdot 
      z(b) \\  
 & =  (\F\circ\RE)(z)(b) 
   \cdot  z(p(\f(\partial_1b), \partial_1b))\\ 
 & = (\F\circ\RE)(z)(b)\cdot u(z)_{\partial_1b}.
\end{align*}
Hence $u(z)\in (z, (\F\circ\RE)(z))$ for any 
$z\in\Zutp$. 
Let $t\in(z_1, z)$
then 
\[
u(z)_a\cdot  t_a =
  t_{\f(a)} \cdot z_1(b(\f(a),a)) =
 (\F\circ\RE)(t)_a \cdot  u(z_1)_a
\]
for any $a\in\Si_0(\Po)$. This means that  the mapping 
$u:\Zutp\ni z \longrightarrow u(z)\in 
(z, (\F\circ\RE)(z))$ is a natural transformation 
between $1_{\Zutp}$ and $(\F\circ\RE)$. 
Finally, note that $u(z)^*\in ((\F\circ\RE)(z),z)$.
Combining this  with the fact that 
$u(z)$ is unitary, we have that $u$ is a natural isomorphism
between $1_{\Zutp}$ and $(\F\circ\RE)$, completing the proof. 
\end{proof}
\subsection{The poset as a basis  for a topological space}
\label{Ad}
Given a topological Hausdorff space $\X$. 
The topics of the previous sections are now investigated 
in the case that $\Po$ is a basis for the topology $\X$ 
ordered under \textit{inclusion} $\subseteq$. This allows us both to show 
the connection between the notions for posets 
and the corresponding topological ones, and to understand 
how topology affects net-cohomology.
\subsubsection{Homotopy}
\label{Ada}
In what follows, by a curve $\ga$ of $\X$ we mean  a continuous function 
from the interval $[0,1]$ into $\X$. We recall that the reverse 
of a curve $\ga$ is the curve $\con{\ga}$ defined as 
$\con{\ga}(t)\defi \ga(1-t)$ for  $t\in[0,1]$.
If $\be$ is a curve such that $\be(1)=\ga(0)$, the composition 
$\ga*\be$ is the curve
\[
(\ga*\be)(t) \defi \left\{
      \begin{array}{ll}
       \be(2t) &  0\leq t\leq 1/2\\
       \ga(2t - 1)&  1/2\leq t\leq 1
      \end{array}
      \right.
\]
Finally, the constant curve $e_x$ is the curve $e_x(t)=x$ 
for any $t\in[0,1]$. 
\begin{df}
\label{Ada:1}
Given a curve $\ga$. A path $p =\{b_n,\ldots,b_1\}$ is said to be a 
\textbf{poset-approximation} of $\ga$ (or simply an \textbf{approximation}) 
if there is a partition $0=s_0 < s_1<\ldots <s_n=1$ of the interval $[0,1]$
such that 
\[
\ga([s_{i-1},s_i])\subseteq |b_i|, \qquad \ga(s_{i-1})\in 
\partial_1b_i, \ \  \ga(s_i)\in \partial_0b_i,
\]
for $i=1,\ldots, n$ (Fig.3). 
By $App(\ga)$ we denote the set of approximations
of $\ga$.
\end{df}
Since $\Po$ is a basis for the topology of $\X$, 
we have that $App(\ga)\ne\emptyset$ for any curve $\ga$. 
It can be easily checked that the approximations
of curves enjoy  the following properties 
\begin{equation}
\label{Ada:2}
\begin{array}{rcl}
p\in App(\ga) & \iff  & \con{p}\in App(\con{\ga})\\
p \in  App(\si), \ q\in  App(\be) & \Rightarrow  & p*q\in App(\si*\be)
\end{array}
\end{equation}
where $\be(1)=\si(0)$, $\partial_0q=\partial_1p$.\\ 

\begin{figure}[!ht]
\begin{center}
     \includegraphics[scale=0.6]{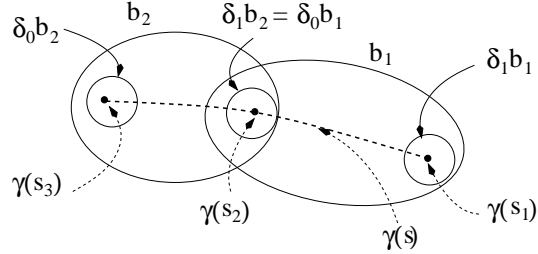}
     \caption{The path $\{b_2,b_1\}$ is an 
    approximation of the curve $\ga$ (dashed).
    The symbol $\delta$ stands for $\partial$.}
\end{center}
\end{figure}
\begin{df}
\label{Ada:3}
Given $p,q\in App(\ga)$,
we say that $q$ is \textbf{finer} than
$p$ whenever  $p=\{b_n,\ldots,b_1\}$ and $q= q_n*\cdots *q_1$
where $q_i$ are paths satisfying 
\[
 |q_i|\subseteq |b_i|, \ \ \partial_0q_i\subseteq \partial_0b_i, \ \ 
  \partial_1q_i\subseteq \partial_1b_i \qquad i=1,\dots, n.
\]
We will write $p\prec q$ to denote that $q$ is a 
finer approximation than $p$ (Fig.4) 
\end{df}
Note that $\prec$ is an order relation in $App(\ga)$.
Since  $\Po$ is a basis for the topology of $\X$, 
$(App(\ga),\prec)$ is directed: that is, for any pair 
$p,q\in App(\ga)$  there exists  
$p_1\in App(\ga)$ of $\ga$ such that $p,q\prec p_1$. 
As already said, we can find an approximation for any curve $\ga$. 
The converse, namely that for a given path $p$ there is 
a curve $\ga$ such that $p$ is an approximation of $\ga$, holds
if the elements of $\Po$ are arcwise connected sets of 
the topological space $\X$.
\begin{figure}[!ht]
\begin{center}
     \includegraphics[scale=0.6]{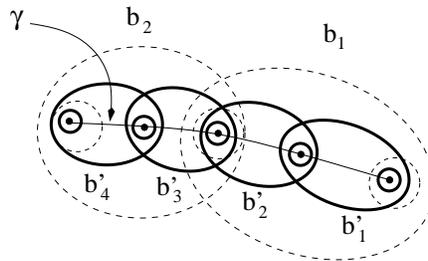}
     \caption{The path $\{b_2,b_1\}$ (dashed) is an 
    approximation of the curve $\ga$. The path $\{b'_4, b_3', b'_2,b'_1 \}$
    (bold) is an approximation of $\ga$ finer than  $\{b_2,b_1\}$}
\end{center}
\end{figure}\\
Concerning the relation between connectedness for  
posets and connectedness for topological spaces, 
in \cite{GLRV} it has been shown that:
if the elements of $\Po$ are 
arcwise connected sets of $\X$,  then  an  open set 
$X\subseteq \X$ is arcwise connected 
in $\X$ if, and only if, the poset $\Po_X$  defined as 
\begin{equation}
\label{Ada:3a}
 \Po_X\defi \{ \dc\in\Po \ | \ \dc\subseteq X\}. 
\end{equation}
is pathwise connected. Note that the set $\Po_X$ is a \textit{sieve}
of $\Po$, namely a subfamily $S$ of $\Po$ such that, 
if $\dc\in S$ and $\dc_1\subseteq \dc$, then $\dc_1\in S$.   
Now, assume that $P$ is a sieve of $\Po$. Then $P$ is 
pathwise connected in $\Po$ if, and only if, the open set $\X_P$
defined as 
\begin{equation}
\label{Ada:3b}
 \X_P \defi \cup \{ \dc \subseteq \X \ | \ \dc\in P\}
\end{equation}
is arcwise connected in $\X$.  
We now turn to analyze simply connectedness.
\begin{lemma}
\label{Ada:3c}
Let $p,q\in \mathrm{P}(a_0,a_1)$ be two approximations of 
$\ga$. Then $p$ and $q$ are homotopic paths.
\end{lemma}
\begin{proof}
It is enough to prove the statement in the case where 
$p\prec q$. So, let $p=\{b_n,\ldots,b_1\}$ and let  
$q=\{q_n,\ldots, q_1\}$  be such the paths  $q_i$ satisfy
\[
 |q_i|\subseteq |b_i|, \ \ \partial_0q_i\subseteq \partial_0b_i, \ \ 
  \partial_1q_i\subseteq\partial_1b_i \qquad i=1,\dots, n.   
\]
Note that for any $i$ the poset formed by $\dc\in\Po$ with 
$\dc\subseteq|b_i|$ is directed. As $|q_i|\subseteq |b_i|$ for any $i$, 
by Proposition \ref{Ab:13}, we have that 
\[
b_1 \sim b'_1*q_1, \ \ \ 
b_i \sim  b'_i*q_i *\con{b'_{i-1}}   \mbox{ for }     i=2,\ldots,n-1, \ \ \
b_n \sim   q_n*\con{b'_{n-1}}, 
\]
where $b'_i$ is a 1-simplex  such that 
\[
\partial_1{b'_i}= \partial_0q_i, \ \  \partial_0{b'_i}= \partial_0b_i,
\ \   |b'_i| = \partial_0b_i
\]
for $i=1,\ldots,n-1$. Hence 
\begin{align*}
p & = b_n*b_{n-1}\cdots b_2*b_1\\
& \sim q_n*\con{b'_{n-1}}* b'_{n-1}*q_{n-1}*  \ldots  *
b'_2*q_2*\con{b'_1}*b'_1*q_1\\ 
& \sim  q_n* b(\partial_0q_{n-1})*q_{n-1}*  \ldots  *
 b(\partial_0q_2)*q_2* b(\partial_0q_1)*q_1 \ \sim \ q,  
\end{align*}
completing the proof.
\end{proof}
\begin{lemma}
\label{Ada:4}
Assume that the elements of $\Po$ are arcwise and simply 
connected subsets of $\X$.
Let $\ga,\be$ be two curves with the same endpoints. 
If there exists a path $p$ such that $p\in App(\ga)\cap App(\be)$, 
then $\ga\sim\be$  
\end{lemma}
\begin{proof}
Since the path $p=\{b_n,\ldots,b_1\}$ is an approximation both 
of $\ga$ and $\be$, 
there are two partitions
$0=s_0 < s_1<\ldots <s_n=1$ and 
$0=t_0 < t_1<\ldots <t_n=1$, 
such that 
\[
\begin{array}{l}
\ga([s_{i-1},s_i])\subseteq |b_i|, \qquad \ga(s_{i-1})\in 
\partial_1b_i, \ \  \ga(s_i)\in \partial_0b_i, \\
\be([t_{i-1},t_i])\subseteq |b_i|, \qquad \be(t_{i-1})\in 
\partial_1b_i, \ \  \be(t_i)\in \partial_0b_i,
\end{array}
\]
for $i=1,\ldots,n$.
Let us define 
\[
\begin{array}{ll}
\ga_i(s) \defi \ga\big(s\cdot (s_i-s_{i-1}) + s_{i-1}\big) & s\in[0,1]\\
\be_i(t) \defi \be\big(t\cdot (t_i-t_{i-1}) + t_{i-1}\big) & t\in[0,1]
\end{array}
\]
for $i=1,\ldots,n$. Note that 
\[
\ga\sim \ga_n*\ldots *\ga_1, \qquad \be\sim \be_n*\ldots *\be_1,
\]
Since  $\ga_i(1),\be_i(1)\in\partial_0b_{i}$ 
and $\partial_0b_{i}$ is arcwise connected subset of $\X$, we can find 
a curve $\si_i$ for  such that 
\[
 \si_i([0,1])\subseteq \partial_0b_{i}, \ \si_i(0)=\ga_i(1), \ 
 \si_i(1)=\be_i(1), \qquad i=1,\ldots n-1.
\]
Let 
\[
\begin{array}{ll}
  \tau_1(t) \defi (\si_1*\ga_1)(t), \\
  \tau_i(t) \defi (\si_i*\ga_i*\con{\si_{i-1}})(t), &   2\leq i \leq n-1,\\
  \tau_n(t) \defi (\ga_n*\con{\si_{n-1}})(t). 
\end{array} 
\]
For $i=1,\ldots,n$, the curve $\tau_i$ is contained in 
$|b_i|$ and has the same endpoints of $\be_i$, thus 
$\tau_i$ is homotopic to $\be_i$ because $|b_i|$ is simply connected. 
Therefore,
\[
\begin{array}{ll}
\ga & \sim \ga_n *\ldots *\ga_2*\ga_1 \\
    & \sim  \ga_n *\con{\si_{n-1}}*\si_{n-1} *\ldots *
    \con{\si_2}*\si_2 *\ga_2* \con{\si_1}*\si_1 *\ga_1\\ 
    & = \tau_n*\cdots *\tau_2*\tau_1
     \ \sim  \ \be_n*\cdots *\be_2*\be_1 \ \sim \ \be,
\end{array}
\]
completing the proof.
\end{proof}
\begin{lemma}
\label{Ada:5}
Assume that the elements of $\Po$ are arcwise and  simply 
connected subsets of $\X$. Let $p,q\in\mathrm{P}(a_0,a_1)$ be 
respectively two approximations of a pair of  curve $\ga$ and $\be$
with the same endpoints. $p$ and $q$ are homotopic if, and only if, 
$\ga$ and $\be$ are homotopic. 
\end{lemma}
\begin{proof}
$(\Rightarrow)$ It is enough to prove the assertion in the case where
$q$ is an elementary ampliation of $p$. So let $p=\{b_n,\ldots,b_1\}$
and $q$ an ampliation of $p$ of the form 
$\{b_n,\ldots,b_{i+1}, \partial_0c,\partial_2c,b_{i-1}, \ldots b_1\}$
where $c\in\amp(b_i)$. Let 
$s_{1},t_{1},s_2,t_2\in[0,1]$ be such that  
$\ga(s_{1}), \be(t_{1}) \in \partial_1b_i$ and 
$\ga(s_2), \be(t_2) \in \partial_0b_i$.  We can decompose 
$\ga\sim \ga_3*\ga_2*\ga_1$ and $\be\sim\be_3*\be_2*\be_1$,
where
\[
\begin{array}{lcl}
\ga_1(s) \defi \ga\big(s\cdot s_{1}\big) & & 
\be_1(t) \defi \be\big(t\cdot t_{1}\big)\\
\ga_2(s) \defi \ga\big(s\cdot (s_2-s_{1}) + s_{1}\big) & & 
\be_2(t) \defi \be\big(t\cdot (t_2-t_{1}) + t_{1}\big) \\
\ga_3(s) \defi \ga\big(s\cdot (1 - s_2) +  s_2\big) & &
\be_3(t) \defi \be\big(t\cdot (1 - t_2) +  t_2\big),
\end{array}
\]
for $s,t\in[0,1]$.  In general 
$\ga_i$ and $\be_i$ might not have the same endpoints. So let 
$\si_1,\si_2$
be two curves such that $\si_1(0)=\ga(s_{1})$ $\si_1(1)=\be(s_{1})$
and $\si_1([0,1])\subseteq \partial_1b_i$, 
and $\si_2(0)=\ga(s_2)$ $\si_1(1)=\be(s_2)$ and 
$\si_2([0,1])\subseteq \partial_0b_i$.  
We now set 
\[
\tau_1 =  \si_1*\ga_1, \ \  \tau_2 =  \si_2*\ga_1*\con{\si_1}, \ \ 
\tau_3 =   \ga_3* \con{\si_2}
\]
Observe that $\ga\sim \tau_3*\tau_2*\tau_1$, and that 
for $i=1,2,3$ the curve  $\tau_i$ has the same endpoints of $\be_i$. 
Furthermore, by construction we have 
\[
\{b_{i-1},\ldots, b_1\}\in App(\tau_1) \cap App(\be_1), \ \ 
\{b_n,\ldots, b_{i+1}\}\in App(\tau_3) \cap App(\be_3), \ \
\]
and that $\tau_2$ and $\si_2$   are contained in
the support of $c$.  So, by Lemma \ref{Ada:4} $\tau_1\sim\be_1$ and 
$\tau_3\sim\be_3$. Moreover $\tau_2\sim\be_2$ because the support of $c$
is simply connected. Hence $\ga\sim\be$, completing the proof.\\ 
$(\Leftarrow)$ Let $h:[0,1]\times [0,1]\longrightarrow\X$ such that 
the curves $\ga_t(s)\defi h(t,s)$ satisfy 
$\ga_0(s)=\ga(s)$, $\ga_{1}(s)= \be(s)$ and 
$\ga_t(0)=x_0$, $\ga_t(1)=x_1$  for any $t\in[0,1]$. For any $t\in[0,1]$
let $p_t\in\mathrm{P}(a_0,a_1)$ be an approximation of $\ga_t$
such that $p_0=p$ and $p_1=q$. Now, let us define 
\[
S_t\defi \{l\in[0,1] \ | \ p_t \mbox{ is an approximation of } \ga_l\}.
\]
$S_t$ is nonempty. In fact $t\in S_t$  because $p_t$ is an approximation 
of $\ga_t$. Moreover $S_t$ is open. To see this assume 
$p_t=\{b_n,\ldots,b_1\}$. By definition of approximation there is a
partition $0=s_0 < s_1<\ldots <s_n=1$ of $[0,1]$
such that 
\[
\ga_t([s_i,s_{i+1}])\subseteq |b_{i+1}|, \qquad \ga_t(s_i)\in 
\partial_1b_{i+1}, \ \  \ga_t(s_{i+1})\in \partial_0b_{i+1}
\]
for $i=0,\ldots, n-1$.  By continuity of $h$ we can find $\eps_i>0$ such
that 
\[
\mbox{ for any } l\in(t-\eps_i, t+\eps_i) \ \ \Rightarrow \ \ 
\left\{
\begin{array}{l}
\ga_l([s_i,s_{i+1}])\subseteq |b_{i+1}| \\
\ga_l(s_i)\in \partial_1b_{i+1} \\
\ga_l(s_{i+1})\in \partial_0b_{i+1}
\end{array} \right.
\]
for any $i=0,\ldots,n-1$. So, if we define 
$\eps\defi \min \{\eps_i \ | \ i\in\{0,\ldots,n-1\} \}$
we obtain that $p_t$ is an approximation of $\ga_l$ for any 
$l\in(t-\eps,t+\eps)$, hence $S_t$ is open in the relative topology 
of $[0,1]$. Now,  for any $t\in[0,1]$, let $I_t\subseteq S_t$ be an open 
interval of $t$. Note that for any $l\in I_t$,  
$p_t$ is a approximation of $\ga_l$. 
By compactness we can find a finite open covering 
$I_{t_0},I_{t_1}, \ldots, I_{t_n}$ of $[0,1]$, where 
$0=t_0 <t_1 < \ldots < t_n=1$. We also have 
$I_{t_i}\cap I_{t_{i+1}} \ne \emptyset$ for any $i=0,\ldots, n-1$.  
This entails that  for any $i=0,\ldots, n-1$ there is $l_i$ such that 
$t_i\leq l_i\leq t_{i+1}$ and that 
$p_{t_i}$, $ p_{t_{i+1}}$ are approximations of $\ga_{l_i}$. 
By Lemma \ref{Ada:3c} we have  that $p_{t_i}$ and $p_{t_{i+1}}$ are
homotopic, completing the proof.    
\end{proof}
\begin{teo}
\label{Ada:6}
 Let $\X$ be a Hausdorff, arcwise connected topological space, and let 
 $\Po$ be a basis for the topology of $\X$ whose elements are arcwise and 
 simply connected subsets of $\X$. Then $\pi_1(\X)\simeq \pi_1(\Po)$.
\end{teo}  
\begin{proof}
 Fix a base 0-simplex $a_0$ and a base point $x_0\in a_0$. Define 
\[
\pi_1(\X,x_0)\ni [\ga]\longrightarrow [p]\in \pi_1(\Po,a_0)
\]
where $p$ is an approximation of $\ga$. By (\ref{Ada:2}) and  
Lemma \ref{Ada:5}, this map is  group isomorphism. 
\end{proof}
\begin{cor}
\label{Ada:7}
 Let $\X$ and $\Po$ be as in the previous theorem. 
 If $\X$ is nonsimply connected, then $\Po$ is not directed under 
 inclusion. 
\end{cor}
\begin{proof}
 If $\X$ is not simply connected, the by the previous theorem 
 $\Po$ is not simply connected. By Proposition \ref{Ab:13}, 
 $\Po$ is not directed.
\end{proof}
\subsubsection{Net-cohomology}
\label{Adb}
Let $\X$ be an arcwise connected, Hausdorff topological space.  
Let $O(\X)$ be  the set of  open subsets 
of $\X$ ordered under inclusion. Assume that $O(\X)$  is equipped 
with a causal disjointness relation $\perp$.
\begin{df}
\label{Adb:1}
We say that $\Po\subseteq O(\X)$ is a \textbf{good index set}
associated with $(\X,\perp)$ if $\Po$ is a basis for the topology 
of $\M$ whose elements are nonempty,
arcwise and simply connected subsets of $\M$  
with a nonempty causal complement. 
We denote by $\mathcal{I}(\X,\perp)$ the collection 
of  good index sets associated with $(\X,\perp)$.
\end{df}
Some observations are in order. \textit{First}, 
note that  $\mathcal{I}(\X,\perp)$ can be  empty. 
However, this does not happen in the applications we have in mind.
\textit{Secondly}, we have used the term  ``good index set'' because 
it is reasonable to assume that any index set of nets local algebras
over $(\X,\perp)$ has to belong to $\mathcal{I}(\X,\perp)$. 
This, to  avoid the ``artificial'' introduction  of  
topological obstructions because, by Theorem \ref{Ada:6}, 
$\pi_1(\Po)\simeq \pi_1(\X)$ for any $\Po\in\mathcal{I}(\X,\perp)$. \\[5pt]
\indent Given $\Po\in\mathcal{I}(\X,\perp)$, let us consider 
an irreducible net of local algebras $\Al_\Po$ defined on a Hilbert 
space $\mathcal{H}_o$. The first aim is to give
an answer to the question, posed at the beginning of this paper,
about the existence of topological obstructions to the triviality in 
$\Bh$ of 1-cocycles. 
To this end, note that if $\X$ is simply connected, 
then by, Theorem \ref{Ada:6}, $\pi_1(\Po)=\mathbb{C}\cdot \mathbbm{1}$.
Hence as a trivial consequence of Theorem \ref{Ac:4}
we have the following 
\begin{cor}
\label{Adb:2}
If $\X$ is simply connected, any 1-cocycle is trivial in $\Bh$, namely 
$\Zup= \Zutp$.
\end{cor}
On the grounds of this result, 
we can affirm that \textit{there might exists 
only a topological obstruction to the triviality in $\Bh$ of 1-cocycles:
the nonsimply connectedness of $\X$}. ``Might'' because we are 
not able to provide here an example of a 1-cocycle which is not trivial
in $\Bh$.\\[5pt]
\indent The next aim is to show that net-cohomology is stable under 
a suitable change of the index set. 
Let us start by observing that the notion 
of a locally relatively connected refinement
of a poset, Definition \ref{Aca:1},  induces an order
relation on $\mathcal{I}(\X,\perp)$. Given 
$\Po_1,\Po_2\in\mathcal{I}(\X,\perp)$,
define 
\begin{equation}
 \Po_1\ordpos \Po_2 \iff \Po_1 \mbox{ is a locally relatively 
                                      connected refinement of  } \Po_2.
\end{equation}
One can easily checks that $\ordpos$ is an order relation 
on $\mathcal{I}(\X,\perp)$. 
\begin{lemma}
\label{Adb:3}
The following assertions hold.\\
(a) Given $\Po\in\mathcal{I}(\X,\perp)$, let $\Po_1$ be   a subfamily 
    of $\Po$. If $\Po_1$ is a basis for the topology 
    of $\X$, then  $\Po_1\in\mathcal{I}(\X,\perp)$ and 
    $\Po_1\ordpos\Po$.\\
(b) $(\mathcal{I}(\X,\perp),\ordpos)$ is a directed poset with  a maximum 
    $\Po_{\mathrm{max}}$.
\end{lemma}
\begin{proof}
(a) follows from  the Definition \ref{Aca:1} and 
    from Lemma \ref{Aca:2}.   
(b) Define 
\[
 \Po_{\mathrm{max}} \defi \{\dc\subseteq \X \ | \dc\in\mathcal{P} 
            \mbox{ for some } \Po\in\mathcal{I}(\X,\perp)\} 
\] 
It is clear that $\Po_{\mathrm{max}}\in\mathcal{I}(\X,\perp)$. By (a),  
we have that $\Po\ordpos \Po_{\mathrm{max}}$ for any 
$\Po\in\mathcal{I}(\X,\perp)$. Hence 
$\Po_{\mathrm{max}}$ is the maximum.
\end{proof}
As an easy consequence of Theorem \ref{Aca:3}, we have the following  
\begin{teo}
\label{Adb:4}
Let $\Al_{\Po_{\mathrm{max}}}$ be an irreducible net,
defined on a Hilbert space $\mathcal{H}_o$,  
and indexed by $\Po_{\mathrm{max}}$. For any pair 
$\Po_1,\Po_2\in\mathcal{I}(\X,\perp)$ the categories 
$\mathcal{Z}^1_t(\Al_{\Po_{\mathrm{max}}|\Po_1})$, 
    $\mathcal{Z}^1_t(\Al_{\Po_{\mathrm{max}}})$ and 
    $\mathcal{Z}^1_t(\Al_{\Po_{\mathrm{max}}|\Po_2})$ are equivalent.
\end{teo}
\begin{oss}
\label{Adb:5}
Some observations on this theorem are in order.\\[3pt]
(1) The Theorem \ref{Adb:4} says that,  once 
a net of local algebras $\Al_{\Po_{\mathrm{max}}}$ is given,  
the category $ \mathcal{Z}^1_t(\Al_{\Po_{\mathrm{max}}})$ is an invariant 
of $\mathcal{I}(\X,\perp)$.\\[3pt] 
(2) Once an irreducible net $\Al_\Po$ indexed 
by an element $\Po\in \mathcal{I}(\X,\perp)$ is given, 
then it is assigned a net indexed by $\Po_{\mathrm{max}}$. 
In fact,  $\Po$ is a basis for the topology of $\X$, 
therefore by defining 
$\A(\dc)\defi (\cup\{\A(\dc_1) \ | \dc_1\in\Po, \ \dc_1\subseteq \dc\})''$,
for any $\dc\in\Po_{\mathrm{max}}$, we obtain 
an irreducible net $\Al_{\Po_{\mathrm{max}}}$  such that 
$\Al_{\Po_{\mathrm{max}}|\Po}= \Al_\Po$.\\[3pt]
(3)  Concerning the applications to the theory of superselection sectors,
we can assume, without loss of generality, the independence of the 
theory of the choice of the index set. 
\end{oss}
\section{Good index sets for  a globally 
hyperbolic spacetime}
\label{B}
In the papers  \cite{GLRV, Rob3} the index set used to study 
superselection sectors in a  globally hyperbolic spacetime $\M$
is the set $\Kr$ of regular diamonds. 
On the one hand, this is a good choice 
because $\Kr\in\mathcal{I}(\M,\perp)$.
But on the other hand,
regular diamonds do not need to have pathwise connected causal complements,
and to this fact are connected several problems 
(see the Introduction). A way to overcome these problems 
is provided  by  Theorem \ref{Adb:4}: 
it is enough to replace $\Kr$ with another good index set 
whose elements have pathwise connected causal complements.
The net-cohomology  is unaffected by this change 
and the mentioned problems are overcome. In this section 
we show that such a good index set 
exists: it is the set $\K$ of diamonds  of $\M$.
The net-cohomology of $\K$ will provide 
us important information for the theory of superselection sectors. 
We want to stress that throughout both this section and in Section \ref{C}, 
by a globally hyperbolic spacetime we will mean a globally hyperbolic 
spacetime with dimension $\geq 3$.
\subsection{Preliminaries on spacetime geometry}
\label{Ba}
We recall some basics on the causal structure of spacetimes 
and establish our notation. Standard references for this topic 
are \cite{One, Wal, EH}.\\[5pt]
\indent A \textit{spacetime} $\M$ consists of 
a Hausdorff, paracompact,  smooth, oriented
manifold  $\M$, with dimension $\geq 3$,  endowed
with  a smooth metric $\g$  with signature $(-,+,+,\ldots,+)$, and with
a time-orientation, that is a  smooth timelike vector field $v$,
(throughout this paper smooth means $C^\infty$).
A  curve $\ga$ in $\M$ is a continuous,  piecewise smooth, 
regular function $\ga:I \longrightarrow \M$,  
where $I$ is a connected subset of $\R$ with nonempty interior. It
is  called timelike, lightlike, spacelike 
if respectively $\g(\dot{\ga},\dot{\ga})<0$, $=0$, $> 0$
all along $\ga$, where $\dot{\ga}=\frac{d\ga}{dt}$.
Assume now that $\ga$ is \textit{causal}, i.e.
a  nonspacelike curve; we can classify it according to the 
time-orientation $v$ as future-directed (f-d) or 
past-directed (p-d) if respectively 
$\g(\dot{\ga}, v) < 0, >0$ all along $\ga$. When  $\ga$ is f-d and 
$\lim_{t\rightarrow\sup I}\ga(t)$ exists
($\lim_{t\rightarrow\inf I}\ga(t)$), then it is said to have 
a future (past) endpoint. Otherwise, it is said to be 
future (past) endless;  $\ga$ is said to be endless if neither  of them
exist. Analogous  definitions are assumed for p-d causal curves.\\
\indent The \textit{chronological future} $\I^+(S)$,
the \textit{causal future} $\J^+(S)$
and the \textit{future domain of dependence } $\D^+(S)$
of a subset $S\subset\M$ are defined as:
\begin{align*}
&\I^+(S)  \defi  \{ x\in\M \ | \ \mbox{there is a f-d timelike curve
    from $S$ to $x$ } \}; \\
&\J^+(S)  \defi   S \cup \{ x\in\M \ | \ \mbox{there is a f-d  causal
    curve from $S$ to $x$ } \}; \\
&\D^+(S) \defi   \{ x\in\M \ | \ \mbox{any p-d endless causal
    curve through $x$ meets $S$ } \}.
\end{align*}
These definitions
have a dual in which ``future'' is replaced by ``past''
and the $+$ by $-$.  So,
we define $\I(S) \defi \I^+(S) \cup  \I^-(S)$,
$\J(S)\defi\J^+(S) \cup  \J^-(S)$ and $\D(S) \defi\D^+(S)\cup\D^-(S)$.  
We recall that: \textbf{1.} $\I^+(S)$ is an open set; 
  \textbf{2.} $\I^+(cl(S)) = \I^+(S)$; 
  \textbf{3.}  $cl(\J^+(S)) = cl(\I^+(S))$ and 
$int(\J^+(S)) = \I^+(S)$\footnote{$cl(S)$ and 
 $int(S)$ denote respectively the closure and the internal part 
                              of the set $S$.}. 
Furthermore,  by (\textbf{2.}) + (\textbf{3.}) we have that 
\textbf{4.}  $cl\big(\J^+(S)\big) =  cl\big(\J^+(cl(S))\big)$.
A subset $S$ of $\M$ is \textit{achronal} (\textit{acausal})
if for any pair $x_1,x_2\in S$ we have $x_1\not\in\I(x_2)$ 
($x_1\not\in\J(x_2)$). Two subsets $S_1,S_2\subseteq \M$, are 
said to be \textit{causally disjoint},  whenever 
\begin{equation}
\label{Ba:0}
S_1\perp S_2 \iff S_1\subseteq \M\setminus \J(S_2)
\end{equation}
A (\textit{acausal}) \textit{Cauchy surface} 
is an achronal (acausal) set $\C$ verifying  $\D(\C)=\M$. 
Any Cauchy surface is a closed, arcwise connected, Lipschitz hypersurface 
of $\M$. Furthermore all the Cauchy surfaces are homeomorphic.
A \textit{spacelike} Cauchy surface is a smooth Cauchy surface 
whose tangent space is everywhere spacelike. It turns out that 
any spacelike Cauchy surface is acausal. \\[5pt] 
\indent A spacetime $\M$ satisfies the 
   \textit{strong causality condition} if 
   the following property is verified for any point $x$ of $\M$: 
   any open neighborhood $U$ of $x$ contains an open neighborhood 
   $V$ of $x$ such that for any  pair 
    $x_1,x_2\in V$ the set $\J^+(x_1)\cap \J^-(x_2)$ is either empty or 
   contained in $V$. The spacetime is said to  be 
   \textit{globally hyperbolic}  if it satisfies the strong causality 
   condition and if for any pair $x_1,x_2\in\M$,  the set 
   $\J^+(x_1)\cap \J^-(x_2)$ is either 
   empty or compact. It turns out that $\M$ is globally hyperbolic 
   if, and only if,  it admits a Cauchy surface.  
   We recall that if $\M$ is a globally hyperbolic spacetime, 
   for any relatively compact  
   set $K$ we have: \textbf{5.} $\J^+(cl(K))$ is closed; 
   \textbf{6.} $\D^+(cl(K))$ is compact;  by the properties 
   \textbf{4.}  and  \textbf{5.} we have that 
   \textbf{7.} $\J^+(cl(K))=cl\big(\J^+(K) \big)$.\\[5pt] 
\indent Although, a globally hyperbolic spacetime $\M$ 
can be continuously ( smoothly ) foliated 
by (spacelike) Cauchy surfaces \cite{BS2}, for our purposes it is enough 
that for any Cauchy surface $\C$ the spacetime $\M$
 admits a foliation ``based'' on $\C$, that is  
 there exists  a 3-dimensional manifold  $\Si$ and a homeomorphism
 $F:\mathbb{R}\times \Si \longrightarrow \M$ such that 
\[
 \Si_t\defi F(t,\Si)  \ \mbox{ are topological hypersurfaces of  } \M, \ 
  \Si_0 = \C,
\] 
but, in general,  for $t\ne 0$ the surface $\Si_t$ need  not be a 
Cauchy surfaces \cite{BS1}.
\begin{lemma}
\label{Ba:1}
Let $\M$ be a globally hyperbolic spacetime. 
Then $\pi_1(\M)$ is isomorphic to $\pi_1(\C)$
for any Cauchy surface of $\M$. Any
curve $\ga:[0,1]\longrightarrow \M$ 
whose endpoints lie in $\C$ is homotopic to a curve and 
lying in $\C\setminus\{x\}$ for any $x\in\M$ with $x\ne\ga(0),\ga(1)$. 
\end{lemma}
\begin{proof}
Let $F$ be the foliation of $\M$ based on $\C$ as  described above. 
Let $(\tau(x),y(x))\defi F^{-1}(x)$ for $x\in\M$. Note that 
\[
 h(t,x)\defi F\left((1-t)\cdot \tau(x),y(x)\right) \qquad t\in[0,1], 
                                         \ \ x\in\M  
\]
is a deformation retract. Hence $\pi_1(\M)$ is isomorphic 
to $\pi_1(\C)$. Let  
$h_1(t,s)\defi h(t,\ga(s))$. Then  curve $\ga(s)=h_1(0,s)$ is 
homotopic to the curve $\be(s)\defi h_1(1,s)$ lying in $\C$.
Given $x\in\M$ with $x\ne\be(1),\be(0)$. It is clear that, 
as  $\C$ is 3-dimensional surface, 
$\be$ is homotopic in $\C$ to a curve 
$\si$ lying in $\C\setminus\{x\}$. 
\end{proof}
Now,  note that the relation $\perp$,
defined by (\ref{Ba:0}), is a causal disjointness relation 
on the poset $O(\M)$ formed by  
the open sets of $\M$ ordered under inclusion.  
\begin{lemma}
\label{Ba:2}
Let $\Po\in\mathcal{I}(\M,\perp)$ (see Section \ref{Adb}).
If $\M$ has compact Cauchy surfaces, then $\Po$ is not directed
under inclusion.
\end{lemma}
\begin{proof} 
Let  $\dc_1,\ldots, \dc_n$ be a finite covering of a Cauchy surface $\C$ 
of $\M$. If $\Po$ were directed then we could find an element
$\dc\in\Po$ with $\dc_1\cup\cdots \cup\dc_n\subseteq \dc$.
Then $\C\subseteq \dc$. But,  
by definition of causal disjointness relation there exists 
$\dc_0\in\Po$ with $\dc\perp \dc_0$. This leads to a contradiction 
because  $\dc_0\subset\M\setminus\J(\dc)\subset 
          \M\setminus\J(\C)=\emptyset$ 
(see Definition \ref{Adb:1}).
\end{proof}
\subsection{The set of diamonds}
\label{Bb}
Consider a globally hyperbolic spacetime $\M$.
We have already observed that the set of regular diamonds $\Kr$
of $\M$ is an element of  the set of indices  $\mathcal{I}(\M,\perp)$
associated with $(\M,\perp)$, 
where  $\perp$ is the relation defined by (\ref{Ba:0}). 
We now introduce the set of diamonds $\K$ of $\M$. We  prove that $\K$
is a locally relatively connected refinement 
of $\Kr$, and that diamonds  have pathwise connected causal
complements. The last part of this section is devoted to study 
the causal punctures of $\K$ induced by points of the spacetime.
\begin{df}
\label{Bb:1}
Given a spacelike Cauchy surface $\C$, we denote by 
$\mathfrak{G}(\C)$ the collection of the open subsets $G$ of $\C$
of the form $\phi(B)$, where 
$(U,\phi)$  is a chart of $\C$ and  
$B$ is an open  ball of 
$\mathrm{R}^3$ with $cl(B) \subset\phi^{-1}(U)$.
We call a 
\textbf{diamond}\footnote{The author is grateful to Gerardo Morsella
for a fruitful discussion on the definition of a diamond of $\M$.}
of $\M$  a subset $\dc$ of the form 
$\D(G)$ where  $G\in\mathfrak{G}(\C)$
for some spacelike Cauchy surface $\C$: $G$ is called the base of $\dc$
while $\dc$ is said to be based on $\C$. We denote by $\K$ the collection
of diamonds  of $\M$.
\end{df}
\begin{prop}
\label{Bb:2}
$\K$ is a basis for the topology of $\M$. Any diamond 
$\dc$ is a relatively compact, arcwise and simply connected,
open subset of $\M$. $\K\in\mathcal{I}(\M,\perp)$ and it 
is a locally relatively connected refinement of $\Kr$
\end{prop}
\begin{proof}
$\K$ is a basis  for the topology of $\M$ because 
$\M$ is foliated by spacelike Cauchy surfaces.
Observe that any $G\in\mathfrak{G}(\C)$, is  an arcwise and  
simply connected,  spacelike hypersurface of $\M$. This entails that,
(see \cite[Section 14, Lemma 43]{One} ) $\D(G)$ is an open subset of 
$\M$. Furthermore,  $\D(G)$ considered as a spacetime, is 
globally hyperbolic and $G$ is a spacelike Cauchy surface of $\D(G)$.
Since $G$ is simply connected, by Lemma \ref{Ba:1}, $\D(G)$ is simply 
connected. Moreover, note that  $G$ is relatively compact in $\C$. 
As $\C$ is closed in $\M$,   $G$ is relatively compact in $\M$. 
By \textbf{6.}, $\D(G)$ is relatively compact in $\M$. Finally,
$\K\subset \Kr$ (see definition of $\Kr$ in \cite{GLRV}).
As $\K$ is a basis for the topology of $\M$, then 
$\K$ is a locally relatively connected  refinement 
of $\Kr$ and $\K\in\mathcal{I}(\M,\perp)$ (see Section \ref{Adb}). 
\end{proof}
The next aim is to show that the  causal complement 
$\dc^\perp$ of a diamond, which is defined as 
\[
\dc^\perp= \{\dc_1\in\K \ | \ \dc_1\perp \dc\} 
\]
(see Section \ref{Aa}) is pathwise connected in $\K$. 
To this end, by  (\ref{Ada:3b}), 
it is enough to prove that 
$\M_{\dc^\perp} = \cup  \{\dc_1\in\K \ | \dc_1\perp \dc\}$ is arcwise 
connected in $\M$ because $\dc^\perp$ is a sieve of $\K$.
\begin{lemma}
\label{Bb:4}
The following assertions hold.\\
(a) $ \M_{\dc^\perp} = 
     \M\setminus cl\big(\J(\dc)\big) = \M\setminus \J\big( cl(\dc) \big)$ 
    for any $\dc\in\K$.\\
(b) If  $\dc=\D(G)$ for $G\in\mathfrak{G}(\C)$, then
    $\M_{\dc^\perp} = \D(\C\setminus cl(G))$.
\end{lemma}
\begin{proof}
(a) By \textbf{7.} we have that 
$\M\setminus \J(cl(\dc)) = \M\setminus cl\big(\J(\dc)\big)$, because 
$\dc$ is relatively compact. 
If $\dc_1\perp\dc$, then $\dc_1\subset int(\M\setminus\J(\dc))$.
This entails that  
$\M_{\dc^\perp}\subseteq \M\setminus \J\big( cl(\dc) \big)$. 
As $\M\setminus \J\big( cl(\dc) \big)$ is an open set and 
$\K$ is basis for the topology of $\M$, for any 
$x\in \M\setminus \J\big( cl(\dc) \big)$
we can find $\dc_1\in\K$ such that $x\in\dc_1$,  
$ \dc_1\subseteq\M\setminus \J\big( cl(\dc) \big)$. Thus
$\dc_1\perp \dc$, and $x\in\M_{\dc^\perp}$,  completing the proof. 
(b) Since $cl(G)$ is compact in $\M$, by (a)  we have 
\[
\M_{\dc^\perp} = \M\setminus cl(\J(\D(G))) 
  = \M\setminus cl(\J(G)) =\M\setminus \J(cl(G)),
\]
where the identity $\J(\D(G))=\J(G)$ has been used. Therefore, 
as $\D(\C\setminus cl(G))\subseteq \M\setminus \J(cl(G))$, 
the inclusion 
$\D( \C\setminus cl(G))\subseteq \M_{\dc^\perp}$
is verified. If
$x\in \M_{\dc^\perp}=\M\setminus \J\big( cl(G)\big)$, then 
any p-d endless causal curve through $x$ 
meets the Cauchy surface $\C$ in $\C\setminus cl(G)$.
Therefore, $x\in \D\big( \C\setminus cl(G)\big)$ and 
$\M_{\dc^\perp}\subseteq  
\D(\C\setminus cl(G))$ which completes the proof.
\end{proof}
\begin{prop}
\label{Bb:5}
The causal complement $\dc^\perp$ of a diamond  $\dc$ is 
pathwise connected in $\K$.
\end{prop}
\begin{proof}
 Let $\dc\in\K$ be of the form $\D(G)$, where 
  $G\in\mathfrak{G}(\C)$ and  $G=\phi(B)$ with $(U,\phi)$ is a chart of $\C$.
  By definition of $\mathfrak{G}(\C)$ there is  an open  ball $B_1$ such that 
  $cl(B)\subset B_1$, $cl(B_1)\subset\phi^{-1}(U)$.   
  As $\phi(B_1)\setminus cl(\phi(B))$ is arcwise connected in $\C$,
  $\C\setminus cl(G)$ is arcwise connected in $\C$. Now,
  by the previous lemma $\M_{\dc^\perp} = \D(\C\setminus cl(G)))$, 
  which  is a globally hyperbolic set with an arcwise connected 
  Cauchy surface $\C\setminus cl(G)$. Therefore, 
  $\M_{\dc^\perp}$ is arcwise connected, hence $\dc^\perp$ is 
  pathwise connected in $\K$. 
\end{proof}
As claimed at the beginning of Section \ref{B}, 
we have established that $\K$ is a locally relatively connected refinement 
of $\Kr$, and that any element of $\K$ has  a pathwise 
connected causal complement. From now on  
we will focus on $\K$, because this will be the index set 
that we will use to study superselection sectors.
\begin{lemma}
\label{Bb:6}
Let $\dc\in\K$ and let $U$ be an open neighborhood 
of $cl(\dc)$. There exist  $\dc_1,\dc_2\in\K$ 
such that $cl(\dc)\subset \dc_1$, $cl(\dc_1)\subset U$, and  
$cl(\dc_2)\subset U$, $\dc_2\perp \dc_1$.
\end{lemma}
\begin{proof}
 Assume that $\dc=\D(G)$ with $G=\phi(B)$, where 
    $(\phi,W)$ is  a chart of a spacelike Cauchy surface $\C$, 
   and $B$ is a ball of $\mathbb{R}^3$ such that  
   $cl(B)\subseteq \phi^{-1}(W)$. As $cl(\dc)\subset U$, 
   $cl(B)$ is contained in the open set $\phi^{-1}(W\cap U)$. 
   Therefore, there exists a ball $B_1$ such that 
   $cl(B) \subset B_1$ and $cl(B_1)\subset \phi^{-1}(W\cap U)$.
   Moreover, the latter inclusion entails that there is a ball 
   $B_2$ such that $cl(B_2)\subset\phi^{-1}(W\cap U)$ and 
   $cl(B_2)\cap cl(B_1)=\emptyset$. Therefore, the diamonds 
   $\dc_1\defi \D(\phi(B_1))$, $\dc_2\defi\D(\phi(B_2))$  verify the 
   property  written in the statement. 
\end{proof}
As a trivial consequence of this lemma we
have that if $U$ is an open neighborhood of $cl(\dc)$, then 
there exist $\dc_1,\dc_3\in\K$ such that 
$cl(\dc),\subset \dc_1$, $cl(\dc_1)\subset U$ and 
$cl(\dc_3)\subset \dc_1$, and $\dc_3\perp \dc$. 
\subsubsection{Causal punctures}
\label{Bba}
The \textit{causal puncture of $\K$} induced by a point  $x\in\M$, is 
the poset $\K_x$ defined as the collection  
\begin{equation}
\label{Bba:1}
\K_x \defi \{\dc\in\K \ | \ cl(\dc)\perp x\}.
\end{equation}
ordered under inclusion, where  $cl(\dc)\perp x$ means that 
$cl(\dc)\subseteq \M\setminus\J(x)$.
The causal puncture $\K_x$ is a sieve of $\K$, hence, 
some properties of $\K_x$ can be deduced by studying  its topological
realization $\M_x\defi \cup\{\dc\in\K \ | \ \dc\in\K_x \}$.
\begin{lemma}
\label{Bba:2} 
Given $x\in\M$, then 
$\M_x = \M\setminus \J(x) = \D(\C\setminus \{x\})$
for some spacelike Cauchy surface $\C$  that meets $x$.
\end{lemma}
\begin{proof}
The set  $\M\setminus \J(x)$ is open. If $y\in\M\setminus \J(x)$,
it follows by the definition of $\K$, that  there is $\dc\in\K$ with 
$y\in\dc$ and $cl(\dc)\subseteq \M\setminus \J(x)$, namely 
$y\in\M_x$. Therefore $\M\setminus \J(x)\subseteq \M_x$. The opposite 
inclusion is obvious, completing the proof of the first identity. 
As $\M$ can be foliated by spacelike Cauchy surfaces, there is 
a spacelike Cauchy surface $\C$  that meets $x$. Now the proof
proceed as  in Lemma \ref{Bb:4}b.
\end{proof}
Considered as a spacetime $\M_x$ 
is globally hyperbolic \cite{Ruz2}. An element $\dc\in\K_x$ 
does not need to be a diamond  of the spacetime $\M_x$. 
However,  $\K_x$ is  a basis for the topology of $\M_x$.
Furthermore as $\M_x$ is arcwise connected, $\K_x$ is  pathwise connected. 
Now, for any $\dc\in\K_x$ we define 
\begin{equation}
\label{Bba:3}
\dc^{\perp}|_{\K_x}  \defi \{ \dc_1\in\K_x \ | \ \dc_1\perp \dc\},
\end{equation}
namely,  the causal complement of $\dc$ in $\K_x$. 
\begin{lemma}
\label{Bba:4}
$\dc^\perp|_{\K_x}$ is pathwise connected in $\K_x$ for any $\dc\in\K_x$.
\end{lemma}
\begin{proof}
Note that $\dc^{\perp}|_{\K_x}$ is a sieve, hence its enough to 
prove that $\cup\{\dc_1\in\K_x \ | \ \dc\perp \dc_1\}$ is 
arcwise connected in $\M$.
Assume that $\dc=\D(G)$ where $G\in\mathfrak{G}(\C)$ for a spacelike 
Cauchy surface $\C$ of $\M$.  
By Lemma \ref{Bb:4}b, 
$\M_{\dc^\perp} = \D(\C\setminus cl(G))$. As $\D(\C\setminus cl(G))$
is a globally hyperbolic spacetime, there is 
a spacelike Cauchy surface $\C_1$ that meets $x$. 
By \cite[Lemma 6]{Ve1} $\C_1\cup cl(G)$ is an acausal Cauchy surface 
of $\M$ that meets $x$. Hence, by \cite[Proposition 3.1]{Ruz2},   
$\C_2\defi (\C_1 \setminus \{x\}) \cup cl(G)$ is 
an acausal Cauchy surface of $\M_x$. In other words, 
$\dc$  is a set of the form $\D(G)$ with  $G\subset\C_2$,  where $\C_2$  
is an acausal Cauchy surface of $\M_x$. 
Now,  as in Proposition \ref{Bb:5}, 
$\C_2\setminus cl(G)$ is arcwise connected in $\M_x$. Furthermore,
$\cup \{ \dc_1 \in \dc^\perp|_{\K_x} \}$ $= \D(\C_2\setminus cl(G))$.
This is an arcwise connected set in $\M_x$, therefore 
$\dc^\perp|_{\K_x}$ is pathwise connected in $\K_x$.
\end{proof}
For any  $\dc\in\K$ with $x\in\dc$, let us define 
\begin{equation}
\label{Bba:4a}
\K_x|_\dc \defi \{\dc_1\in\K_x \ | \ \dc_1\subseteq \dc\},
\end{equation}
Note that $\K_x|_\dc$ is a sieve of $\K$.
\begin{lemma}
\label{Bba:4b}
Let $\dc\in\K$ with $x\in\dc$. Then $\K_x|_\dc$ is pathwise 
connected.
\end{lemma}
\begin{proof}
$\dc$ is a globally hyperbolic spacetime, therefore
there is a spacelike Cauchy surface $\C$ of $\dc$ that meets $x$.
Our aim is to show that 
$\D(\C\setminus \{x\}) = \cup \{ \dc_1\in\K_x|_\dc\}$. 
If this holds, since 
$\K_x|_\dc$ is sieve and  $\D(\C\setminus \{x\})$ is arcwise connected,
by (\ref{Ada:3b}),  $\K_x|_\dc$ is pathwise connected. 
We obtain the proof of this equality in two steps. 
First.  By Lemma \ref{Bba:2} we have that 
$\cup \{\dc_1 \in  \K_x|_\dc \}$ is contained in the open set 
$(\M\setminus \J(x))\cap \dc$. 
For any $x_1\in (\M\setminus \J(x))\cap \dc$, let 
$\dc_1\in\K$ with $cl(\dc_1)\subseteq (\M\setminus \J(x)) \ \cap \ \dc$. 
This entails that, $x_1\in \cup \{\dc_1 \in  \K_x|_\dc\}$. Therefore,
\[
\cup \{\dc_1 \in  \K_x|_\dc \}=(\M\setminus \J(x))\cap \dc. \qquad(*) 
\]
Secondly, note that  
$\D(\C\setminus \{x\})\subseteq (\M\setminus \J(x))\cap \dc$, because
$\C\setminus\{x\}\subseteq (\M\setminus \J(x))\cap \dc$. 
Let $x_2\in (\M\setminus \J(x))\cap \dc$. Then, any f-d endless causal 
curve through $x_2$ meets $\C$ in $\C\setminus \{x\}$,
therefore $x_2\in \D(\C\setminus \{x\})$ and 
$\D(\C\setminus \{x\}) =(\M\setminus \J(x))\cap \dc$. This and $(*)$,
entail that $\cup \{\dc_1 \in  \K_x|_\dc \}=
\D(\C\setminus \{x\})$, completing the proof.
\end{proof}
As the last issue of this section, consider the set 
$\K_x\times\K_x$ and endow it with    
the order relation defied as 
\[
 (\dc_1,\dc_2) \leq (\dc_3,\dc_4) \ \iff \ \dc_1\subseteq \dc_3 \mbox{ and }
\dc_2\subseteq \dc_4
\]
The graph 
\begin{equation}
\label{Bba:5}
\K^\perp_x \defi \{ (\dc_1,\dc_2)\in \K_x\times\K_x \ | \ 
   \dc_1\perp \dc_2 \}
\end{equation}
of the relation $\perp$ in $\K_x$ is pathwise connected in 
$\K_x\times\K_x$. First note that $\K_x\times\K_x$ is a basis for the topology
of $\M_x\times\M_x$, and that $\K^\perp_x$ is a sieve in  $\K_x\times\K_x$. 
Now,  the set $\cup \{ (\dc_1,\dc_2)\in \K^\perp_x\}$  
is equal to 
$\M^\perp_x\defi \{(x_1,x_2)\in \M_x\times\M_x \ | \ x_1\perp x_2 \}$. 
As observed $\M_x$ is globally hyperbolic. 
By \cite[Lemma 2.2]{GLRV},  $\M^\perp_x$ is an arcwise connected set of 
$\M_x\times \M_x$. 
By (\ref{Ada:3b}), $\K^\perp_x$ is pathwise connected in 
$\K_x\times\K_x$.
\subsection{Net-cohomology}
\label{Bc}
Before studying the net-cohomology 
of $\K$, it is worth showing how the topological properties 
of the spacetime stated in Lemma \ref{Ba:1} are codified in the 
poset structure of $\K$.  
\begin{lemma} 
\label{Bc:1}
The following properties hold. \\
(a) $\pi_1(\K)\simeq \pi_1(\M)\simeq\pi_1(\C)$ for any Cauchy surface 
     $\C$ of $\M$.\\
(b) Consider a path  $p\in\mathrm{P}(a_0)$ where $a_0\in\Si_0(\K)$ is, 
as a diamond,  based on a spacelike Cauchy surface $\C_0$. 
Let $x\in\C_0$ such that 
$cl(a_0) \cap x = \emptyset$. Then $p$ is homotopic 
to a path $q=\{b_n,\ldots,b_1\}\in\mathrm{P}(a_0)$ such that 
$|b_i|$, as a diamond,  is based on $\C_0$ and $|b_i|\cap x=\emptyset$
for any $i$. 
\end{lemma}
\begin{proof}
(a) follows from Theorem \ref{Ada:6} and from Lemma \ref{Ba:1}. 
(b) As observed in Section \ref{Ad}, since the  elements of $\K$ are 
arcwise connected sets of $\M$,  there exists
a curve $\ga:[0,1]\longrightarrow \M$, with 
$\ga(0)=\ga(1)\in a_0\cap \C_0$, and such that $p\in App(\ga)$. 
By Lemma \ref{Ba:1} $\ga$ is homotopic to a closed 
curve $\be$  lying  in $\C_0\setminus\{x\}$. 
This allows us to find a path $q\in App(\be)$ such that 
the elements $q$, as diamond,  are based on $\C_0$. 
Lemma \ref{Ada:5} completes the proof.
\end{proof}
Let $\Al_\K$ be an irreducible  net of local algebras defined on a  
Hilbert space $\mathcal{H}_o$. Let $\Zu$ be the set of 1-cocycles 
of $\K$ with values on $\Al_\K$ and let use denote by $\Zut$ those 
elements  of $\Zu$ which are trivial in $\Bh$.  As a trivial 
application of Corollary \ref{Adb:2}, we have that
if $\M$ is simply connected,   then $\Zu=\Zut$. 
This result answers  the  question posed at the beginning of 
this paper, saying that \textit{the compactness of the Cauchy surfaces
of the spacetime is not a topological obstruction
to the  triviality in $\Bh$ of 1-cocycles}. As already observed, 
the only possible obstruction in this sense is the nonsimply connectedness of 
the spacetime.\\
\indent The next proposition will turn out to be fundamental 
for the theory of superselection sectors
because it provides a way to prove triviality in $\Bh$ of 1-cocycles 
on an arbitrary  globally hyperbolic
spacetime.
\begin{prop}
\label{Bc:3}
Assume that $z\in\Zu$ is path-independent on $\K_x$ for  any point
$x\in\M$. Then $z$ is path-independent on $\K$, therefore $z\in\Zut$.
\end{prop}
\begin{proof}
Let $p\in\mathrm{P}(a_0)$ and let $\C_0$ be the Cauchy surface 
where $a_0$ is based.  Let us take $x\in\C_0$ 
such that $cl(a_0)\cap x= \emptyset$.
By Lemma \ref{Bc:1}b $p$ is homotopic 
to a path $q\in \mathrm{P}(a_0)$ whose elements are based on 
$\C_0\setminus\{x\}$. This means that $q\in\K_x$.
$z(p)=z(q)=\mathbbm{1}$ because $p$ and $q$ are homotopic and because 
$z$ is path-independent on  $\K_x$ for any $x\in\M$. 
\end{proof}
\section{Superselection sectors}
\label{C}
We begin the study of the superselection sectors 
of a net of local observables on an arbitrary  
globally hyperbolic spacetime  $\M$, with dimension $\geq 3$. 
We start by describing the setting in which we study superselection
sectors.  Afterwards we explain the strategy  we will follow, 
which consists in deducing  the global properties
of superselection sectors from   the local ones.   
We refer the reader to the appendix for all 
the categorical notions used in this section.\\[5pt]
\indent Let $\K$ be the set of diamonds of $\M$. 
We consider an irreducible  net 
$\Al_\K:\K\ni\dc\longrightarrow \A(\dc)\subseteq\Bh$ of local algebras 
defined on a fixed infinite dimensional separable 
Hilbert space $\mathcal{H}_o$. We assume that $\Al_\K$ satisfies 
the following two properties. \\[5pt]
$\bullet$ \textit{Punctured Haag duality}, that means that  
\begin{equation}
\label{C:1}
\A(\dc_1) \ = \ \cap\big\{ \A(\dc)' \ | \ 
		\dc\in\K_x, \  \dc \perp \dc_1 \} \qquad \dc_1\in\K_x
\end{equation}
for any $x\in\M$, where $\K_x$
is the causal puncture of $\K$ induced  by $x$ (\ref{Bba:1}).\\[5pt]
$\bullet$ The \textit{Borchers property}, that means that 
given  $\dc\in\K$ there is $\dc_1\in\K$ with $\dc_1\subset \dc$ such that 
for any orthogonal  projection $E\in\A(\dc_1)$, $E\ne 0$ there exists
an isometry $V\in\A(\dc)$ such that $V\cdot V^* = E$\\[5pt]
Let $\Zut$ be the $\mathrm{C}^*-$category of 1-cocycles of $\K$,
trivial in $\Bh$, with values in $\Al_\K$.  Then, the   
\textit{superselection sectors} are the 
equivalence classes $[z]$ of the irreducible 
elements $z$ of $\Zut$. From now on, our aim will be to prove 
that $\Zut$ is a tensor $\mathrm{C}^*-$category with a symmetry, 
left-inverses, 
and that any object with finite statistics has conjugates.  
Note, that by the Borchers property, $\Zut$ is closed under direct sums 
and subobjects.\\[5pt]
\indent We now discuss the differences between our setting and that used in 
\cite{GLRV, Rob3}. \textit{First},  we have used the set 
of diamonds $\K$, instead of the set of regular diamonds $\Kr$, 
as index set of the net of local algebras.
\textit{Secondly}, we assume punctured Haag duality while 
in the cited papers the authors assume  \textit{Haag duality}, 
that is 
\begin{equation}
\label{C:3}
\A(\dc_1) \ = \ \cap\big\{ \A(\dc)' \ | \ \dc\in\K, \  \dc \perp \dc_1 \},
\end{equation} 
for any $\dc_1\in\K$. Punctured Haag duality was 
introduced in \cite{Rob3}. Both the existence of models satisfying 
punctured Haag duality  and the relation of this property 
to other properties of $\Al_\K$ have been shown  in
\cite{Ruz2}.  
It turns out  that punctured Haag duality entails 
Haag duality and that $\Al_\K$  is 
\textit{locally definite}, namely 
\begin{equation}
\label{C:4}
\mathbb{C}\cdot \mathbbm{1} =  \cap\big\{ \A(\dc) \ | \ \dc\in\K, 
\ x\in  \dc\}. 
\end{equation} 
for any $x\in\M$. The reason why we assume punctured Haag duality 
will become clear in the next section. 
\begin{oss}
It is worth observing that in \cite{Ruz2}, punctured Haag duality 
has been shown  for the net of local 
algebras $\mathcal{F}_{\Kr}$, indexed by the set of regular diamonds, 
and associated with the free Klein-Gordon field in the representation
induced by quasi-free Hadamard states. One might wonder 
if this property  holds also for 
the net of fields $\mathcal{F}_{\Kr|\K}$  obtained by restricting
$\mathcal{F}_{\Kr}$ to $\K$. 
The answer is yes, because the 
net $\mathcal{F}_{\Kr}$ is  additive\footnote{
The net $\Al_{\Po_\mathrm{max}}$ is additive, 
if given $\dc\in\Po_\mathrm{max}$ and a covering  $\cup_i\dc_i =\dc$, then 
$\A(\dc) = (\cup_i \A(\dc_i))''$.}. 
As observed in Section \ref{B}, 
$\K\in\mathcal{I}(\M,\perp)$ and $\K\ordpos\Kr$. Then, 
it can be easily checked
that punctured Haag duality for  $\mathcal{F}_{\Kr}$ entails
punctured Haag duality for  $\mathcal{F}_{\Kr|\K}$.
\end{oss}
\subsection{Presheaves and the strategy for studying 
                       superselection sectors}
\label{Ca}
The way we study superselection sectors resembles  a standard argument 
of differential geometry.
To prove the existence 
of global objects, like for instance 
the affine connection in a Riemannian manifold, 
one first shows that these objects exist locally, afterwards
one checks that these local constructions
can be glued together to form an object defined over all the manifold. 
Here, the role of the manifold
is played by the category $\Zut$ and the objects that we want 
to construct  are  a tensor product, a symmetry and a conjugation.
To see what categories play the role of 
``charts'' of $\Zut$  some preliminary notions are necessary.
The  $\mathrm{C}^*-$\textit{presheaf associated} with $\Al_{\K}$ is  
the correspondence $\K\ni\dc\longrightarrow \A(\dc^\perp)$
which associates the
$\mathrm{C}^*-$algebra $\A(\dc^\perp)$ with any $\dc\in\K$, 
where $\A(\dc^\perp)$ is the algebra associated with  
the causal complement of $\dc$ (see Section 
\ref{Aa}).
The \textit{stalk} in a point $x$ is 
the $\mathrm{C}^*-$algebra  
\begin{equation}
\label{Ca:0}
\A^\perp(x) \defi \big( \cup\{\A(\dc^\perp)  \ | \ 
		 x\in \dc \}\big)^{-\norm{ \ }}.
\end{equation}
Note that $\A^\perp(x)$  is also equal to the $\mathrm{C}^*-$algebra generated 
by  the algebras $\A(\dc)$ for  $\dc\in\K_x$. The correspondence 
\[
\Al_{\K_x}:\K_x\ni\dc\longrightarrow \A(\dc)\subseteq \A^\perp(x)
\]
is a net of local algebras over the poset $\K_x$.
By local definiteness and punctured Haag duality, it can be easily verified 
that the net $\Al_{\K_x}$ is irreducible and satisfies Haag duality. 
Furthermore, $\Al_{\K_x}$ inherits from $\Al_\K$ the Borchers property.
Now, let $\Zutx$  be the $\mathrm{C}^*-$category of the 1-cocycles of $\K_x$, 
trivial in $\Bh$, with values in $\Al_{\K_x}$. Observe that 
the category $\Zut$ is connected to $\Zutx$ by a covariant 
functor  defined as 
\begin{equation}
\label{Ca:1}
\begin{array}{rclcl}
 \Zut& \ni & z\longrightarrow z\upharpoonright 
   \Si_1(\K_x)& \in & \Zutx   \\
  (z,z_1) & \ni & t \longrightarrow 
 t\upharpoonright 
   \Si_0(\K_x) & \in & 
    (z\upharpoonright\Si_1(\K_x), z_1\upharpoonright \Si_1(\K_x)).
\end{array}
\end{equation}
This is a faithful functor that we call  \textit{the restriction functor} 
to $\K_x$. Then, the categories $\Zutx$ play the role of ``charts'' of 
$\Zut$. In the following we first prove the existence of a 
tensor product, a symmetry,  left inverses and conjugates in $\Zutx$.
Afterwards we will prove that all these constructions can be glued,
leading to corresponding notions on $\Zut$.\\[5pt]
\indent We now explain the reasons why we choose the categories 
$\Zutx$ as ``charts'' of $\Zut$. 
\textit{First}, because  studying $\Zutx$  is very similar 
to studying superselection sectors 
in Minkowski space \cite{Rob2}. As observed  the net 
$\Al_{\K_x}$ is irreducible and 
verifies the Borchers property and Haag duality.
Furthermore, the point $x$ plays for $\K_x$  the same role 
that the  spatial infinite plays for the set of double cones 
in the Minkowski space. In fact, $\K_x$ admits an asymptotically 
causally disjoint sequence of diamonds  ``converging'' to $x$ 
(see Section \ref{Cbb}).
\textit{Secondly}, by Proposition \ref{Bc:3}
the mentioned gluing procedure, that we now explain,  works well.\\[5pt]
\indent A collection $\{z_x\}_{x\in\M}$ of 1-cocycles 
$z_x\in\Zutx$  is said to be \textit{extendible to $\Zut$},  if 
there exists a  1-cocycle  $z$ of $\Zut$ 
such that $z\upharpoonright\Si_1(\K_x) = z_x$ for any  $x\in\M$. 
It is clear  that if there exists an extension, 
then it is  unique.
\begin{prop}
\label{Ca:2}
The collection $\{z_x\}_{x\in\M}$, where $z_x\in\Zutx$, 
is extendible to $\Zut$ if, and only if, 
for any $b\in\Si_1(\K)$ the relation 
\begin{equation}
\label{Ca:3}
z_{x_1}(b) = z_{x_2}(b) 
\end{equation}
is verified  for any pair  $x_1,x_2\in\M$  with 
$|b|\in \K_{x_1}\cap\K_{x_2}$. 
\end{prop}
\begin{proof}
The implication ($\Rightarrow$) is trivial.
($\Leftarrow$) For any  $b\in\Si_1(\K)$,  we define 
\[
z(b) \defi z_x(b) \mbox{ for some  } x\in\M \mbox{ with } |b|\in\K_x
\]
The definition does not depend on the chosen point $x$. 
Clearly $z(b)\in\A(|b|)$ because $z_x(b)\in\A(|b|)$. Furthermore,
given $c\in\Si_2(\K)$ let $x\in\M$ with $|c|\in\K_x$. Then 
$z(\partial_0c)\cdot  z(\partial_2c)$ 
$ = z_x(\partial_0c)\cdot z_x(\partial_2c)$ 
$= z_x(\partial_1c) = z(\partial_1c)$, 
showing that $z$ verifies the 1-cocycle identity. What remains to be shown
is that $z$ is trivial in $\Bh$. It is at  this point 
that Proposition \ref{Bc:3} intervenes in the proof. In fact, for any
$x\in\M$, $z$ is path-independent on $\K_x$  because
$z\upharpoonright \Si_0(\K_x) = z_x$ and $z_x$ is path-independent on
$\K_x$. Then, the proof follows 
from Proposition \ref{Bc:3}. 
\end{proof} 
An analogous notion of extendibility can be given for  arrows.
Consider   $z,z_1\in\Zut$. A collection 
$\{t_x\}_{x\in\M}$, where 
$t_x\in(z\upharpoonright\Si_1(\K_x),z_1\upharpoonright\Si_1(\K_x))$ 
in $\Zutx$, is said to be \textit{extendible} to $\Zut$ 
if there exists an arrow $t\in(z,z_1)$ such that 
$t\upharpoonright\Si_0(\K_x)= t_x$ for any $x\in\M$.
Also in this case, if the extension $t$ exists, then it is unique.
\begin{prop}
\label{Ca:4}
Let $z,z_1\in\Zut$. The  collection 
$\{t_x\}_{x\in\M}$, where 
$t_x\in(z\upharpoonright\Si_1(\K_x),z_1\upharpoonright
     \Si_1(\K_x))$, 
is  extendible to $\Zut$ if, and only if, 
for any $a\in\Si_0(\K)$ the relation 
\begin{equation}
\label{Ca:5}
(t_{x_1})_a= (t_{x_2})_a 
\end{equation}
is verified for any pair of points $x_1,x_2$ with $a\in\K_{x_1}\cap\K_{x_2}$.
\end{prop}
\begin{proof}
Also in this case the implication $(\Rightarrow)$ is trivial. 
$(\Leftarrow)$  For any $a\in\Si_0(\K)$ define 
\[
t_a \defi (t_{x})_a \mbox{ for some  } x\in\M \mbox{ with }a\in\K_x 
\]
The definition does not depend on the chosen point $x$. Clearly 
$t_a\in\A(a)$. Given $b\in\Si_1(\K)$, let us take $x\in\M$ with $|b|\in\K_x$. 
Then, $t_{\partial_0b}\cdot z(b)$ $ = (t_x)_{\partial_0b}\cdot z(b)$ 
$=z_1(b)\cdot (t_x)_{\partial_1b}$ $=z_1(b)\cdot t_{\partial_1b}$,
completing the proof.
\end{proof}
In the following we will refer to (\ref{Ca:3}) (\ref{Ca:5})
as \textit{gluing conditions}. 
\subsection{Local theory}
\label{Cb}
We begin  the study of superselection structure of 
the category  $\Zutx$. 
Our first aim is to show that to  1-cocycles
of $\Zutx$ there correspond  endomorphisms
of the algebra $\A^\perp(x)$ which are  localized and transportable,
in the sense of DHR analysis.
This is a key result for 
the local theory because will allow us to introduce in a very 
easy way the tensor product on $\Zutx$ and all the rest will proceed 
likewise to \cite{Rob2}.\\[5pt] 
\indent  The usual procedure used 
to define endomorphisms associated with 1-cocycles \cite{Rob2, GLRV, Rob3}, 
does not work in this case. This procedure leads to 
an  endomorphism  of the net $\Al_{\K_x}$, but it is not clear 
whether this is extendible to an endomorphism of $\A^\perp(x)$:
since $\K_x$ might not be  directed,  $\A^\perp(x)$ might not be  
the $\mathrm{C}^*-$inductive limit of $\Al_{\K_x}$.  
This problem can be overcome  by  applying, in a suitable way,  
a different procedure which makes use of the underlying 
presheaf structure \cite{Ruz1}.
Given $z\in\Zutx$, fix  $a\in\Si_0(\K_x)$. 
For any diamond $\dc\in\K$ with $x\in \dc$, define 
\begin{equation}
\label{Cb:1}
y^z_\dc(a)(A) \defi z(p)\cdot A\cdot z(p)^* \qquad A\in\A(\dc^\perp)
\end{equation}
where $p$ is path in $\K_x$ such that $\partial_1p\subset \dc$ and
$\partial_0p = a$. This definition does not depend on the path chosen 
and  on the choice of the starting point $\partial_1p$, as the following lemma
shows.
\begin{lemma}
Let $z\in\Zutx$ and let $\dc\in\K$ with $x\in\dc$. 
Let $p,q$ be two paths in $\K_x$ with $\partial_0p=\partial_0q$
and $\partial_1p,\partial_1q\subseteq \dc$. Then 
$z(p)\cdot A\cdot z(p)^*=z(q)\cdot A\cdot z(q)^*$ for any 
$A\in\A(\dc)$.
\end{lemma}
\begin{proof}
Note that $z(p)\cdot A\cdot z(p)^*=
z(q)\cdot z(\con{q}*p)\cdot A\cdot z(\con{q}*p)^*\cdot z(q)^*$,
for any $A\in\A(\dc^\perp)$.  
$\con{q}*p$ is a path in $\K_x$ whose endpoints are contained 
in $\dc$. This means that the endpoints of $\con{q}*p$ 
belong to $\K_x|_{\dc}$, see (\ref{Bba:4a}).  
As $\K_x|_{\dc}$ is pathwise connected, Lemma
\ref{Bba:4b},  we can find a path $q_1$ in $\K_x|_{\dc}$ with the same
endpoints of $\con{q}*p$. By path-independence we have 
that $z(\con{q}*p)=z(q_1)$. But $z(q_1)\subseteq\A(\dc)$ 
because the support $|q_1|$ is contained in $\dc$. Therefore 
$z(\con{q}*p)\cdot A = A\cdot z(\con{q}*p)$ for any
$A\in\A(\dc^\perp)$, completing the proof. 
\end{proof}
Therefore,  if we take $\dc_1\in\K$ with 
$x\in \dc_1\subseteq \dc$,  
then
$y^z_{\dc_1}(a)\upharpoonright \A(\dc^\perp)  = y^z_\dc(a)$.
This means that the collection 
\begin{equation}
\label{Cb:1b}
y^z(a)\defi \{y^z_\dc(a) \ | \ \dc\in\K, \ x\in \dc\}
\end{equation}
is a morphism of  the presheaf 
$\{\dc_1\in\K, \ x\in\dc_1\}\ni\dc\longrightarrow \A(\dc^\perp)$.
It then follows that 
$y^z(a)$ is extendible to an endomorphism of $\A^\perp(x)$ 
(see the definition of $\A^\perp(x)$ (\ref{Ca:0})).
\begin{lemma}
\label{Cb:2}
The following
properties hold:\\
(a) $y^z(a):\A^\perp(x)\longrightarrow \A^\perp(x)$ is a unital endomorphism;\\
(b) $y^z(a) \upharpoonright \A(a_1) = id_{\A(a_1)}$ 
    for any $a_1\in\Si_0(\K_x)$ with $a_1\perp a$;\\ 
(c) if $p$ is a path, then 
    $z(p)\cdot  y^z(\partial_1p)(A) =  y^z(\partial_0p)(A)\cdot z(p)$   
    for  $A\in\A^\perp(x)$;\\
(d) if $t\in(z,z_1)$, then $t_a \cdot y^z(a)(A) = y^{z_1}(a)(A)\cdot t_a$  
    for $A\in\A^\perp(x)$;\\
(e) $y^z(a)(\A(a_1))\subseteq \A(a_1)$ for any $a_1\in\K_x$ with 
     $a\subseteq a_1$. 
\end{lemma}
\begin{proof}
(a) is obvious from the Definition (\ref{Cb:1}). 
(b) Let $\dc\in\K$ with $x\in\dc$ and $\dc\perp a_1$. 
    Given  $A\in\A(a_1)$, it follows from the definition of $y^z(a)$
    that 
    $y^z(a)(A)= y^z_\dc(a)(A) = z(p)\cdot A\cdot z(p)^*$, 
    where $p$ is a path of $\K_x$ with
    $\partial_1p\subset\dc$,  $\partial_0p=a$. Hence 
    $\partial_1p, \partial_0p\perp a_1$. As the causal complement 
    of $a_1$ is pathwise connected 
    in $\K_x$ (Lemma \ref{Bba:4}), the proof follows by (\ref{Aa:2}).
   $(c)$ and $(d)$ follow by  routine calculations. (e) 
   follows by (b) because $\Al_{\K_x}$ fulfils Haag duality. 
\end{proof}
Note that $\{y^z(a) \ | \ a\in\Si_0(\K_x)\}$ is a collection 
of endomorphisms of the algebra $\A^\perp(x)$  
which are localized and transportable 
in the same sense of the DHR analysis:
Lemma \ref{Cb:2}b  says that $y^z(a)$ \textit{localized in} $a$; 
Lemma \ref{Cb:2}c  says that $y^z(a)$ is \textit{transportable} 
to any $a_1\in\Si_0(\K_x)$. 
\subsubsection{Tensor structure}
\label{Cba}
The tensor product on $\Zutx$ is defined by means 
of the localized and transportable endomorphisms of $\A^\perp(x)$
associated with 1-cocycles. To this end  some  preliminaries are 
necessary. Let 
\begin{equation}
\label{Cba:1}
\begin{array}{rll}
  z(p)\times z_1(q) & \defi z(p)\cdot y^z(\partial_1p)(z_1(q)), &
	p,q\mbox{ paths in } \K_x, \\
  t_a\times s_{a_1}  & \defi  t_a \cdot 
  y^z(a)(s_{a_1}),  &    a,a_1\in\Si_0(\K_x),
\end{array}
\end{equation}
for any $z,z_1,z_2,z_3\in\Zutx$, $t\in (z,z_2)$ and  $s\in(z_1,z_3)$.  
The tensor product in $\Zutx$, that we will define later, is a particular
case of $\times$.
\begin{lemma}
\label{Cba:2}
Let $z,z_1,z_2,z_3\in\Zutx$, and let $t\in (z,z_2)$, $s\in (z_1,z_3)$.
The following relations hold:\\
 \phantom{A} (a) \  $(t_{\partial_0p}\times s_{\partial_0q})\cdot 
       z(p)\times z_1(q)  = 
 z_2(p)\times z_3(q)\cdot (t_{\partial_1p}\times s_{\partial_1q})$;\\
  \phantom{A} (b) \  $z(p_2*p_1)\times z_1(q_2*q_1)  =  
     z(p_2)\times z_1(q_2)\cdot z(p_1)\times z_1(q_1)$,\\
for any   $p,q$, $p_2*p_1$, $q_2*q_1$ paths in $\K_x$.
\end{lemma}
\begin{proof}
(a) By using  Lemma \ref{Cb:2}c and 
Lemma \ref{Cb:2}d   we have 
\begin{align*}
t_{\partial_0p}\times s_{\partial_0q} \cdot 
         z(p)\times z_1(q) &  = t_{\partial_0p} \cdot y^z(\partial_0p) 
   (s_{\partial_0q})\cdot z(p) \cdot y^z(\partial_1p)(z_1(q))\\ 
  & =  y^{z_2}(\partial_0p)(s_{\partial_0q}) \cdot z_2(p)\cdot 
         t_{\partial_1p} \cdot 
      y^z(\partial_1p)(z_1(q)) \\
  & =  z_2(p)\cdot y^{z_2}(\partial_1p)(s_{\partial_0q}) \cdot  
      y^{z_2}(\partial_1p)(z_1(q)) \cdot t_{\partial_1p} \\
  & =  z_2(p)\cdot y^{z_2}(\partial_1p)(z_3(q))\cdot 
       y^{z_2}(\partial_1p)(s_{\partial_1q})  
       \cdot t_{\partial_1p} \\
  & =  z_2(p)\times z_3(q) \cdot 
        t_{\partial_1p} \cdot y^{z}(\partial_1p)(s_{\partial_1q})\\
  & =  z_2(p)\times z_3(q) \cdot t_{\partial_1p} \times s_{\partial_1q}.
\end{align*}
(b) By Lemma \ref{Cb:2}c, we have 
\begin{align*}
 z(p_2*p_1)\times z_1(q_2&*q_1)   =\\  
& =  z(p_2*p_1) \cdot  y^z(\partial_1p_1) (z_1(q_2*q_1))\\ 
 & =  z(p_2) \cdot z(p_1)  \cdot y^z(\partial_1p_1)(z_1(q_2)) 
     \cdot y^z(\partial_1p_1)( z_1(q_1))\\
 &  =  z(p_2) \cdot y^z(\partial_1p_2)(z_1(q_2)) \cdot 
        z(p_1)\cdot y^z(\partial_1p_1)(z_1(q_1)) \\
 &  = z(p_2) \times z_1(q_2) \cdot 
        z(p_1)\times z_1(q_1),
\end{align*}
where the  equality $\partial_0p_1=\partial_1p_2$ has been 
used. 
\end{proof}
We now are ready to introduce the tensor product. Let us define
\begin{equation}
\label{Cba:4}
\begin{array} {rcll}
  (z\otimes z_1)(b) & \defi &  z(b)\times z_1(b), &
	b\in\Si_1(\K_x), \\
  (t\otimes s)_a  & \defi &  t_a \times s_a  &  a\in\Si_0(\K_x),
\end{array}
\end{equation}
for any $z,z_1,z_2,z_3\in\Zutx$,  $t\in (z,z_1)$ and  $s\in(z_2,z_3)$. 
\begin{prop}
\label{Cba:5}
$\otimes$ is a tensor product in $\Zutx$. 
\end{prop}
\begin{proof}
First, we prove that if  $z,z_1\in\Zutx$, then 
$z\otimes z_1\in\Zutx$. By Lemma \ref{Cb:2}e
we have that $(z\otimes z_1)(b)\in\A(|b|)$.
Given $c\in\Si_2(\K_x)$, by applying Lemma \ref{Cba:2}b 
with respect to the path 
$\{\partial_0c,\partial_2c\}$ we have 
\[
(z\otimes z_1)(\partial_0c)\cdot  (z\otimes z_1)(\partial_2c)  = 
z(\partial_0c)\cdot z(\partial_2c) \times z_1(\partial_0c)\cdot z_1(\partial_2c) =(z\otimes z_1)(\partial_1c),
\]
proving that $z\otimes z_1$  satisfies the 1-cocycle identity. 
By Lemma \ref{Cba:2}b  it follows that 
$ (z\otimes z_1) (b_n)\cdots (z\otimes z_1) (b_1)$
$=  z(p)\cdot  y^z(\partial_1p)(z_1(p))$, 
for any path $p=\{b_n,\ldots,b_1\}$. 
Therefore as $z$ and $z_1$ are path-independent in $\K_x$,
$(z\otimes z_1)$ is path independent in $\K_x$. Namely,
$z\otimes z_1\in\Zutx$. If $t\in(z,z_2)$ and $s\in(z_1,z_3)$,
then by Lemma \ref{Cba:2}a  it follows that  
$t\otimes s\in (z\otimes z_1, z_2\otimes z_3)$. The rest of the properties 
that $\otimes$ has to satisfy to be a tensor product in $\Zutx$ 
can be easily checked.
\end{proof}
\subsubsection{Symmetry and Statistics}
\label{Cbb}
The following lemma is fundamental for the existence of a symmetry.
\begin{lemma}
\label{Cbb:1}
Let $p,q$ be a pair of paths in $\K_x$ with 
$\partial_ip\perp\partial_iq$  for  $i=0,1$. Then 
$z(p) \times z_1(q) = z_1(q) \times  z(p)$.
\end{lemma}
\begin{proof}
As $\K^\perp_x$ is pathwise connected (see \ref{Bba:5}) 
there are in $\K_x$ two paths $p_1 =\{b_{j_n}\ldots b_{j_1}\}$ and 
$q_1 =\{b_{k_n}\ldots b_{k_1}\}$  such that  
$|b_{j_i}|\perp|b_{k_i}|$ for $i=1,\ldots, n$
and  $\partial_1p_1=\partial_1p$, 
$\partial_0p_1=\partial_0p$ and 
$\partial_1q_1=\partial_1q$, $\partial_0q_1=\partial_0q$.
By path-independence  $z(p_1)=z(p)$ and $z_1(q_1)= z(q)$.
By Lemma \ref{Cba:2}b, we have 
\begin{align*}
z(p) \times  z_1(q) &    = z(p_1)  \times  z_1(q_1) \\
& =  z(b_{j_n}) \times z_1(b_{k_n})\cdot \cdots \cdot z(b_{j_1}) \times z_1(b_{k_1})\\ 
 & =   
 z(b_{j_n}) \cdot  y^z(\partial_1b_{j_n}) (z_1(b_{k_n})) 
 \cdot \cdots \cdot z(b_{j_1})\cdot y^z(\partial_1b_{j_1}) (z_1(b_{k_1}))\\ 
& =   z(b_{j_n})\cdot z_1(b_{k_n})\cdot \cdots \cdot 
            z(b_{j_1})\cdot z_1(b_{k_1})\\
& =   z_1(b_{k_n})\cdot z(b_{j_n})\cdot \cdots \cdot 
           z_1(b_{k_1})\cdot z(b_{j_1})\\
& = z_1(b_{k_n}) \cdot  y^{z_1}(\partial_1b_{k_n}) (z(b_{j_n})) 
 \cdot \cdots \cdot z_1(b_{k_1})\cdot y^{z_1}(\partial_1b_{k_1})(z(b_{j_1}))\\ 
& =  z_1(q_1) \times  z(p_1) = 
   z_1(q) \times z(p), 
\end{align*}
where the localization property of the endomorphisms 
$y^z(b_{j_i})$, $y^{z_1}(b_{k_i})$
has been used (Lemma \ref{Cb:2}b).  
\end{proof}
\begin{teo} 
\label{Cbb:2}
There exists a symmetry $\eps$ in $\Zutx$
defined as
\begin{equation}
\label{Cbb:2a}
\eps(z,z_1)_a = z_1(q)^*  \times z(p)^* \cdot z(p)  \times  z_1(q),
\qquad a\in\Si_0(\K_x) 
\end{equation}
where $p,q$ are two paths  with  
$\partial_0p\perp\partial_0 q$ and   
$\partial_1p=\partial_1 q=a$.
\end{teo}
\begin{proof}
First we prove that the r.h.s of (\ref{Cbb:2a}) is independent of
the choice of the paths $p,q$. So, let $p_1,q_1$ be two paths in $\K_x$ 
such that $\partial_1p_1=\partial_1q_1=a$
and $\partial_0p_1\perp \partial_0q_1$. 
Let  $q_2 \defi q*\con{q_1}$  and $p_2\defi p*\con{p_1}$. 
By Lemma \ref{Cba:2}b  we have 
\begin{align*}
  z_1(q)^* & \times   z(p)^* \cdot z(p)\times z_1(q)
     = \\ 
    & =  z_1(q*\con{q_1}*q_1)^* \times z(p*\con{p_1}*p_1)^* 
        \cdot z(p*\con{p_1}*p_1)\times z_1(q*\con{q_1}*q_1) \\
    & = (z_1(q_1)^*\cdot z_1(q_2)^* \times z(p_1)^*\cdot z(p_2)^*) 
              \cdot (z(p_2)\cdot z(p_1)\times z_1(q_2)\cdot z_1(q_1))\\
   & = z_1(q_1)^*\times z(p_1)^* \cdot z_1(q_2)^* \times z(p_2)^* 
              \cdot z(p_2)\times z_1(q_2) \cdot z(p_1)\times z_1(q_1).
\end{align*}
Note that $\partial_ip_2\perp\partial_iq_2$ for $i=0,1$. 
By Lemma \ref{Cbb:1}  we have that 
$z(p_2)\times z_1(q_2)= z_1(q_2)\times z(p_2)$. Therefore
\[
 z_1(q)^* \times z(p)^*  \cdot  z(p)\times z_1(q)
    =  z_1(q_1)^* \times z(p_1)^*
     \cdot z(p_1)\times z_1(q_1)
\]
which proves our claim. We now  prove that 
$\eps(z,z_1)\in (z\otimes z_1, z_1\otimes z)$. Let $b\in\Si_1(\K_x)$ 
and let $p,q$ be two paths with $\partial_1p=\partial_1q=\partial_0b$
and $\partial_0p\perp \partial_0q$. By Lemma \ref{Cba:2}b  we have 
\begin{align*}
\eps(z,z_1)_{\partial_0b}\cdot (z\otimes z_1)(b) & 
 =   z_1(q)^*  
  \times z(p)^*\cdot z(p) \times z_1(q)\cdot 
    (z\otimes z_1)(b) \\ 
& =  z_1(q)^*  \times z(p)^* \cdot 
     (z(p)\cdot z(b)  \times  z_1(q)\cdot z_1(b))\\
& = (z_1\otimes z)(b) \cdot  (z_1(q_1)^*  
  \times z(p_1)^* ) \cdot  (z(p_1)\times z_1(q_1))\\
& = (z_1\otimes z)(b)\cdot \eps(z,z_1)_{\partial_1b}
 \end{align*}
where $p_1= p*b$ and $q_1=q*b$ and it is trivial to check 
that $p_1$ and $q_1$ satisfy the properties written in the 
statement. Given $t\in(z,z_2)$, $s\in(z_1,z_3)$, and two paths 
$p,q$ with $\partial_1p=\partial_1q=a$, $\partial_0p\perp \partial_0q$, 
by Lemma \ref{Cba:2} we have 
\begin{align*}
\eps(z_2,z_3)_{a}\cdot (t\otimes s)_a  & 
 =    z_3(q)^*  \times z_2(p)^* \cdot  z_2(p) \times z_3(q) \cdot 
       (t\otimes s)_a  \\
 & =   z_3(q)^*  \times z_2(p)^* \cdot  (t_{\partial_0p}\times s_{\partial_0q})
        \cdot z(p)\times z_1(q)\\
 & =   z_3(q)^*  \times z_2(p)^* \cdot  
    (s_{\partial_0q}\times t_{\partial_0p})
                \cdot z(p)\times z_1(q)\\
 & =  (s_a\times t_a)\cdot 
  z_1(q)^*  \times z(p)^*       
                \cdot z(p)\times z_1(q)\\
& =   (s\otimes t)_a \cdot \eps(z,z_1)_{a}, 
\end{align*}
where $t_{\partial_0p}\times s_{\partial_0q} = 
t_{\partial_0p}\cdot y^z(\partial_0p)(s_{\partial_0q}) =
t_{\partial_0p}\cdot s_{\partial_0q} =
s_{\partial_0q} \cdot t_{\partial_0p} = 
s_{\partial_0q} \times t_{\partial_0p}$, because 
$\partial_0p\perp \partial_0q$. The rest of the properties 
can be easily checked. 
\end{proof}
Now, in order to classify the statistics of the irreducible elements
of $\Zutx$ we have to prove the existence of left inverses (see Appendix).  
To this end, consider a sequence $\{\dc_n\}_{n\in\mathbb{N}}$ 
of diamonds of $\K$ such that 
\[
 x\in\dc_n, \ \forall n\in\mathbb{N}, \ \dc_{n+1}\subsetneq \dc_n, \ 
 \cap_{n\in\mathbb{N}} \dc_n = \{x\}.
\]
For any $n$ let us take $a_n\in\Si_0(\K_x)$ such that 
$a_n\subset \dc_n$. We get in this way 
an \textit{asymptotically causally disjoint sequence} 
$\{a_n\}_{n\in\mathbb{N}}$:
for any $a\in\Si_0(\K_x)$ there exists $k(a)\in\mathbb{N}$ such that 
for any $n\geq k(a)$ we have 
$a_n\perp a$. This is enough to prove the existence of left inverses.  
Following \cite{GLRV, Rob3}, given $z\in\Zutx$ and $a\in\Si_0(\K_x)$, 
let $p_n$ be a path from $a$ to $a_n$. Let 
\begin{equation}
\label{Cbb:3}
\phi^z_a(A)\defi \lim_n  \ z(p_n)\cdot A\cdot z(p_n)^*, \qquad A\in\A^\perp(x),
\end{equation}
be a Banach-limit over $n$.
$\phi^z_a:\A^\perp(x)\longrightarrow \Bh$ 
is a positive linear map and,  it can be easily checked, that   
\begin{equation}
\label{Cbb:4}
\phi^z_a(A \cdot y^z(a)(B))  =  \phi^z_a(A) \cdot B, \qquad 
  A,B\in\A^\perp(x), 
\end{equation}
and that for any $b\in\Si_1(\K_x)$ we have
\begin{equation}
\label{Cbb:5}
\phi^z_{\partial_0b}(z(b)\cdot A\cdot z(b)^*)  =   \phi^z_{\partial_1b}(A) 
   \qquad  A\in\A^\perp(x)
\end{equation}
\begin{prop}
Given $z\in\Zutx$ and $t\in(z\otimes z_1,z\otimes z_2)$, let 
\begin{equation}
\phi^z_{z_1,z_2}(t)_a \defi \phi^z_a(t_a),  \qquad a\in\Si_0(\K_x),
\end{equation}
where $\phi^z_a$ is defined by 
(\ref{Cbb:3}). Then, the collection 
$\phi^z \defi \{\phi^z_{z_1,z_2} \ | \ z_1,z_2\in\Zutx \}$ 
is a left inverse of $z$.  
\end{prop}
\begin{proof}
Following \cite{Rob3}, by using (\ref{Cbb:4}) and (\ref{Cbb:5}) one can easily
show that 
$\phi^z_{z_1,z_2}(t)_{\partial_0b} \cdot z_1(b)$ 
$= z_2(b)\cdot \phi^z_{z_1,z_2}(t)_{\partial_1b}$
for  $t\in(z\otimes z_1,z\otimes z_2)$. Let 
$\dc\in\K_x$  with $\dc\perp a$, for any $B\in\A(\dc)$ we have that 
\begin{align*}
\phi^z_{z_1,z_2}(t)_a\cdot B  & = 
\phi^z_a(t_a)\cdot B =  \phi^z_a(t_a \cdot y^z(a)(B)) \\
 &  =  
 \phi^z_a(t_a \cdot B)    
  = \phi^z_a(B\cdot t_a) =  B\cdot \phi^z_a(t_a) = B\cdot 
\phi^z_{z_1,z_2}(t)_a.
\end{align*}
Hence $\phi^z_{z_1,z_2}(t)_a\in\A(a)$ because 
$\Al_{\K_x}$ satisfies Haag duality. This entails that 
$\phi^z_{z_1,z_2}(t)\in (z_1,z_2)$. The other 
properties of left inverses can be easily checked (see \cite{Rob3}).
\end{proof}
An object of $\Zutx$ is said to have \textit{finite statistics} 
if it admits a standard left inverse, namely a left inverse $\phi^z$
such that 
\[
 \phi^z_{z,z}(\eps(z,z))\cdot \phi^z_{z,z}(\eps(z,z)) = c\cdot \mathbbm{1} 
   \mbox{ with } c > 0
\] 
In the opposite case $z$ is said to have  \textit{infinite statistics}. 
The type of the statistics  is an invariant of the 
equivalence class of objects. 
Let $\Zutx_\f$ be the full subcategory of $\Zutx$ whose objects have
finite statistics. 
$\Zutx_\f$ is closed under direct sums, subobjects and  tensor products. 
Furthermore, any object of $\Zutx_\f$ is a finite direct sums
of irreducible objects. From now on we focus on $\Zutx_\f$, because 
the finiteness of the statistics is a necessary condition for the existence
of conjugates (see Appendix \ref{X}). 
\subsubsection{Conjugation} 
\label{Cbc}
The proof of  the existence of conjugates in $\Zutx_\f$ is equivalent 
to  proving  that any simple object 
has conjugates (see Appendix \ref{X}). Recall that an object $z\in\Zutx_\f$ 
is said to be \textit{simple} whenever 
\[
\eps(z,z) = \chi(z)\cdot 1_{z\otimes z}, \mbox{ where }\chi(z)\in\{1,-1\}.
\] 
Simplicity is a property of the equivalence class, and 
it turns out to be equivalent to that fact that $z^{\otimes_n}$ is 
irreducible for any $n\in\mathbb{N}$, where $z^{\otimes_n}$ is the n-fold
tensor product of $z$. 
\begin{lemma}
\label{Cbc:3}
Let $z$ be  a simple object. Then
$\chi(z) \cdot z(b)$ $=  y^z (\partial_0b)(z(b))$ 
$= y^z (\partial_1b)(z(b))$,
for any 1-simplex $b$ with  $\partial_1b\perp\partial_0b$.
\end{lemma}
\begin{proof}
Consider  the  1-simplex  
$b(\partial_1b)$ degenerate to $\partial_1b$ and recall 
that any 1-cocycle  evaluated on 
a degenerate 1-simplex is equal to $\mathbbm{1}$, Lemma \ref{Ac:2}a. 
By the defining relation  of $\eps$, (\ref{Cbb:2a}),
we have
\[
  \eps(z,z)_{\partial_1b}
  = z(b(\partial_1b))^* \times z(b)^*   \cdot 
    z(b)\times z(b(\partial_1b))  =                       
  y^z (\partial_1b)(z(b)^*)\cdot  z(b),
\]
Since $\chi(z) \cdot \mathbbm{1}= \eps(z,z)_{\partial_1b}$, we have 
$\chi(z) \cdot z(b) = y^z (\partial_1b)(z(b))$. 
The other identity follows by replacing, in this  reasoning,
$b$ by $\con{b}$.
\end{proof}
\begin{prop}
\label{Cbc:4}
Let $z$  be a simple object. Then,
$y^z(a):\A^\perp(x) \longrightarrow \A^\perp(x)$ is an automorphism,   
for any $a\in\Si_0(\K_x)$.
\end{prop}
\begin{proof}
Let $\dc\in\K$ with $x\in \dc$  and $\dc \perp a$.
As the causal complement of $a$ in  $\K_x$ is pathwise 
connected, Lemma \ref{Bba:4}, there is a path $q$ of the  form
$b*p$, where $b$ is a 1-simplex 
such that $\partial_0b=a$ and $\partial_1b\perp a$; $p$ is a path 
satisfying 
\[
\partial_1p\subset \dc, \ \ \partial_1p\perp x, \ \ \partial_0p = \partial_1b,
\ \ |p|\perp a
\]
Now, observe that by Lemma \ref{Cb:2}b we have that 
$y^z(a)(z(p))=z(p)$ and that 
$y^z(\partial_1p)(A)= A$  for any $A\in\A(\dc^\perp)$.
By using these relations and the previous lemma, 
for any $A\in\A(\dc^\perp)$ we have 
\begin{align*}
   y^z (a)& (A)  =
    z(q) \cdot y^z(\partial_1p)(A) \cdot z^*(q) 
   = z(b)\cdot z(p) \cdot  A \cdot z^*(p)\cdot z(b)^*  \\ 
  & = z(b)\cdot y^z(a)(z(p)) \cdot  A \cdot y^z(a)(z(p)^*)\cdot z(b)^*  \\ 
  & =  \chi(z)\cdot y^z(a)(z(b))\cdot y^z(a)(z(p)) \cdot A \cdot 
      y^z(a)(z(p)^*) \cdot   \chi(z)\cdot y^z(a)(z(b)^*)\\
  &  = y^z(a)(z(b)\cdot z(p)) \cdot A \cdot  
        y^z(a)((z(b)\cdot z(p))^*)
\end{align*}
That is $y^z(a)\big(z(q)^* \cdot A\cdot z(q)\big) $ $= A$ for any 
$A\in\A(\dc^\perp)$. This means that 
$\A^\perp(x)\subseteq y^z(a)(\A^\perp(x))$,  
that entails that $y^z(a)$ is an automorphism of $\A^\perp(x)$. 
\end{proof}
Assume that $z$ is a simple object of $\Zutx$. Let us denote 
by $y^{z-1}(a)$ the inverse of $y^{z}(a)$. Clearly,
$y^{z-1}(a)$ is an automorphism of $\A^\perp(x)$ localized in $a$. 
Let
\begin{equation}
\label{Cbc:5}
\con{z}(b)  \defi   y^{z-1}(\partial_0b)(z(b)^*), \qquad  b\in\Si_1(\K_x).
\end{equation}
We claim that $\con{z}$ is the conjugate object of $z$. The proof
is achieved in two steps.
\begin{lemma}
\label{Cbc:7}
Let $z$ be a simple object. Then 
$\con{z}(p)$ $=  y^{z-1}(\partial_0p)(z(p)^*)$
$=y^{z-1}(\partial_1p)(z(p)^*)$,
for any path $p$ in $\K_x$.
\end{lemma}
\begin{proof}
Within this proof, to save space,    we will omit 
the superscript $z$ from $y^z(a)$ and $y^{z-1}(a)$.
First we prove the  relations written above in the case that $p$ is a 
1-simplex $b$. For any $A\in\A^\perp(x)$ 
we have
\begin{align*}
\con{z}(b)\cdot   y^{-1}(& \partial_1b  )(A)  
      = y^{-1}(\partial_0b)(z(b)^*)\cdot  y^{-1}(\partial_1b)(A) \\
   &  =   y^{-1}(\partial_0b) \big( z(b)^*  \cdot  
     y(\partial_0b)\big( y^{-1}(\partial_1b)(A)\big)\big)\\
  &  =  y^{-1}(\partial_0b) \big( 
      y(\partial_1b)\big( y^{-1}(\partial_1b)(A)\big)  \cdot
     z(b)^* \big) 
     =  y^{-1}(\partial_0b)  ( A\cdot z(b)^*)  \\
   &  = y^{-1}(\partial_0b) ( A) \cdot \con{z}(b)
\end{align*}
Using this  relation we obtain
$\con{z}(b)\cdot  y^{-1}(\partial_1b) (z(b))$
  $= y^{-1}(\partial_0b) (z(b))\cdot \con{z}(b)$ 
  $=  y^{-1}(\partial_0b) (z(b)) 
  \cdot y^{-1}(\partial_0b)(z(b)^*)$ $  = \mathbbm{1}$,
completing the first part of the proof. We now proceed by  induction: 
let $p=\{b_n,\ldots,b_1\}$ and assume 
that the statement holds for the path $q=\{b_{n-1},\ldots,b_1\}$, then
\begin{align*}
\con{z}(p) & = \con{z}(b_n) \cdots \con{z}(b_1)  
        = \con{z}(b_n)\cdot y^{-1}(\partial_0q)
             ( z(q)^*)\\ 
       & =  y^{-1}(\partial_1q)
             (z(q)^*) \cdot \con{z}(b_n) = 
            y^{-1}(\partial_0b_n)
             ( z(q)^*) \cdot \con{z}(b_n) \\
       & =  y^{-1}(\partial_0b_n)
             ( z(q)^*) \cdot  y^{-1}(\partial_0b_n)
             ( z(b_n)^*)
         = y^{-1}(\partial_0p)( z(q)^*\cdot  z(b_n)^*)\\ 
       & =  y^{-1}(\partial_0 p)(z(p)^*).
\end{align*}
The other relation is obtained in a similar way. 
\end{proof}
\begin{lemma}
\label{Cbc:8}
Let $z$ be a simple object of $\Zutx$. 
Then $\con{z}\in\Zutx$ and  is a conjugate object of $z$. 
\end{lemma}
\begin{proof}
By Lemma \ref{Cb:2}e we have that $\con{z}(b)\in\A(|b|)$
for any $b\in\Si_1(\K_x)$. Let $c\in\Si_2(\K_x)$, then 
\begin{align*}
\con{z}(\partial_0c) & \cdot \con{z}(\partial_2c)  =
 \  y^{z-1}(\partial_{00}c) (z(\partial_0c)^*) \cdot
    y^{z-1}(\partial_{02}c) (z(\partial_2c)^*) \\
  & =  y^{z-1}(\partial_{00}c) 
    \big( \ z(\partial_0c)^*  \cdot
  y^z(\partial_{00}c)(
  y^{z-1}(\partial_{02}c) (z(\partial_2c)^*)) \ 
 \big)\\
  & =  y^{z-1}(\partial_{00}c) \big( \ 
  y^z(\partial_{10}c)( 
  y^{z-1}(\partial_{02}c) (z(\partial_2c)^*)) \cdot  z(\partial_0c)^* \ 
 \big)\\
  & =  y^{z-1}(\partial_{00}c) \big(
             z(\partial_2c)^*  \cdot  z(\partial_0c)^* \big)
   = y^{z-1}(\partial_{00}c) (z(\partial_1c)^*) \\
  & = y^{z-1}(\partial_{01}c) (z(\partial_1c)^*)
  = \con{z}(\partial_1c) 
\end{align*}
Where the relations $\partial_{00}c= \partial_{01}c$
$\partial_{10}c= \partial_{02}c$ have been used. Finally by 
Lemma \ref{Cbc:7} $\con{z}$ is path-independent in $\K_x$ because 
$z$ is path-independent in $\K_x$.  Therefore, 
$\con{z}$ is trivial in $\Bh$, thus  is an object
of $\Zutx$. Now we have to prove that $\con{z}$ is the conjugate
object of $z$ (see definition in Appendix). We need a preliminary 
observation. Let $y^{\con{z}}(a)$ be the endomorphisms 
of $\A^\perp(x)$ associated with $\con{z}$. Then 
\begin{equation}
\label{Cbc:9}
y^{\con{z}}(a) =   y^{z-1}(a), \qquad \forall a\in\Si_0(\K_x).
\end{equation}
In fact, let $\dc\in\K$ with $x\in\dc$ and $\dc\perp a$. 
Let $p$ be path in $\K_x$ with $\partial_1p\subset \dc$
and $\partial_0p=a$.
For any $A\in\A(\dc^\perp)$ we have
$y^{\con{z}}(a)(A)$  $= \con{z}(p)\cdot A \cdot  \con{z}(p)^*$ 
   $=  \con{z}(p)\cdot  y^{z-1}(\partial_1p)(A) \cdot \con{z}(p)^*$
   $=   y^{z-1}(a)(A)$,
which proves (\ref{Cbc:9}). Now, by (\ref{Cbc:9})  
and by Lemma \ref{Cbc:7} we have
\[
\begin{array}{l}
(z\otimes \con{z})(b)   =  
 z(b)\cdot 
 y^z(\partial_1b) ( y^{z-1}(\partial_1b)(z(b)^*)) 
  =  z(b)\cdot z(b)^* = \mathbbm{1} \\
(\con{z}\otimes z)(b)   =  
\con{z}(b)\cdot 
 y^{\con{z}}(\partial_1b)\big( z(b)\big) = 
  y^{z-1}(\partial_1b)(z(b)^*)\cdot  y^{z-1}(\partial_1b)(z(b)) = 
\mathbbm{1}
\end{array}
\]
So, if we take  $r=\con{r}=\mathbbm{1}$, then 
$r$ and $\con{r}$ satisfy  the conjugate  equations for 
$z$ and $\con{z}$, completing the proof. 
\end{proof}
According to the discussion made at the beginning of this section 
we have 
\begin{teo}
\label{Cbc:11}
Any object of $\Zutx_{\f}$  has conjugates.
\end{teo}
\subsection{Global theory} 
\label{Cc}
We now turn back to study $\Zut$. The aim of this section 
is to show that all the constructions we have made in the 
categories $\Zutx$ can be glued together and 
extended to  corresponding constructions on $\Zut$.\\[5pt] 
\indent Given $z\in\Zut$ let us denote by $y^z_x(a)$ 
the morphism of the algebra $\A^\perp(x)$ associated with the restriction 
$z\upharpoonright \Si_1(\K_x) \in\Zutx$ (\ref{Cb:1b}). For any  
$a\in\Si_0(\K)$   we define 
\begin{equation}
\label{Cc:1}
y^{z}(a) \defi \{y^z_x(a) \ | \ x\in\M, \mbox{ with }  a\in\K_x  \}.
\end{equation}
We call $y^z(a)$ a \textit{morphism of stalks} 
because it is compatible with the presheaf structure, 
that is given $\dc\in\K$, for any pair of points
$x,x_1\in\dc$ we have
$y^z_x(a)\upharpoonright \A(\dc^\perp) = 
y^z_{x_1}(a)\upharpoonright \A(\dc^\perp)$.
This is an easy consequence of the following 
\begin{lemma}[Gluing Lemma]
\label{Cc:2}
Let $z\in\Zut$ and $\dc\in\K$. Then 
\begin{equation}
\label{Cc:3}
y^z_{x_1}(a)\upharpoonright \A(\dc) = y^z_{x_2}(a)\upharpoonright \A(\dc), 
\end{equation}
for any pair $x_1,x_2\in\M$ with $\dc\in\K_{x_1}\cap\K_{x_2}$.
Let  $p$ be a path in $\K$ for which  there exist a pair of points
$x_1,x_2\in\M$ with $|p|\subset\K_{x_1}\cap \K_{x_2}$. Then 
$y^z_{x_1}(a)(z(p))= y^z_{x_2}(a)(z(p))$. 
\end{lemma}
\begin{proof}
By (\ref{Cb:1}) and 
(\ref{Cb:1b}), for $A\in\A(\dc)$ we have that 
$y^z_{x_i}(a)(A)  = z(p_i)\cdot A\cdot z(p_i)^*$, for $i=1,2$,
where $p_i$ is a path in $\K_{x_i}$ such that $\partial_0p_i=a$,
$\partial_1p_i=a_i$ and $a_i$ is contained in some diamond
$\dc_i$ such that $x_i\in\dc_i$ and $\dc_i\perp \dc$ for i=1,2.
Note that $\con{p_2}*p_1$ is a path from $a_1$ to  $a_2$ and that 
$a_1,a_2\perp \dc$. By (\ref{Aa:2}) we have 
\begin{align*}
y^z_{x_1}(a)(A) & = z(p_1)\cdot A\cdot z(p_1)^*\\ 
     & = z(p_2)\cdot z(\con{p_2}*p_1)\cdot A\cdot z(\con{p_2}*p_1)^* \cdot
        z(p_2)^* \\
    &  = z(p_2)\cdot A\cdot  z(p_2)^* 
     = y^z_{x_2}(a)(A) 
\end{align*}
for any $A\in\A(\dc)$, which proves  (\ref{Cc:3}). Now, let 
$p$ and $x_1,x_2$ be as in the statement.
By applying (\ref{Cc:3}) we have 
\begin{align*}
y^z_{x_1}(a)(z(p)) & = y^z_{x_1}(a)(z(b_n))\cdots y^z_{x_1}(a)(z(b_1))\\
 & = y^z_{x_2}(a)(z(b_n))\cdots y^z_{x_2}(a)(z(b_1)) 
  = y^z_{x_2}(a)(z(p)),
\end{align*} 
completing the proof. 
\end{proof}
We have called this lemma \textit{the gluing lemma}  because it will allow 
us to extend  to $\Zut$ the constructions that we have made on the 
categories $\Zutx$. As a first application 
of this fact, we define the tensor product on $\Zut$.
For any $z,z_1\in\Zut$ and for any pair $t,s$ of arrows of $\Zut$ we define
\begin{equation}
\begin{array}{ll}
\label{Cc:4}
(z\otimes z_1)(b)  \defi  (z\otimes_x {z_1})(b),&   \qquad b\in\Si_1(\K),\\
(t\otimes s)_a\defi   (t\otimes_{x_1}  s)_a  & \qquad a\in\Si_0(\K) 
\end{array}
\end{equation}
for some $x\in\M$ with $|b|\in\K_x$  and for some $x_1\in\M$ with 
$a\in\K_{x_1}$, where $\otimes_x$ is the tensor product in  $\Zutx$. 
\begin{lemma}
\label{Cc:5}
$\otimes$ is a tensor product on $\Zut$. 
\end{lemma}
\begin{proof}
By the gluing lemma 
we have $(z\otimes_x z_1)(b)$ 
$=z(b)\cdot y^z_x(\partial_1b)(z_1(b))$ 
$= z(b)\cdot y^z_{x_1}(\partial_1b)(z_1(b))$ $=(z\otimes_{x_1} z_1)(b)$,
for any pair of points $x,x_1$ with 
$|b|\in\K_{x}\cap\K_{x_1}$. Therefore, by Proposition \ref{Ca:2} 
we have that $(z\otimes z_1)\in \Zut$. Now, let $t\in(z,z_2)$ and 
let $s\in(z_1,z_3)$. Note that  the gluing lemma 
entails that $(t\otimes_{x_1}  s)_a$ $= t_a \cdot y^z_x(a)(s_a)$ 
$=t_a \cdot y^z_{x_1}(a)(s_a)$ for any pair of points $x,x_1$ 
with $a\in\K_{x}\cap\K_{x_1}$.  
By Proposition \ref{Ca:4} we have that  
$t\otimes s\in (z\otimes z_1, z_2\otimes z_3)$. 
The other properties of the tensor product 
can be easily checked. 
\end{proof}
\begin{oss}
\label{Cc:5a}
As an easy consequence of Lemma \ref{Cc:5}, we have that 
the tensor product $\widehat{\otimes}$ 
introduced in \cite{GLRV} is well defined 
in $\Zut$: namely, $z\widehat{\otimes} z_1 \in\Zut$ (see Introduction). 
It is enough to observe that the restriction 
of $y^z_x(a)$ to the algebras $\A(\dc)$ with $\dc\perp x$ and 
$a\subseteq \dc$, is equal to the morphism of the net 
associated with $z$,  introduced in that paper, and used 
to define the tensor product. This entails that  
$z\widehat{\otimes} z_1 = z\otimes z_1$,
where $\otimes$ is the tensor product (\ref{Cc:4}).
\end{oss}
We conclude this section, by 
generalizing, to an arbitrary globally hyperbolic spacetime,
\cite[Theorem 30.2]{Rob3} which holds for globally hyperbolic
spacetimes with noncompact Cauchy surfaces.
\begin{prop}
\label{Cc:6}
The restriction functor (\ref{Ca:1}) is a full and faithful tensor functor.
\end{prop}
\begin{proof}
It is clear that the restriction functor is a faithful tensor functor.
So, we have to prove that this functor is full. 
To begin with, recall the construction made in  \cite[Theorem 30.2]{Rob3}. 
Let $t_{x_0}$ be an element of  
$(z\upharpoonright\Si_1(\K_{x_0}) , 
z_1\upharpoonright\Si_1(\K_{x_0}))$ in $\mathcal{Z}^1_t(\Al_{\K_{x_0}})$.
For $a\in\Si_0(\K)$, we define
\begin{equation}
 t_a\defi z_1(p)\cdot (t_{x_0})_{a_0}\cdot z(p)^*
\end{equation} 
where $a_0$ is a 0-simplex in $\K_{x_0}$ and $p$ is path from 
$a_0$ to $a$. The definition does not depend on the chosen path
non on the choice of $a_0$ in $\K_{x_0}$. This entails that 
$t_a= (t_{x_0})_a$ for any $a\in\Si_0(\K_{x_0})$. Furthermore,
given $b\in\Si_1(\K)$ and a path $p$ from $a_0$ to $\partial_0b$
we have
\begin{align*}
t_{\partial_0b}\cdot z(b)& 
= z_1(p)\cdot (t_{x_0})_{a_0}\cdot z(p)^*\cdot z(b) \\
  &   = z_1(b) \cdot z_1(p_1)\cdot (t_{x_0})_{a_0}\cdot z(p_1)^* = 
        z_1(b)\cdot  t_{\partial_1b},
\end{align*}
where $p_1= \con{b}*p$. 
So, what remains to be proved
is that $t_{a}\in\A(a)$ for any $a\in\Si_0(\K)$.
Our proof starts from this point. First we prove that 
if $x_1\perp x_0$, then  $t_a\in\A(a)$ for $a\in\Si_0(\K_{x_1})$. Take 
$a_1\in\Si_0(\K_{x_1})$ with  $a_1\perp a$.  
Since  $\Si_0(\K_{x_1})$ admits an asymptotically
causally disjoint sequence (Section \ref{Cbb}), 
and since  $x_0\perp x_1$, we can find $a_2\in \K_{x_1}\cap \K_{x_0}$ 
with $a_2\perp a_1$.  Therefore, 
$t_a = z_1(p_1)\cdot (t_{x_0})_{a_2}\cdot z(p_1)^*$.
where $p_1$ is a path from $a_2$ to $a$. Note that  $a_2$ and $a$
belong to the causal complement $a_1^\perp|_{\K_{x_1}}$  
of $a_1$ in $\K_{x_1}$ and that $a_1^\perp|_{\K_{x_1}}$  
is pathwise connected in $\K_{x_1}$ (Lemma \ref{Bba:4}). 
Since $z,z_1$ are path-independent,
we can assume
that $p_1$ is contained in  $a_1^\perp|_{\K_{x_1}}$.
Hence for any $A\in\A(a_1)$ we have 
\begin{align*}
t_a\cdot A & = z_1(p_1)\cdot (t_{x_0})_{a_2}\cdot z(p_1)^*\cdot A 
    = z_1(p_1)\cdot (t_{x_0})_{a_2}\cdot A\cdot z(p_1)^* \\ 
   & = z_1(p_1)\cdot A\cdot (t_{x_0})_{a_2}\cdot z(p_1)^* = A\cdot t_a.
\end{align*}
Namely $t_a\in \A(a_1)'$ for any $a_1\in\K_{x_1}$ such that $a_1\perp a$.
Since $\Al_{\K_{x_1}}$ satisfies Haag duality we have $t_a\in\A(a)$. 
Now, if $x_n$ is a generic point of $\M$, observe that 
we can find a finite sequence of points 
$x_1,\ldots,x_{n-1}$ such that 
$x_0\perp x_1, \ x_1\perp x_2,\ldots, x_{n-1}\perp x_n$.
From this observation 
the proof of the fullness of the restriction functor follows.
\end{proof} 
Given $z\in\Zut$, 
we know that if  $z\upharpoonright\Si_1(\K_x)$ is irreducible in $\Zutx$ 
for some $x\in\M$, then $z$ is irreducible. 
The converse is an easy consequence of Proposition \ref{Cc:6}, 
namely if $z$ is irreducible, then $z\upharpoonright\Si_1(\K_x)$ 
is irreducible in $\Zutx$ for any $x\in\M$. Finally, note that 
Proposition \ref{Cc:6} is a strengthening of Proposition \ref{Ca:4}.   
\subsubsection{Symmetry,  statistics and conjugation}
\label{Cca}
We conclude our analysis of $\Zut$. First we prove the existence
of a symmetry in $\Zut$. Afterwards, we prove the existence of 
left inverses and define the category of objects with 
finite statistics. Finally, we prove that this category has conjugates.\\[5pt]
\indent Let $\eps_x$ denote  the symmetry of the category $\Zutx$.  
\begin{lemma}
\label{Cca:1}
There exists a unique symmetry $\eps$ in $\Zut$ such that
given  $z,z_1\in\Zut$ and $a\in\Si_0(\K)$,  then 
$\eps(z,z_1)_a = \eps_x(z, z_1)_{a}$ 
for any $x\in\M$ with $x\perp a$.
\end{lemma}
\begin{proof}
Let $z,z_1\in\Zut$. For any $a\in\Si_0(\K)$, define 
\begin{equation}
\label{Cca:2}
\eps(z,z_1)_a \defi \eps_x(z, z_1)_{a} 
\end{equation}
for some $x\in\M$ with $x\perp a$. The uniqueness follows 
once we have shown that (\ref{Cca:2}) defines
a symmetry in $\Zut$. To this 
end we prove that (\ref{Cca:2}) is  independent of the 
chosen $x$. Let $x_1\in\M$ with $x_1\perp a$. 
This means that  $a$ is contained in the open set 
$\M\setminus (\J(x) \cup  \J(x_1))$. 
There exists a 1-simplex $b$  with 
$cl(|b|)\subset\M\setminus (\J(x) \cup \J(x_1))$ (this is equivalent 
to $|b|\in\K_x\cap\K_{x_1}$)  and  $\partial_1b=a$ 
and $\partial_0b\perp a$ (see observation below the Lemma \ref{Bb:6}). 
Note that the paths $b$ and 
$b(a)$ satisfy  the assumptions
in the definition (\ref{Cbb:2a}). By the gluing lemma we have 
\begin{align*}
\eps_x(z,z_1)_{a} & = 
z_1(b(a))^*  \times_x z(b)^* \cdot z(b)  \times_x z_1(b(a)) \\
& = z_1(b(a))^*  \times_{x_1} z(b)^* \cdot z(b)  \times_{x_1} z_1(b(a))
= \eps_{x_1}(z,z_1)_a
\end{align*}
which proves our claim. By Proposition \ref{Ca:4} we have that 
$\eps(z,z_1)\in (z\otimes z_1, z_1\otimes z)$.  
The remaining properties can be easily checked.
\end{proof}
We now turn to prove the existence of left inverses in $\Zut$.
Let $\phi^z_x$ be a left inverse of the restriction 
$z\upharpoonright\Si_1(\K_x)$ in $\Zutx$  for $x\in\M$.
For any  $t\in (z\otimes z_1, z\otimes z_2)$, we define
\begin{equation}
\label{Cca:2a}
  \phi^z_{z_1,z_2}(t)_a \defi z_2(p)\cdot  
  (\phi^z_x)_{z_1, z_2}(t)_{a_0}\cdot z_1(p)^*,\qquad a\in\Si_0(\K), 
\end{equation}
where $a_0\in\Si_0(\K_x)$ and $p$ is a path from 
$a_0$ to $a$. By the same argument used in  Proposition \ref{Cc:6} we  
have that   $\phi^z_{z_1,z_2}(t)\in (z_1, z_2)$ for any 
$t\in (z\otimes z_1, z\otimes z_2)$. Furthermore, 
as $\phi^z_x$ is a left inverse of $z\upharpoonright \Si_1(\K_x)$, 
one can easily check that $\phi^z$ is a left inverse of $z$. Therefore any 
element of $\Zut$ has left inverses.
\begin{prop}
\label{Cca:3} 
 Let $z\in\Zut$. \\
 (a) If $z$ has finite statistics, then 
    $z\upharpoonright \Si_1(\K_x)$ has finite 
    statistics in $\Zutx$ for any $x\in\M$, and if  $z$ is irreducible,  
    then $\la(z) = \la_x(z)$ for any  $x\in\M$, where 
    $\la(z)$ and $\la_x(z)$ are the statistics parameters 
    of $z$ and $z\upharpoonright \Si_1(\K_x)$ respectively 
    (see Appendix \ref{X}).\\
 (b) If for some $x\in\M$, 
    $z\upharpoonright \Si_1(\K_x)$ has finite 
    statistics in $\Zutx$, then $z$ has finite  statistics, and, 
    if $z\upharpoonright \Si_1(\K_x)$ is irreducible, then 
    $\la(z) = \la_x(z)$. 
\end{prop}
\begin{proof}
(a) If $z$ has finite statistics, then $z$ admits a standard left 
inverse $\phi^z$. Clearly $\phi^z$  is a left inverse also for 
the restriction $z\upharpoonright \Si_1(\K_x)$ for any $x\in\M$. 
Because of Lemma \ref{Cca:1} 
$\eps(z,z)_a = \eps_x(z,z)_a$ for $a\in\Si_0(\K_x)$. Hence
\[
\phi^z_{z,z}(\eps(z,z))_a = \phi^z_{z,z}(\eps_x(z,z))_a, \qquad (*)
\] 
which entails that $\phi^z$ is a standard left inverse of 
$z\upharpoonright \Si_1(\K_x)$ for any $x\in\M$. 
If $z$ is irreducible, then, by Proposition \ref{Cc:6}, 
$z\upharpoonright \Si_1(\K_x)$ is irreducible
in $\Zutx$ for any $x\in\M$.  By $(*)$ we have 
$\la(z)=\la_x(z)$ for any $x\in\M$. (b) 
Let $\phi^z_x$ be  a  left inverse of  $z\upharpoonright \Si_1(\K_x)$,
and let $\phi^z$ the left inverse of $z$, associated with 
$\phi^z_x$,  defined by (\ref{Cca:2a}). 
Let 
$\phi^z_x$ be  a standard left inverse of  $z\upharpoonright \Si_1(\K_x)$,
and let $\phi^z$ be  the left inverse of $z$ defined by (\ref{Cca:2a}). 
Then 
\[
 \phi^z_{z,z}(\eps(z,z))_a =   z(p)\cdot  
 (\phi^z_x)_{z,z}(\eps_x(z,z))_{a_0}\cdot z(p)^*, \qquad (**)
\]
which implies that $\phi^z$ is a standard left inverse of $z$. 
If $z\upharpoonright\Si_1(\K_x)$ is irreducible, then 
$z$ is irreducible. Therefore by 
$(**)$ we have that $\la(z) = \la_x(z)$, completing the proof. 
\end{proof}
Let $\Zut_\f$ be the full subcategory 
of $\Zut$ whose objects have finite statistics.
\begin{teo}
\label{Cca:4}
Any object $\Zut_\f$ has conjugates.
\end{teo}
\begin{proof}
As can be seen from  Appendix \ref{X}, it is sufficient to prove that 
the theorem holds in the case of simple objects.
Thus, let $z$ be a simple object of $\Zut$. By 
Proposition \ref{Cca:3}, any restriction $z\upharpoonright\Si_1(\K_x)$ is a 
simple object. Let $y^z(a)= \{ y^z_x(a) \ | \ x\in\M, \ a\in\K_x\}$
be the morphism of stalks associated with $z$. $y^z(a)$ 
is in fact an  automorphism of stalks, because by  
Proposition \ref{Cbc:4}
any $y^z_x(a)$ is an automorphism of $\A^\perp(x)$. Clearly 
also the inverse $y^{z-1}_x(a)$  of $y^{z}_x(a)$ is an automorphism 
of $\A^\perp(x)$. Given $a\in\Si_0(\K)$, let 
\[
y^{z-1}(a)\defi \{ y^{z-1}_x(a) \ | \ x\in\M, \ a\in\K_x\}.
\]
(From this definition, up to this moment, we can assert neither 
that the gluing 
lemma is applicable  to $y^{z-1}(a)$ nor that $y^{z-1}(a)$ is an 
automorphism of stalks. Both these properties will be 
a consequence of the fact that $y^{z-1}(a)= y^{\con{z}}(a)$,
where $\con{z}$ will be defined below). Now, we prove 
that for $y^{z-1}(a)$ a weaker 
form of the gluing lemma holds, namely 
given $\dc\in\K$ with $a\subseteq \dc$ we have 
\[
y^{z-1}_{x_1}(a)\upharpoonright \A(\dc) =  
    y^{z-1}_{x_2}(a)\upharpoonright \A(\dc) \qquad (*)
\]
for any pair of points $x_1,x_2$ with $\dc\in\K_{x_1}\cap \K_{x_2}$.
In fact, by using  
the gluing lemma for  $y^z(a)$ we have 
\begin{align*}
y^{z-1}_{x_1}(a)(A) &  
 =y^{z-1}_{x_1}(a)(y^{z}_{x_2}(a)( y^{z-1}_{x_2}(a)(A)))\\
   &   =  y^{z-1}_{x_1}(a)(y^{z}_{x_1}(a)( y^{z-1}_{x_2}(a)(A)))
               =  y^{z-1}_{x_2}(a)(A)
\end{align*}
for any $A\in\A(\dc)$, which proves $(*)$. Within this proof 
we have used the identities: 
$y^z_x(a)(\A(\dc))=\A(\dc)$,  
$y^{z-1}_x(a)(\A(\dc))=\A(\dc)$ for any $\dc\in\K_x$ with 
$a\subseteq \dc$. Both the  identities derive from the Lemma \ref{Cb:2}e, 
and from the fact that $y^z_x(a)$ is an automorphism of $\A^\perp(x)$.
Now, recall that the conjugate 
$\con{z}_x$  of  $z\upharpoonright \Si_1(\K_x)$ in $\Zutx$
is defined as $\con{z}_x(b)={y^z_x}^{-1}(\partial_0b)(z(b)^*)$ 
(\ref{Cbc:5}). Given $b\in\Si_1(\K)$, by applying $(*)$ we have that  
\[
\con{z}_{x_1}(b) = y^{z-1}_{x_1}(\partial_0b)(z(b)^*) 
= y^{z-1}_{x_2}(\partial_0b)(z(b)^*) =\con{z}_{x_2}(b)
\]
for any pair of points  $x_1,x_2$ with $|b|\in\K_{x_1}\cap \K_{x_2}$. 
Therefore, by defining 
\[
\con{z}(b)\defi \con{z}_x(b)\qquad b\in\Si_1(\K)
\]
for some point $x$ with $|b|\in\K_x$,
by Proposition \ref{Ca:2} we have that $\con{z}\in\Zut$. Furthermore, by  
(\ref{Cbc:9}) we have $y^{z-1}(a) = y^{\con{z}}(a)$, where
$y^{\con{z}}(a)$ is the morphism of stalks associated with 
$\con{z}$. To prove that $\con{z}$ is the conjugate of $z$ it is enough to 
observe that for any $b\in\Si_1(\K)$ we have 
\[
(\con{z}\otimes z)(b) =  (\con{z}_x \otimes_x z_x)(b) 
 = \mathbbm{1}, \ \ \   (z\otimes \con{z})(b) =  (z_x\otimes_x \con{z}_x)(b) 
 = \mathbbm{1}
\]
(see within the proof of Lemma \ref{Cbc:8}) for some $x\in\M$ with 
$|b|\in\K_x$. By defining 
$r=\con{r}=\mathbbm{1}$ we have that $r$ and $\con{r}$ satisfy 
the conjugate equations for $z$ and $\con{z}$, completing the proof.
\end{proof}
\section{Concluding remarks}

(1) The topology of the spacetime affects the net-cohomology of posets. 
    We have shown  that the poset, used as index set 
    of a net of local algebras, is nondirected when  the spacetime 
    is either nonsimply connected or has compact Cauchy surfaces.
    In the former case, furthermore, there might exist  
    1-cocycles which are nontrivial in $\Bh$. 
    In spite of these facts the structure of superselection sectors of 
    DHR-type is the same as in the case of the 
    Minkowski space (as one can expect because of the sharp 
                 localization): 
    sectors define a $\mathrm{C}^*-$category  in 
    which the charge structure manifests itself by 
    the existence of a tensor product,
    a symmetry, and a conjugation. 
    An aspect of the theory,  not covered by 
    this paper, and that deserves further investigation 
    is the reconstruction of the net of local fields
    and of the gauge group from the net of local observables  and 
    the superselection sectors. The mathematical machinery 
    developed in \cite{DR} to prove the reconstruction theorem 
    in the Minkowski space does not apply  as it stands 
    when the index set of the net of local observables 
    is nondirected. \\[3pt] 
    (2) In Section \ref{A} we presented net-cohomology 
    in terms of abstract posets. The intention is to provide 
    a general framework  for the theory of 
    superselection sectors. In particular, we also 
    hope to find applications in the study of sectors which might 
    be induced by the nontrivial topology of spacetimes. 
    It  has been shown in \cite{AS} that  the topology of Schwartzschild  
    spacetime, a space whose second homotopy group is nontrivial, 
    might induce superselection sectors. However,  as observed earlier, 
    it is not possible, up until now, to apply the ideas of 
    DHR-analysis to these sectors since their localization properties 
    are not known. However the results obtained in this paper allow 
    us  to make some speculations in the case that the spacetime is 
    nonsimply connected: the existence of 1-cocycles 
    nontrivial in $\Bh$, 
    might be related to the existence of superselection sectors 
    induced by the nontrivial topology of the spacetime.  In fact, these
    cocycles define  nontrivial representations of the fundamental group 
    of the spacetime (theorems  \ref{Ac:4} and \ref{Ada:6}). However, 
    what is missing in this interpretation  is the proof that these 
    1-cocycles are associated with representations of the net of
    local observables. We foresee  to approach  this problem 
    in the future. 
    Finally, we believe that this framework could be suitably 
    generalized for applications  in the context of the generally locally 
    covariant quantum field theories \cite{BFV}, \cite{BR}.  \\[3pt] 
\noindent (3)  Some  techniques introduced in this paper 
present analogies with techniques adopted  to  study  
superselection sectors of conformally covariant 
theories on the circle $S^1$.
In these theories, the spacetime is the circle $S^1$;
the index set for the net of local observables 
is the set   $\mathcal{J}$ of the open 
intervals of $S^1$; the causal disjointness relation is 
the disjointness: given $I,J\in\mathcal{J}$, 
then,  $I\perp J$  if $I\cap J=\emptyset$. 
The analogies arise because,  referring to Section \ref{A}, the poset formed 
by $\mathcal{J}$ with the inclusion 
order relation, is nondirected, pathwise connected, and  
nonsimply connected. It is usual in these theories 
to restrict the study of 
superselection sectors to the spacetime $S^1/\{x\}$ 
for $x\in S^1$, i.e.  the causal puncture of $S^1$ in $x$  
(see for instance \cite{BMT, FRS, GL}); the same idea has been 
used in \cite{Bau} to study superselection sectors over 
compact spaces\footnote{We stress that in the papers \cite{Bau,BMT,FRS,GL},
the authors puncture the spacetime in order 
to obtain a directed set of indices.
Our aim is different since $\K_x$ is in general nondirected.
Indeed, $\K_x$ has an asymptotically 
causally disjoint sequence of diamonds ``converging to $x$'' 
(see Section \ref{Cbb}) which is sufficient for the analysis of 
 the categories $\Zutx$.}.
The punctured Haag duality is strictly related to strong 
additivity (see \cite{KLM} and references therein).  
Finally, in \cite{FRS} in order 
to prove that endomorphisms of the net are extendible to  
the universal $\mathrm{C}^*-$algebra, the authors'  
need to check the invariance 
of these extensions for homotopic paths 
(this definition of a homotopy of paths
is a particular case of that given in \cite{Rob3} p.322).\\[3pt]
 \noindent (4) The way we define the first homotopy group 
of a poset is very similar to some constructions in algebraic 
topology. We are referring to 
the edge paths group of a simplicial complex \cite{ST} and to 
the first homotopy group of a Kan complex \cite{May}. 
Although similar they are different. Indeed, 
the simplicial set $\Si_*(\Po)$ of a poset $\Po$ 
is not  a simplicial complex. Furthermore, if $\Po$ is not directed, 
then $\Si_*(\Po)$ is  not a  Kan complex. 
\appendix
\numberwithin{equation}{section}

\section{Tensor $\mathrm{C}^*-$categories}
\label{X}
We give some basics definitions and results on tensor 
$\mathrm{C}^*$-categories. References for this appendix 
are \cite{Mac, LR}.\\[8pt]
Let $\mathcal{C}$ be a category. 
We denote by $z,z_1,z_2,\ldots $  the objects of the category
and the set of the arrows between $z,z_1$ by $(z,z_1)$.
The composition of arrows is indicated by ``$\cdot$'' and
the unit arrow of $z$ by $1_z$. \\[5pt]
\textbf{Tensor $\boldsymbol{\mathrm{C}^*}-$categories - }
A category $\mathcal{C}$ is said to be a $\mathrm{C}^*$-category if
the set of the arrows between two objects $(z,z_1)$ is a complex
Banach space and the composition between arrows is bilinear;
there should be an adjoint, that is an involutive contravariant functor $*$
acting as the  identity on the objects and  the norm should satisfy the
$\mathrm{C}^*$-property,
namely  $\norm{r^{*} r} \ = \ \norm{r}^2$ for each
$r\in(z,z_1)$. Notice, that if $\mathcal{C}$ is a $\mathrm{C}^*$-category
then $(z,z)$ is a $\mathrm{C}^*$-algebra for
each $z$. \\
\indent Assume that $\mathcal{C}$ is  a $\mathrm{C}^*$-category. An arrow
$v\in(z,z_1)$ is said to be  an isometry if $v^* \cdot v=1_z$;
a unitary, if it is an isometry and $v\cdot v^*=1_{z_1}$.
The property of  admitting  a unitary arrow, defines an equivalence
relation on the set of the objects of the category. We denote
by the symbol $[z]$ the unitary equivalence class of the object
$z$. An object $z$ is said to be irreducible if
$(z,z)=\mathbb{C}\cdot 1_z$. $\mathcal{C}$ is said to be closed
under  subobjects if for each orthogonal projection
$e\in(z,z)$, $e\ne 0$ there exists an isometry
$v\in(z_1,z)$ such that $v\cdot v^{*}  =  e$.
$\mathcal{C}$ is said to be closed under  direct sums,
if given $z_i \ i=1,2 $ there exists an
object $z$ and two isometries $w_i\in(z_i,z)$  such that
$w_1  \cdot w^{*}_1  +  w_2 \cdot w^{*}_2 \ = \ 1_z$. \\
\indent A \textit{strict tensor $\mathrm{C}^*$-category}
(or \textit{tensor $\mathrm{C}^*$-category}) is a
$\mathrm{C}^*$-category $\mathcal{C}$ equipped with a tensor product,
namely  an associative bifunctor
$\otimes : \mathcal{C}\times\mathcal{C}\longrightarrow\mathcal{C}$
with a unit $\io$, commuting with $*$, bilinear on the arrows
and satisfying the exchange property, i.e.
$(t\otimes s)\cdot (t_1\otimes s_1)  =  t\cdot t_1 \otimes s\cdot s_1$
when the composition of the arrows is defined.\\[5pt]
\indent From now on, we assume that
$\mathcal{C}$ is a tensor $\mathrm{C}^*$-category
closed under direct sums, subobjects, and that 
the identity object $\io$ is irreducible. \\[5pt]
\textbf{Symmetry and left inverses - } 
A {\em symmetry} $\eps$ in the tensor $\mathrm{C}^*$-category
$\mathcal{C}$ is a map
$\mathcal{C}\ni z_1,z_2\longrightarrow\eps(z_1,z_2)\in
(z_1\otimes z_2, z_2\otimes z_1)$
satisfying the relations:
\[
\begin{array}{rlrl}
(i)   & \eps(z_3,z_4)\cdot t\otimes s \ = \ s\otimes t\cdot\eps(z_1,z_2)\\
(ii)  &   \eps(z_1,z_2)^* \ = \ \eps(z_2,z_1)  \\
(iii) &  \eps(z_1,z_2\otimes z) \ = \ 1_{z_2}\otimes\eps(z_1,z)\cdot
                     \eps(z_1,z_2)\otimes 1_{z}  \qquad\qquad\qquad\qquad \\
(iv) &   \eps(z_1,z_2)\cdot\eps(z_2,z_1) \ = \ 1_{z_2\otimes z_1}
\end{array}
\]
where $t\in(z_2,z_4), s\in(z_1,z_3)$.
By $ii)- iv)$ it follows that
$\eps(z,\io)=\eps(\io,z) = 1_z$ for any $z$. In this paper by the
{\em left inverse} of an object $z$ is we mean a set of nonzero linear maps
$\phi^z  =  \{ \phi^z_{z_1,z_2}:
 (z\otimes z_1, z\otimes z_2)\longrightarrow(z_1,z_2) \}$ satisfying
\[
\begin{array}{rl}
(i) &  \phi^z_{z_3,z_4}(1_z\otimes t \cdot r \cdot 1_z\otimes s^*) =
  t\cdot  \phi^z_{z_1,z_2}(r)\cdot s^*, \qquad \qquad \qquad \\
(ii) &   \phi^z_{z_1\otimes z_3, z_2\otimes z_3} (r\otimes 1_{z_3})   =
   \phi^z_{z_1,z_2}(r) \otimes 1_{z_3},\\
(iii) &   \phi^z_{z_1,z_1}(s_1^*\cdot s_1) \geq 0,\\
(iv)  &  \phi^z_{\io,\io}(1_z) =\mathbbm{1}, 
\end{array}
\]
where $t\in(z_1,z_3)$, $s\in(z_2,z_4)$,  
$r\in(z\otimes z_1, z\otimes z_2)$ and $s_1\in (z\otimes z_1, z\otimes z_1)$.
\textbf{Statistics - }
From now on we assume that 
$\mathcal{C}$ has a symmetry $\eps$ and that any object of $\mathcal{C}$
has left inverses. An object $z$ of $\mathcal{C}$ is said to have 
\textit{finite} statistics if it admits a \textit{standard left inverse}, 
that is a left inverse $\phi^z$ satisfying the relation   
\[
\phi^z_{z,z}(\eps(z,z))\cdot \phi^z_{z,z}(\eps(z,z)) = c\cdot 1_z 
 \mbox{ with } c>0
\]
The full subcategory $\mathcal{C}_\f$ of $\mathcal{C}$ whose objects 
have finite statistics, is closed under direct sums, subobjects,
tensor products, and equivalence. Any object of $\mathcal{C}_\f$ is 
direct sums of irreducible objects. Given  an irreducible 
object $z$ of $\C_\f$ and a left inverse $\phi^z$ of $z$, we have 
\[
\phi^z_{z,z}(\eps(z,z)) = \la(z)\cdot 1_z 
\]
It turns out  that $\la(z)$ is an invariant of the 
equivalence class of $z$, called 
the \textit{statistics parameter}, and it is the product of two invariants: 
\[
\la(z) = \chi(z)\cdot d(z)^{-1} \ \mbox{ where } \ \chi(z)\in\{1,-1\}, \ \ 
  d(z)\in\mathbb{N}
\] 
The possible statistics of $z$ are classified by the 
\textit{statistical phase} $\chi(z)$ distinguishing para-Bose $(1)$ 
and para-Fermi $(-1)$ statistics and by the \textit{statistical dimension}
$d(z)$ giving the order of the parastatistics. Ordinary 
Bose and Fermi statistics correspond to $d(z)=1$. The objects 
with $d(z)=1$ are called \textit{simple objects}. The following
properties are equivalent (\cite{Ruz1}):
\[
z \mbox{ is simple }  \ \iff  \ \eps(z,z)= \chi(z) \cdot 1_{z\otimes z} \ 
 \iff \ z^{\otimes_n} \mbox{ is irreducible  } \forall n\in\mathbb{N}.
\]
\textbf{Conjugation - } 
An object $z$ has \textit{conjugates} if there exists 
an object $\con{z}$ and a pair of arrows $r\in(\io,\con{z}\otimes z)$, 
$\con{r}\in(\io, z\otimes\con{z})$  satisfying the \textit{conjugate 
equations} 
\[
\con{r}^*\otimes 1_z\cdot 1_z\otimes r = 1_z, \ \ 
r^*\otimes 1_{\con{z}}\cdot 1_{\con{z}}\otimes \con{r} = 1_{\con{z}}.
\]
Conjugation is a property stable under, subobjects, direct sums, tensor
products and, furthermore, it is stable under equivalence.  
It turns out  that 
\[
 z \mbox{ has conjugates} \Rightarrow  z \mbox{ has finite statistics}. 
\]
The full subcategory of objects with finite statistics $\C_\f$
has conjugates if, and only if, each object with statistical dimension 
equal to one has conjugates (see \cite{DHR2, Kun}).  
First we observe that if each 
irreducible object of $\C_\f$ has conjugates,
then any object of $\C_\f$ has conjugates, because any object of $\C_\f$
is a finite direct sum of irreducibles, and because conjugation 
is stable under direct sums. Secondly, note that if $z$ 
is an irreducible object with statistical dimension $d(z)$, then  
there exist a pair of isometries 
$v\in(z_0,z^{\otimes_{d(z)}})$ and $w\in(z_1,z^{\otimes_{d(z)-1}})$ 
where $z_0$ is a  simple object. 
Assume given $\con{z_0}$ and a pair of arrows $s,\con{s}$ which solve 
the conjugate equations for $z_0$ and $\con{z_0}$. 
Let $\phi^z$ is a standard left inverses of $z$. Setting
\[
\begin{array}{rl}
 \con{z} & \defi z_1 \otimes \con{z_0};\\
 \con{r} &\defi  d(z)^{1/2}\cdot (1_{z}\otimes w^*\otimes 1_{\con{z_0}}) \cdot 
    v\otimes 1_{\con{z_0}}\cdot \con{s};\\
  r & \defi d(z)\cdot  \phi^z_{\io,\con{z}\otimes z}(\con{r}\otimes 1_z),
\end{array}
\]
one can easily show that $r,\con{r}$ solve the conjugate equations for 
$z$ and $\con{z}$ (\cite{DHR2}).\\[8pt] 
{\small \textbf{Acknowledgements.} I would like to thank John E. Roberts
  for his constant support throughout this work, and 
  Daniele Guido for fruitful discussions and helpful suggestions. 
  The present paper has been improved by comments and suggestions 
  of the anonymous referees: I am grateful with them.
  I wish to thank the II Institute of Theoretical Physics
  of the University of Hamburg for the kind 
  hospitality, 12-19, January 2004, and for fruitful discussions 
  on the topics studied in this paper.   
  Finally, I would like to thank my family for their support 
  before and during this work.}


\end{document}